\documentclass[graybox]{svmult}

\usepackage{verbatim}
\usepackage{type1cm}        
\usepackage{makeidx}         
\usepackage{graphicx}        
\usepackage{multicol}        
\usepackage[bottom]{footmisc}

\usepackage{tikz}
\usetikzlibrary{shapes,arrows}
\usetikzlibrary{intersections}
\usetikzlibrary{mindmap}

\usepackage{newtxtext}       %
\usepackage[varvw]{newtxmath}       

\usepackage[inline]{enumitem}
\usepackage{hyperref}
\usepackage{nicematrix}
\usepackage{geometry}
\usepackage{caption}
\usepackage{subcaption}
\usepackage{siunitx}
\usepackage{booktabs}
\usepackage{bm}
\usepackage{mathtools}
\usepackage{gensymb}
\usepackage[ruled,vlined]{algorithm2e}
\usepackage{tikz}
\usepackage{xcolor}
\usepackage{longtable}
\usepackage[table]{xcolor}
\usepackage{multirow}

\usepackage{pgfplots}

\usetikzlibrary{external}

\makeindex             

\begin{document}

\title*{RIS-Aided Localization and Sensing}
\author{Dimitris Kompostiotis\orcidID{0009-0001-2589-806X} and\\ Dimitris Vordonis \orcidID{0000-0002-4093-0494} and\\ Konstantinos D. Katsanos\orcidID{0000-0003-1894-5216} and\\
Florin-Catalin Grec\orcidID{0009-0002-2944-9613} and\\
Vassilis Paliouras\orcidID{0000-0002-1414-7500} and\\ George C. Alexandropoulos\orcidID{0000-0002-6587-1371}}
\institute{Dimitris Kompostiotis \at Electrical and Computer Engineering Department, University of Patras, Panepistimiopolis Rio-Patras, 26504 Patras, Greece, \email{d.kompostiotis@ac.upatras.gr}
\and 
Dimitris Vordonis \at Electrical and Computer Engineering Department, University of Patras, Panepistimiopolis Rio-Patras, 26504 Patras, Greece, \email{d.vordonis@ac.upatras.gr}
\and Konstantinos D. Katsanos \at Department of Informatics and Telecommunications, National and Kapodistrian University of Athens, Panepistimiopolis Ilissia, 16122 Athens, Greece, \email{kkatsan@di.uoa.gr}
\and Florin-Catalin Grec \at European Space Agency (ESA), Keplerlaan 1, 2201AZ Noordwijk, Netherlands, \email{florin-catalin.grec@esa.int}
\and
Vassilis Paliouras \at Electrical and Computer Engineering Department, University of Patras, Panepistimiopolis Rio-Patras, 26504 Patras, Greece, \email{paliuras@upatras.gr}
\and George C. Alexandropoulos \at Department of Informatics and Telecommunications, National and Kapodistrian University of Athens, Panepistimiopolis Ilissia, 16122 Athens, Greece \email{alexandg@di.uoa.gr}}
%
%
\maketitle
\abstract{High-precision localization and environmental sensing are essential for a new wave of applications, ranging from industrial automation and autonomous systems to augmented reality and remote healthcare. Conventional wireless methods, however, often face limitations in accuracy, reliability, and coverage, especially in complex non-line-of-sight (NLoS) environments. Reconfigurable Intelligent Surfaces (RISs) have emerged as a key enabling technology, offering dynamic control over the radio propagation environment to overcome these challenges. This chapter provides a comprehensive overview of RIS-aided localization and sensing, bridging fundamental theory with practical implementation. The core principles of the RIS technology are first described detailing how programmable metasurfaces can intelligently combat blockages, enhance signal diversity, and create virtual line-of-sight (LoS) links. The chapter then reviews a range of application scenarios where RISs can offer significant improvements. A significant portion of the chapter is dedicated to algorithmic methodologies, covering beam sweeping protocols, codebook-based techniques, and advanced optimization and machine learning strategies for both localization and sensing. To validate the theoretical concepts in real-world conditions, recent experimental results using an RIS prototype are detailed, showcasing the technology's efficacy and illustrating key performance trade-offs.}

\section{Introduction}
\label{sec:Chapter_Outline}
Future wireless systems are envisioned to provide advanced localization and sensing services in addition to ultra-high data rate communications tasks. RISs~\cite{8741198} constitute an emerging  low-cost hardware technology that enables dynamic manipulation of the radio propagation environment through intelligently reflecting, refracting, and/or scattering incident electromagnetic waves~\cite{9765815}. By properly deploying RISs in a wireless environment of interest~\cite{APK23}, metasurfaces can either enhance/boost or even enable localization and sensing, an advantage that stems from their capability to add extra degrees of freedom in the system by controlling reflections. Based on the operating mode and hardware architecture~\cite{10596064}, an RIS can be utilized to provide favorable scattering conditions by enabling virtual LoS paths, performing beamforming toward desired positions, or providing diversity gains in the signal. For example, a receiving RIS~\cite{9053976,R-RIS} can serve as an extra anchor point in the reference system that may act either as a receiver of reference signals or even as a transmitter of localization beacons~\cite{10230036}. Although such an architecture, and similar ones (e.g., ~\cite{10352433,SORIS}), comes with increased manufacturing costs and power consumption compared to purely reflective RIS, the receiving or transmitting capabilities may be endowed only to a limited yet sufficient number of elements, or they may be deactivated for the most part of the operation. 

According to the relevant literature~\cite{APK23}, the RIS technology has been predominantly considered for the following use cases and scenarios: 
\begin{itemize}
    \item \textbf{Combating LoS blockages via reflection links:} Regardless of the localization system used, the estimation error is affected largely by the signal-to-noise (SNR) of the beacon signals. As a result, in cases where the Base Station (BS) - User Equipment (UE) link (for bistatic/monostatic cases) is attenuated due to severe blockages, it is reasonable to expect degraded localization performance. Therefore, by deploying RISs to enhance signal strength between the system's nodes~\cite{RDK21, RDK22}, performance gains may be obtained even without configuring the RIS to specifically operate to serve localization objectives. In the cases where the blockage areas can be known a priori, even statically configured reflectors may be deployed (an example for subTHz frequencies is discussed in~\cite{11078147}). On the downside, the calculations involved by each system may need to take into consideration the fact that the signal arrives via a reflected path (and thus longer, and with different observed angle of arrival) to the end nodes.
    \item \textbf{Diversity gains through beam management:} Popular localization, mapping, or Simultaneous Localization and Mapping (SLAM) approaches may rely on channel estimation or, more generally, require large and diverse samples of receive signals. Exploiting the reconfiguration capabilities of reflecting RISs, such measurement diversity may be provided at low cost and with short latency to the underlying algorithmic procedures~\cite{zheng2022survey, ZSA21}, and diversity gains may be more pronounced when the RIS is endowed with large number of unit elements. The diversity gains may be further enhanced if algorithmic beam management procedures are implemented to intelligently control the RIS phase shifts. Radio-localization approaches commonly involve the use of purposely-designed directive beams that scan the target area to localize the user or map passive objects. The quality of such beams depends on the available hardware capabilities and, predominantly, on the number of antennas. Since RISs are comprised of numerous phase shifting elements, they are capable of constructing narrow and accurate beamforming patterns, and can thus aid in such endeavors. Specifically, both transmitting and reflecting RISs may be controlled to perform beam sweeping or hierarchical codebook practices to direct reference signals toward different geographical sectors, while receiving RISs may be used to determine the direction of the incoming pilots. 
    \item \textbf{Self-localization in the absence of access points:} The RIS technology is able to offer new deployment paradigms for self-localization without the need for widely scaled infrastructure. Concretely, when the positions of multiple RISs are known to the UE at the time of operation, beacon signals may be transmitted toward the surfaces, whose reflected components may be collected back at the device to determine information such as Angle of Arrival (AoA) and Doppler shifts~\cite{HHM22}. By utilizing multiple surfaces conveniently placed around the target area, the user is able to recover their position. This methodology concurrently offers increased privacy levels, as pilot signals are not received by any other entities, apart from the participating user.
    \item \textbf{Receiving reference signals in the absence of anchor points:} Similarly to the previous case, BS-absent self-localization systems may be implemented when the installed RISs are endowed with sensing capabilities. In such cases, multiple collection points exist in the system, which may offer increased performance and lower overhead in terms of signal exchanging. It is also possible for a monostatic setup to be implemented where a single RISs with a limited number of partially connected Radio-frequency Chains (RFCs) is used as a receiver~\cite{HFA23}, so that different sub-regions of the metasurface are using different receive configurations simultaneously, playing the role of multiple ``virtual'' RISs.
    \item \textbf{Leveraging near-field characteristics:} The near-field region, which depends on the wavelength and the RIS's size, enables novel methods to enhance localization and mapping capabilities. When a UE is within the near-field region of a RIS, conventional far-field assumptions no longer apply, and its precise position, along with the positions of scatterers, can be determined by taking advantage of the wavefront curvature~\cite{WHD20,RKI23,BDS21,PJW22,KDA23}. This remains possible even when there is an obstruction that affects the direct path between the UE and the RIS~\cite{WHD20,KDA23}. 
    \item \textbf{Energy efficient operations:} RIS technology operates predominantly through passive reflection of signals, which inherently consumes minimal energy. Unlike conventional techniques that rely on powered relay nodes or active signal generation, RIS achieves its purpose without a significant energy footprint~\cite{8741198}. This translates to longer operational duration and reduced operational costs.
    \item \textbf{Reduced interference:} One of the distinguishing features of RIS is its reduced interference footprint. As it passively modulates the wireless environment, RIS ensures minimal disruption to other communication systems operating in proximity. This is a stark contrast to traditional methods which might inadvertently cause interference, thereby affecting the integrity of neighboring networks~\cite{9693982}. 
    \item \textbf{Dynamic environment configurations:} The inherent adaptability of RIS allows it to dynamically configure the environment to optimize signal propagation paths. This is crucial in ever-changing urban landscapes where building structures, vehicular movement, and other obstacles constantly alter the wireless communication landscape. Traditional methods often struggle to adapt in real-time, making RIS a game changer in this domain. 
    \item \textbf{Scalability:} The RIS technology is not limited to specific applications or environments. Its modular nature ensures that it can easily integrate with a plethora of applications, ranging from indoor positioning to large-scale urban mapping. Furthermore, as demands increase, RIS systems can be seamlessly scaled, ensuring consistent performance.
    \item \textbf{Cost effectiveness:} From an infrastructural standpoint, RIS also offers cost benefits. Traditional localization methods might require multiple antennas, extensive infrastructure, or even powered relay nodes. RIS, with its passive operation and adaptive capabilities, reduces the need for such extensive setups, leading to cost savings in both installation and maintenance.
\end{itemize}

This chapter commences with a summary of the latest discussions on localization and sensing the use cases in the framework of sixth Generation (6G) of wireless networks. Then, the fundamental principles of RISs and their relevance to localization and sensing tasks are discussed. Techniques and algorithms that leverage RISs to enhance positioning accuracy, improve signal coverage, and reduce multipath-induced errors are presented. Passive sensing paradigms are examined, with an emphasis on how RISs can support real-time scene reconstruction, device-free localization, and robust tracking under NLoS conditions. In addition, the chapter addresses critical implementation challenges, including synchronization, channel estimation, power constraints, and hardware limitations. It also discusses system-level trade-offs between accuracy, energy consumption, latency, and cost. Finally, the chapter concludes with an outlook on future research directions, highlighting the need for standardized RIS-based localization frameworks and the integration of machine learning to adapt to dynamic and heterogeneous environments.

\subsection{Use Cases for Localization and Sensing}
The upcoming 6G of wireless mobile networks aim to enable intelligent wireless interactions ubiquitously, spanning people-to-people-, people-to-machine-, and machine-to-machine-type connections; a surge of innovative Internet of Things (IoT) applications has emerged~\cite{3gppTS872,wymeersch2022localisation,10463676}. These include advancements in healthcare, smart cities, precision agriculture, process automation, modular and flexible assembly lines, education, security, entertainment, and immersive communications, all contributing to the vision of universal connectivity. However, the stringent demands of these applications, particularly in terms of localization and sensing accuracy, present major challenges for next generation wireless networks. It thus becomes essential to identify specific applications and use cases that rely on even more precise localization and sensing. And consequently, a structured gap analysis is needed: first, to determine whether a gap exists between the current capabilities and the future requirements of these applications; and second, to define the nature and root causes of this gap. The outcome of such an analysis should finally propose feasible solutions to bridge this disparity between present and future technological states.

\begin{figure}[t]
    \centering
    \begin{tikzpicture}[
    mindmap,
    concept color = gray!20, 
    every node/.style = {concept, text centered}, 
    grow cyclic,
    level 1/.append style = {
        level distance = 4cm,
        sibling angle = 45
    },
    level 2/.append style = {
        level distance = 3.5cm,
        sibling angle = 30}
]
\node {Use Cases of\\ RIS-Aided Localization and Sensing}
    child {node [concept color = blue!20] {e-Health}
    }
    child {node [concept color = green!30] {Industrial}
            }
    child {node [concept color = pink!30] {Smart Cities}
            }
    child {node [concept color = orange!30] {Aerial}
            }
    child {node [concept color = magenta!30] {Education}
            }
    child {node [concept color = olive!30] {Environment}
            }
     child {node [concept color = lime!30] {Entertainment and Gaming}
            }
    child {node [concept color = purple!30] {Security}
            };
\end{tikzpicture}
\vspace{1cm}
    \caption{Indicative use cases benefiting from RIS-aided localization and sensing.}
    \label{fig:Level1_apps}
\end{figure}
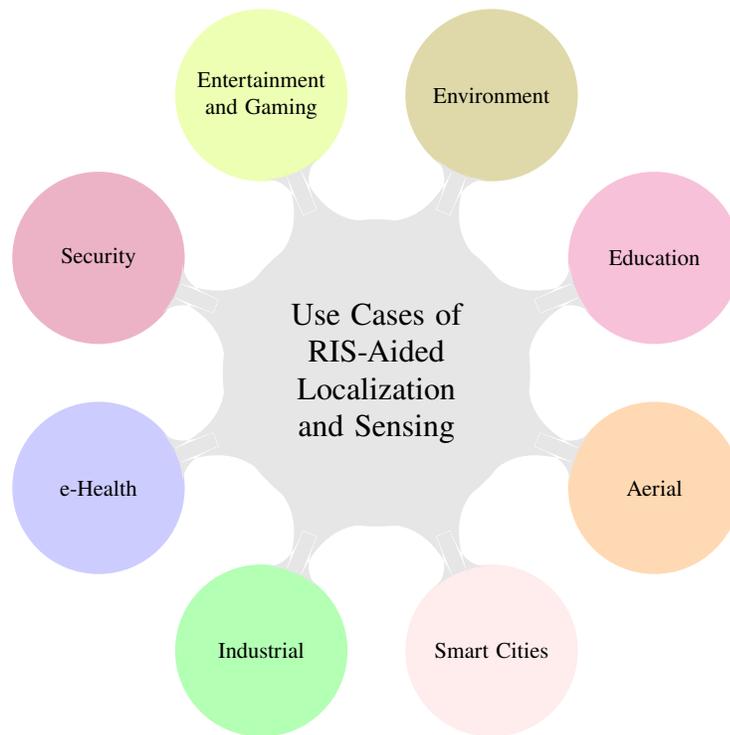

The applications and use cases that benefit from improved localization and sensing are mentioned in various research projects and Third Generation Partnership Project (3GPP) documents ~\cite{XZ22,3gppTS104,3gppTS186,3gppTS22,3gppTS261,3gppTS263,3gppTS840,3gppTS855,3gppTS857,3gppTS862,3gppTS872}, and they are categorized according to their underlying theoretical principles. This initial classification is depicted in Fig.~\ref{fig:Level1_apps}.
In specific domains, like location-based services and e-Health~\cite{mohan2021telesurgery,3gppTS872}, where real-time systems provide immediate medical interventions, high-precision localization is essential to enable advanced applications, regardless of whether they are deployed indoors or outdoors. Similarly, in industrial environments~\cite{3gppTS104,3gppTS261,3gppTS872}, accurate asset tracking is critical for identifying equipment and monitoring the movement of mobile elements, such as forklifts or components on the assembly line. Comparable accuracy demands are found in transportation and logistics, including rail systems, road networks, and the use of Unmanned Aerial Vehicles (UAVs)~\cite{3gppTS22,3gppTS261,3gppTS872}. Also applications involving Vehicle-to-Everything (V2X)~\cite{3gppTS186,3gppTS261} communications benefit from improved localization and sensing precision. In safety-critical scenarios, such as Autonomous Guided Vehicles (AGVs) in manufacturing or UAV navigation—ultra,-precise positioning is indispensable. Autonomous driving, for instance, requires real-time three-dimensional (3D) sensing to determine relative distances between vehicles and obstacles, forming an accurate model of the surrounding environment. Likewise, Vehicle-to-Vehicle (V2V) communications demand sub-meter localization to enhance communication efficiency. Finally, localization and sensing of enhanced accuracy enable more effective and precise integration of the digital with the physical environment for Augmented Reality (AR) applications~\cite{3gppTS263,3gppTS261,3gppTS872}. In this context, numerous applications can be enhanced, including those in the fields of healthcare and remote medical monitoring, education through AR-based learning, and entertainment, where new, more interactive games are being developed.

As already described in Section~\ref{sec:Chapter_Outline}, RISs, when properly configured have the capability to enhance both localization accuracy and environmental sensing, in both indoor and outdoor settings. Therefore, it is useful to examine their potential contribution to various applications, which fall under the categories shown in Fig.~\ref{fig:Level1_apps}, and especially those that require high-precision localization and sensing. Some indicative applications, from the use cases demonstrated in Fig.~\ref{fig:Level1_apps} are shown in Table~\ref{t:list-of-apps} and analyzed in the following paragraphs.

\begin{table}[t]
\centering
\caption[List of second-level apps in this document]{Indicative RIS-aided localization and sensing applications belonging in the uses cases described in Fig.~\ref{fig:Level1_apps}\label{t:list-of-apps}\\[0.5\baselineskip]}
\scalebox{0.95}
{
\begin{NiceTabular}{l p{9cm}l}
\CodeBefore
\rowcolor{blue!20}{2-5}
\rowcolor{green!20}{6-7}
\rowcolor{pink!20}{8-9}
\rowcolor{orange!20}{10-11}
\rowcolor{olive!20}{12-13}
\rowcolor{purple!20}{14}
\Body
\toprule
    & \textbf{Application/Use case}  & \textbf{Ref.} \\
    \hline
 1 & \textbf{Remote healthcare using AR} & \cite{3gppTS263,mohan2021telesurgery,sivaganesan2021robotics,wymeersch2022localisation}  \\
  \hline
  2 & \textbf{Health monitoring} & \cite{3gppTS261,wymeersch2022localisation}  \\
  \hline
 3 & \textbf{Person and medical equipment monitoring in hospitals} &  \cite{3gppTS872}\\
\hline
 4 & \textbf{Patient monitoring outside hospitals} & \cite{3gppTS872}\\
\hline
 5 &\textbf{Asset and tool tracking in vertical domains} & \cite{3gppTS104,3gppTS261} \\
\hline
 6 & \textbf{Tool assignment in flexible, modular assembly area in smart factories} & \cite{3gppTS104,3gppTS261}\\
 \hline
  7 & \textbf{Position-based handovers} &  \cite{PKH00,ZYW21,SSK20,LOCUS_D2.6} \\
  \hline
 8 & \textbf{Autonomous driving} &  \cite{3gppTS186,3gppTS261} \\
  \hline
 9 & \textbf{Accurate positioning to support UAV missions and operations} & \cite{3gppTS22}  \\
  \hline
 10 & \textbf{Transport and inspection by drones for medical purposes} &  \cite{3gppTS872} \\
  \hline
 11 & \textbf{Waste management and collectors} & \cite{3gppTS872}  \\
  \hline
    12 & \textbf{Earth monitoring} & \cite{3gppTS872,wymeersch2022localisation}  \\
  \hline
   13 & \textbf{Automatic public security} & \cite{wymeersch2022localisation}  \\
\bottomrule
\end{NiceTabular}}
\end{table}

\subsubsection*{Remote Healthcare and Health Monitoring}
Remote healthcare enabled by AR tools demands high precision localization and sensing~\cite{3gppTS263,mohan2021telesurgery,sivaganesan2021robotics,wymeersch2022localisation}. AR integrates digital content into the user's real-world environment in real time, enabling interactive experiences through devices such as smartphones, tablets, AR glasses, or headsets. Unlike Virtual Reality (VR), AR overlays virtual elements onto the physical world, requiring accurate tracking, sensing, and low latency communications to ensure seamless alignment. While technologies like Global Positioning System (GPS) and depth sensors support this functionality, limitations in certain environments, such as urban canyons or indoor settings, highlight the need for more robust localization methods. Radio-based localization and sensing, particularly in the context of 6G networks, offer promising solutions. RISs are expected to significantly enhance positioning accuracy and reduce latency, by beneficially shaping the wireless propagation environment\cite{KDA23,wymeersch2022localisation}, thereby advancing AR capabilities. The Key Performance Indicators (KPIs) for AR applications vary by use case and criticality. For instance, remote healthcare scenarios, such as telesurgery, demand ultra-low latency and sub-millimeter precision~\cite{wymeersch2022localisation,3gppTS263,mohan2021telesurgery,sivaganesan2021robotics}, whereas less critical applications, like virtual consultations, require more relaxed KPIs~\cite{wymeersch2022localisation,3gppTS263,mohan2021telesurgery}. These variations emphasize the need for adaptive network solutions to meet the diverse demands of AR in future wireless systems. 

Another important application is health monitoring~\cite{wymeersch2022localisation,3gppTS261}. Precision medicine represents a burgeoning initiative focused on personalized disease treatment and prevention. Personalized treatment relies on gathering health-related data and administering medicine through tiny wireless robots~\cite{wymeersch2022localisation,3gppTS261,sivaganesan2021robotics} that navigate within the soft tissues of the human body. 6G connectivity can be harnessed to collect sensor-based measurements for monitoring patients' vital signs. The ability to individually track and monitor patients enhances their security and ensures that appropriate medical assistance is provided based on their specific health condition or ailment; for example, perform various tasks such as hyperthermia, cauterisation, taking biopsies, stimulation of nerves, and drug delivery. These type of use cases  leverage the envisioned sensing and localization capabilities of 6G. Since inside the human body satellite-based signals for localization and sensing are not efficient enough, the new possibility of location estimation and
sensing using wireless networks' physical layer signals, i.e., radio localization/sensing, is becoming increasingly urgent. To this end, RISs are coming more and more to the forefront~\cite{KDA23,wymeersch2022localisation,9721205}, due to the capability to contribute significantly in the localization and sensing accuracy. In order to perform all of the above functions inside the body, tiny devices together with their high precision positioning are required. Considering also the fact that the response of the sensors (e.g., through the delivery of medication from the sensor) needs to be immediate, it can be concluded that tight KPIs for such types of applications are necessary.

Of major importance is also the use case of person and medical equipment monitoring inside and outside hospitals~\cite{3gppTS872}. In hospital environments, particularly in psychiatric and geriatric wards, real-time tracking of patients, caregivers, and critical medical equipment is essential for ensuring safety and operational efficiency~\cite{3gppTS872}. Accurate localization is crucial for preventing incidents such as patient elopement or wandering, as well as for enabling rapid access to emergency equipment during critical situations. These needs become even more pressing in large healthcare facilities that span multiple buildings and outdoor areas. For such complex settings, RIS-enabled smart wireless environments offer promising enhancements to real-time location systems. By improving signal propagation and mitigating interference, RISs can significantly boost localization accuracy in both indoor and outdoor hospital environments. This enables prompt detection of patient movements and faster retrieval of vital medical equipment, thereby enhancing emergency responsiveness and overall care quality. Real-time location tracking is also essential outside hospital settings for patients with severe health conditions who require continuous monitoring. In this case, RISs can significantly improve tracking accuracy and coverage, enhancing patient safety both inside and outside hospital environments.

\subsubsection*{Industrial Use Cases}
Improved radio localization and sensing can also be beneficial for industry-related use cases~\cite{3gppTS104,3gppTS261}, by contributing to critical tasks such as tool and asset tracking to vertical domains, and tool assignment in flexible modular assembly areas in smart factories. A vertical domain is a particular industry or group of enterprises in which similar products or services are developed, produced, and provided. Automation in vertical domains refers to the control of processes, devices, or systems by automatic means, i.e., using technology and software to streamline and optimize various tasks and processes within an organization. It can include the use of robotics, software applications, and other tools to improve efficiency, reduce errors, and save time and resources in various industries and domains. Low power high accuracy positioning is an integral part of a considerable number of applications, and especially industrial ones. The total energy needed for a specific operation time for such a low power high accuracy positioning-optimized IoT device~\cite{3gppTS104,3gppTS261} is a combination of energy for positioning (varies depending on the used positioning method), energy for communication/synchronization, and a difficult-to-predict factor that takes into account additional losses, e.g. security, power management, and self-discharge of batteries. Examples of target applications for low power high accuracy positioning are asset tracking in process automation, tracking of vehicles, and tool tracking.

Many public organizations have made a large investment into gaining and deploying a Geospatial Information System (GIS) to track their spatial assets, maintain historical records and sustain an accurate inventory. Process automation in the context of asset and tool tracking uses technology and software solutions to monitor and manage physical assets within an organization efficiently and accurately. Assets can include equipment, machinery, vehicles, tools, inventory, or any other valuable items critical to business operations. Asset tracking automation provides real-time visibility into asset location, status, and usage, leading to improved asset utilization, reduced operational costs, and enhanced asset security. According to~\cite{3gppTS104,3gppTS261}, the horizontal localization accuracy requirement of this use case is about $2-3$~m. And the battery lifetime (or the minimum operation time) for the device performing localization-performing is defined as $>6$ months.

Moreover, in flexible modular industrial environments, tool assignment refers to the dynamic allocation of tools and equipment to specific workstations or zones, aiming to optimize operational efficiency, productivity, and safety~\cite{3gppTS104,3gppTS261}. This process must account for task-specific requirements, ergonomic workspace layout, and risk minimization. The inherent modularity of such assembly areas demands continuous reconfiguration of workstation zones, necessitating precise localization and sensing of both tools and the surrounding environment. Unlike conventional asset tracking, tool assignment involves higher accuracy demands~\cite{3gppTS104,3gppTS261} due to its direct impact on the production process and product quality. To address the limitations of systems based on Global Navigation Satellite System (GNSS) in indoor industrial settings, radio-based localization and sensing technologies are employed, offering robust performance in environments with obstructions such as walls and machinery. Furthermore, RISs may enhance the accuracy and reliability of these systems by exploiting advanced propagation control mechanisms. Tool assignment thus represents a critical evolution of traditional tracking applications, enabling agile manufacturing, quality assurance, and responsiveness to market fluctuations. This use case requires a horizontal positioning accuracy better than $30$~cm and the localization device must support a minimum operational lifetime of $18$ months, as defined in the 3GPP documents~\cite{3gppTS104,3gppTS261}.

\subsubsection*{Smart Cities}
In the context of smart cities, an important feature for more efficient wireless communication operations constitute the position-based handovers~\cite{PKH00,ZYW21,SSK20,LOCUS_D2.6}. Handover is a pivotal mechanism in wireless communication systems that ensures uninterrupted connectivity by transferring an ongoing session of a mobile device from one cell, or base station, to another as it moves through the coverage area. A common impairment to handover performance is the ``ping-pong effect,'' where a device repeatedly oscillates between neighboring cells, particularly affecting users at cell edges. Mitigating this phenomenon requires careful tuning of cell configurations and handover algorithms. Accurate, real-time positioning of mobile users is instrumental in this effort, enabling informed handover triggers, reducing unnecessary handovers, and lowering signaling overhead. RISs present a promising solution to these challenges. By dynamically controlling the reflection of electromagnetic waves, RISs can enhance signal-to-noise ratio, alleviate multipath distortions, and suppress inter-cell interference. These improvements bolster localization accuracy, which is critical for timely and precise handover decisions, while adapting in real time to changing environmental conditions. For the position-based handover use case, the target performance requirements include a localization accuracy of $1$–$5$~m and service latency below $1$~second.

Another immersive application is the autonomous driving~\cite{3gppTS186,3gppTS261}. This paradigm, also known as self-driving, refers to a vehicle's ability to operate and navigate without human intervention. It relies on a combination of advanced technologies, sensors, and algorithms to control the vehicle's movement, make decisions, and respond to the environment. These technologies include: \begin{enumerate*}[label= (\alph*)]
\item sensors such as Light Detection And Ranging (LiDAR), radar, cameras, and ultrasonic sensors that are equipped on autonomous vehicles. These sensors provide real-time data about the vehicle's surroundings, including the positions of other vehicles, pedestrians, road signs, and road conditions.
\item Machine learning algorithms that process the sensor data and make decisions based on it. These algorithms enable the vehicle to interpret its environment, plan its route, and make driving decisions, such as accelerating, braking, and steering.
\item Connectivity to assist autonomous vehicles to communicate with each other and with infrastructure through V2V and vehicle-to-infrastructure (V2I) communication systems. This connectivity can enhance safety and traffic management.
\item Sensing and localization, since high-definition maps and GPS data are used to accurately locate the vehicle within its environment. This is essential for precise and safe navigation. To this end, the technology RIS is expected to contribute both by enabling localization and sensing in areas where it was previously impossible, and also by enhancing existing technologies with the diversity it offers.
\end{enumerate*}

Currently, most autonomous vehicles involve some automation but still require driver supervision. Achieving full autonomy is a complex challenge that involves not only technology but also regulatory, legal, and safety considerations. The development of autonomous driving technology has the potential to improve road safety, reduce traffic congestion, and enhance mobility for people who cannot drive due to age, disability, or other reasons. However, it also raises important questions and challenges related to safety, liability and ethics. Therefore, autonomous driving is an application that requires the cooperation of many individual technologies, and because each of the data sources (sensors) that cooperate to achieve its operation have different accuracy, update rate, resolution, and many other parameters, it is still difficult to assess all its KPIs. 

\subsubsection*{Aerial-Related Use Cases}
Accurate positioning is also a critical enabler for the effective deployment of UAVs~\cite{3gppTS872,3gppTS22} across a wide range of mission-critical applications, including surveillance, search and rescue, and precision agriculture. These applications demand precise location awareness for navigation, autonomous operation, data collection, and coordinated activity among multiple platforms. Key requirements for UAV positioning arise from the need for reliable navigation and path planning, high resolution sensing and geo-referencing, autonomous decision-making, collaborative operations, and responsive emergency actions~\cite{3gppTS872,3gppTS22}. To meet the latter qualitative requirements, advanced localization technologies are necessary, particularly in dynamic and cluttered environments where GNSS performance may degrade. RISs have emerged as a promising solution enhancing localization accuracy through optimized signal reflection and propagation. By leveraging their beamforming capabilities and high spatiotemporal resolution potential~\cite{RIS_potential}, RISs can mitigate multipath interference and support precise tracking of individual UAVs or swarms. Moreover, while current KPIs for UAV localization primarily reflect service-level needs, such as video streaming or mapping, more stringent KPIs for real-time navigation and operational safety are anticipated, especially in dense swarm deployments. Additionally, since conventional cellular infrastructure is designed for ground-level users, RIS deployment targeting higher elevation angles, offers a path toward seamless aerial connectivity without necessitating a complete network redesign~\cite{10817446}. In this context, RIS-enabled localization holds the potential to significantly improve the reliability, responsiveness, and safety of UAV operations, supporting their expanding role in both civilian and industrial domains.

Furthermore, in the context of aerial-related use cases, the integration of UAVs, or drones, into medical service operations~\cite{3gppTS872} is revolutionizing healthcare delivery by enabling fast, secure, and efficient transport of critical medical supplies. Drones offer a highly effective solution for intra- and inter-hospital logistics, particularly within large hospital campuses and between geographically distributed healthcare facilities. Their utility spans a range of tasks, including the transportation of medications, medical equipment, and emergency supplies, addressing logistical bottlenecks and enhancing inter-hospital collaboration~\cite{3gppTS872}. In emergency response scenarios, such as sudden cardiac arrests, drones enable rapid delivery of life-saving equipment like automated external defibrillators while providing real-time video feeds to emergency responders, thereby improving patient outcomes. In addition to transport functions, drones are also employed for inspection and monitoring of extensive medical infrastructure, enhancing operational safety and maintenance efficiency. These drone operations, particularly those conducted at low altitudes, demand continuous and reliable mobile network connectivity, as well as high precision localization and sensing to ensure secure and traceable deliveries, especially in dense urban environments. To this end, RISs emerge as a critical enabler, offering enhanced localization accuracy and robust communication support. Strategically deployed RISs improve UAV navigation, facilitate obstacle avoidance, and ensure efficient coordination between drones and healthcare facilities. In addition, RISs strengthen drone-to-network communication links, ensuring real-time data exchange and situational awareness.

\subsubsection*{Environmental Use Cases}

The waste management~\cite{3gppTS872,wymeersch2022localisation} and earth monitoring~\cite{3gppTS872,wymeersch2022localisation} sectors are undergoing significant transformations driven by the need for greater efficiency, sustainability, and data-driven decision-making. Traditional waste collection methods, based on fixed schedules and static routes, often lead to operational inefficiencies, increased costs, and negative environmental impacts. To address these challenges, municipalities are increasingly deploying smart waste management systems incorporating sensor-equipped bins. These connected bins communicate their fill levels in real time, enabling dynamic route optimization and significantly reducing unnecessary collections, fuel consumption, and greenhouse gas emissions. The RIS technology emerges as a powerful enabler in this domain, enhancing localization and sensing services for both waste bins and collection vehicles. By providing accurate, real-time positioning, RISs can facilitate efficient navigation and routing, while contributing to the sustainability goals of modern waste management. The system must achieve localization accuracy of less than $2$~m, respectively, with high availability and minimal energy consumption impact. In parallel, earth monitoring -central to achieving the United Nations’ sustainable development goals- relies on large-scale deployment of low-cost, biodegradable sensing devices that collect environmental data, such as climate metrics, biodiversity indicators, and ice sheet movement. Accurate localization of these devices is essential for reliable geo-tagging and meaningful data interpretation. RISs have the potential to play a vital role in enabling precise positioning in challenging environments where line-of-sight communication is compromised. Their reconfigurability and energy efficiency~\cite{10693440} support reliable data collection under dynamic conditions, further aligning with ecological sustainability.

Both latter use cases underscore the transformative potential of RISs in enhancing positioning accuracy, communication reliability, and system efficiency. The technology's adaptability and low power requirements make it an ideal solution for scalable deployment in environmentally sensitive and operationally demanding applications. 

\subsection*{Security}
Automatic public security in the realm of 6G wireless technology entails leveraging radio signals across various frequency bands to detect and pinpoint threats, especially in high frequencies, such as weapons, explosives, and hazardous materials in public environments~\cite{wymeersch2022localisation}. This state-of-the-art approach capitalizes on the predictable behavior of radio signals to enable autonomous threat detection and situational awareness in public spaces. Given the critical nature of such applications, highly accurate localization and sensing services are essential. Real-time and precise information about the position and movement of potential threats is vital for effective threat identification and response. Moreover, the system must be capable of differentiating between dangerous and harmless objects, further emphasizing the importance of precise localization for reliable detection.

RISs are particularly well-suited for supporting this application. Not only do they enhance the performance of localization and sensing functions~\cite{Kim2023RISRadar,11161901,Zhang2025Joint}, but they also contribute to physical-layer security~\cite{9501003,10289918,10143983,Xu2023Reconfiguring,10283801,Rexhepi2025Blinding}. The RIS technology strengthens security by actively controlling how electromagnetic waves propagate. Specifically, it can reflect signals constructively toward authorized users and destructively toward unauthorized listeners, thereby minimizing the risk of eavesdropping. This selective reflection ensures that sensitive communications remain secure. In addition, RISs can suppress interference from jamming sources, further reinforcing the security mechanisms of the system. RISs also improve localization accuracy, supporting more robust sensing capabilities. Their ability to operate in real time adds another layer of effectiveness to the public security infrastructure. Altogether, integrating RIS technology with 6G networks is poised to transform public safety, enabling more intelligent and responsive security systems.

According to performance benchmarks described in~\cite{wymeersch2022localisation}, the system aims to achieve a location (range) accuracy of $1$~cm, ensuring highly precise threat detection. The same $1$~cm range resolution allows for fine-grained object or individual detection within the monitored area. To accommodate dynamic security environments, the system must support an unambiguous range of up to $120$~m. Sub-degree angular resolution enables accurate direction finding of potential threats. Additionally, the system should be capable of tracking moving targets at incoming/outcoming speeds of $30$~km/h, with a velocity resolution of $0.5$~m/s. With an update rate of $0.1$~millisecond, the system delivers near-instantaneous location data. To ensure high reliability, the system needs to be designed with an availability target of $99.99\%$.

\subsection{Impact of RIS Radiation Pattern on Localization and Sensing}
A systematic examination of the geometric factors that influence the behavior and performance of RISs is presented in this Section. The discussion begins with the fundamental geometry-based trade-offs that govern RIS array structure, including element placement, array-response characterization under far- and near-field conditions, and the implications of dual-polarized architectures. These considerations establish the physical foundations upon which RIS beamforming and signal manipulation capabilities depend.

Subsequently, the inter-element spacing trade-off is analyzed to elucidate how element separation affects beamforming fidelity, susceptibility to mutual coupling, and the emergence of grating lobes. By contrasting continuous and quantized phase configurations, the inherent limitations and practical constraints associated with realistic RIS hardware are highlighted. This analysis clarifies the conditions under which improved angular resolution or enhanced robustness can be achieved through appropriate geometric design choices.

Finally, the impact of RIS size and element count on reflection beam characteristics is investigated. The evolution of array gain, main-lobe sharpness, and side-lobe behavior is assessed for both linear and rectangular array configurations, under continuous and finite-resolution phase control. The resulting trends demonstrate how scaling the RIS aperture influences directional concentration of radiated power and determines the effectiveness of beam-steering mechanisms.

Collectively, the subsections in this part provide a structured evaluation of how RIS geometry shapes fundamental beamforming properties. The insights obtained establish the basis for understanding design constraints, performance trade-offs, and practical considerations relevant to RIS-assisted communication, localization, and sensing systems.

\subsubsection{Geometry-based Trade-offs}
The RIS under consideration consists of $N = N_HN_V$ reflective elements arranged on a two-dimensional rectangular grid, with $N_H$ and $N_V$ elements along the horizontal and vertical axes, respectively. Each element occupies an area $A = d_H d_V$, where $d_H$ and $d_V$ denote the horizontal and vertical dimensions. Assuming that the RIS is placed at the $y$–$z$ plane (i.e., at $x=0$), the $n$th element (with $n \in [1, N]$) is indexed in a row-wise manner, and its spatial coordinates are given by
\begin{equation}
    \mathbf{u}_n = \begin{bmatrix}0, \; i(n)d_H, \; j(n)d_V\end{bmatrix}^{T},
    \label{e:ris_location}
\end{equation}
where the index mappings $i(n) = \operatorname{mod}(n-1, N_H)$ and $j(n) = \lfloor(n-1)/N_H\rfloor$ represent the horizontal and vertical indices of element $n$, respectively. When a plane wave of wavelength $\lambda$ impinges on the RIS with azimuth angle $\varphi$ and elevation angle $\theta$, the array response vector can be expressed as~\cite[Sec.~7.3]{bjornson2017massive}:
\begin{equation}
    \mathbf{a}(\varphi, \theta) = \begin{bmatrix}
    e^{-j\mathbf{k}^\mathsf{T}(\varphi, \theta)\mathbf{u}_1}, \ldots, e^{-j\mathbf{k}^\mathsf{T}(\varphi, \theta)\mathbf{u}_N}
    \end{bmatrix}^{T},
    \label{eq:array_response_vec}
\end{equation}
with the wave vector $\mathbf{k}(\varphi, \theta) \in \mathbb{R}^{3 \times 1}$ defined as:
\begin{equation}
    \mathbf{k}(\varphi, \theta) = \frac{2\pi}{\lambda}
    \begin{bmatrix}
    \cos(\theta)\cos(\varphi), \; \cos(\theta)\sin(\varphi), \; \sin(\theta)
    \end{bmatrix}^{T}.
    \label{eq:wave_vector}
\end{equation}

It should be emphasized that the plane-wave assumption holds under far-field conditions, where the signal propagation distance is significantly larger than the physical aperture of the RIS. In near-field regimes, the curvature of the wavefront cannot be ignored, and the array response is instead defined using spherical propagation principles~\cite{tse_2005,rahal2023performance}. For a transmitter located at position $\mathbf{p}$ and a RIS element at $\mathbf{u}_n$, the array response vector under near-field conditions is given by
\begin{equation}
\mathbf{a}(\mathbf{p}) =
\left[
e^{-\frac{j2\pi}{\lambda}\left(\|\mathbf{p} - \mathbf{u}_1\| - \|\mathbf{p} - \mathbf{p}_{\text{RIS}}\|\right)},
\ldots,
e^{-\frac{j2\pi}{\lambda}\left(\|\mathbf{p} - \mathbf{u}_N\| - \|\mathbf{p} - \mathbf{p}_{\text{RIS}}\|\right)}
\right]^T,
\label{e:nf_array_response_vec}
\end{equation}
where $\mathbf{p}_{\text{RIS}}$ denotes the geometric center of the RIS. Therefore, the general array response is dependent on either the relative angles (far-field) or the spatial distances (near-field) between the source and receiver. Importantly, since the RIS is designed for a specific carrier frequency $f_c$, the element dimensions $d_H$ and $d_V$ are fixed, and the array response varies with frequency due to changes in wavelength $\lambda$.

In realistic propagation environments, electromagnetic waves undergo phenomena such as scattering, diffraction, and reflection, which can alter their polarization states, a process known as channel depolarization. A widely accepted metric for quantifying this effect is the cross-polarization discrimination (XPD) ratio~\cite{9497725}, which is defined as
\begin{equation}
\text{XPD} = \frac{1-a}{a},
\end{equation}
where $a$ denotes the fraction of power that changes polarization. In the ideal case where $a=0$, the wave retains its original polarization entirely. Dual-polarized RISs are well-suited to counter polarization loss, as they can capture energy regardless of the incident wave's polarization. This makes them preferable to uni-polarized RISs in high-performance and power-sensitive applications. Since the RIS platform considered in this work is dual-polarized, the following analysis and simulations assume dual-polarization support.

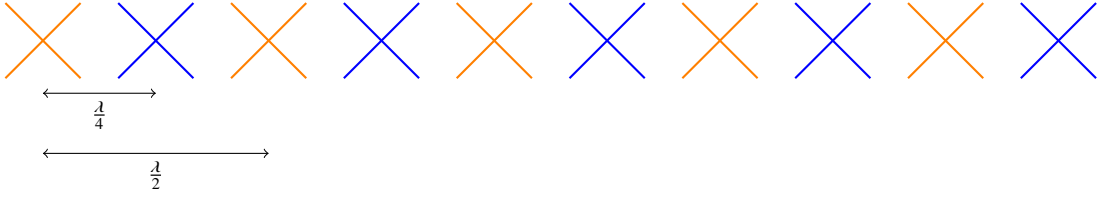
\begin{figure}[!t]
    \centering
    \begin{tikzpicture}
  \foreach \x in {0,1,2,3,4,5,6,7,8,9}
    \foreach \y in {0}
    {
      \begin{scope}[shift={(\x*1.5cm,\y*1.5cm)}]
      \ifnum\x=0
        \draw[orange, thick] (0,0) -- (1,1);
        \draw[orange, thick] (0,1) -- (1,0);
       \fi 
        \ifnum\x=1
        \draw[blue, thick] (0,0) -- (1,1);
        \draw[blue, thick] (0,1) -- (1,0);
       \fi 
       \ifnum\x=2
        \draw[orange, thick] (0,0) -- (1,1);
        \draw[orange, thick] (0,1) -- (1,0);
       \fi 
        \ifnum\x=3
        \draw[blue, thick] (0,0) -- (1,1);
        \draw[blue, thick] (0,1) -- (1,0);
       \fi 
       \ifnum\x=4
        \draw[orange, thick] (0,0) -- (1,1);
        \draw[orange, thick] (0,1) -- (1,0);
       \fi 
        \ifnum\x=5
        \draw[blue, thick] (0,0) -- (1,1);
        \draw[blue, thick] (0,1) -- (1,0);
       \fi 
       \ifnum\x=6
        \draw[orange, thick] (0,0) -- (1,1);
        \draw[orange, thick] (0,1) -- (1,0);
       \fi 
        \ifnum\x=7
        \draw[blue, thick] (0,0) -- (1,1);
        \draw[blue, thick] (0,1) -- (1,0);
       \fi 
       \ifnum\x=8
        \draw[orange, thick] (0,0) -- (1,1);
        \draw[orange, thick] (0,1) -- (1,0);
       \fi 
        \ifnum\x=9
        \draw[blue, thick] (0,0) -- (1,1);
        \draw[blue, thick] (0,1) -- (1,0);
       \fi 
        
        \ifnum\x<1
          \pgfmathsetmacro\dist{sqrt(1)} 
          \draw[<->, thin] (0.5, -0.2) -- node[below] {$\frac{\lambda}{4}$} (2, -0.2);
        \fi
         \ifnum\x=1
          \pgfmathsetmacro\dist{sqrt(2)} 
          \draw[<->, thin] (-1, -1) -- node[below] {$\frac{\lambda}{2}$} (2, -1);
        \fi
      \end{scope}
    }
\end{tikzpicture}
    \caption{Dual-polarized ULA RIS model \#1.}
    \label{fig:dp_ris_model_v1}
\end{figure}

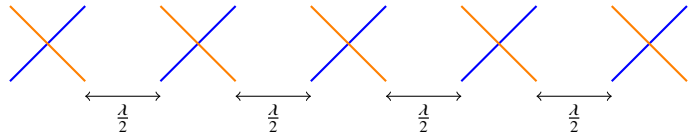
\begin{figure}[!t]
\centering
\begin{tikzpicture}
  \foreach \x in {0,1,2,3,4}
    \foreach \y in {0}
    {
      \begin{scope}[shift={(\x*2cm,\y*2cm)}]
        \draw[blue, thick] (0,0) -- (1,1);
        \draw[orange, thick] (0,1) -- (1,0);
        
        \ifnum\x<4
          \pgfmathsetmacro\dist{sqrt(2)} 
          \draw[<->, thin] (1, -0.2) -- node[below] {$\frac{\lambda}{2}$} (2, -0.2);
        \fi
      \end{scope}
    }
\end{tikzpicture}
\vspace{0.5cm}
 \caption{Dual-polarized ULA RIS model \#2.}
    \label{fig:dp_ris_model_v2}
\end{figure}

There are two common architectural models for dual-polarized RISs~\cite{ramezani2023broad,ramezani2023dual}, illustrated in Figs.~\ref{fig:dp_ris_model_v1} and~\ref{fig:dp_ris_model_v2}. In the first model, separate elements are used for each polarization, with spatial separation constraints that ensure that elements of the same polarization are not more than $\lambda/2$ apart. The second model assumes the coexistence of both polarizations within a single element spaced by at most $\lambda/2$.

Although smaller inter-element spacings can improve beamforming fidelity, they also increase mutual coupling effects, which in turn limit independent control over element phases~\cite{bjornson2021optimizing}. Simulations show that both dual-polarization models yield equivalent performance; thus, the first model is adopted for subsequent analyses, consistent with the prior literature.

Nevertheless the effects of mutual coupling, phase-dependent amplitude responses, and frequency-dependent reflection coefficients of the RIS can influence the final radiation and beam patterns, and these are initially considered negligible for the current Section. This simplification, supported only by certain practical implementations and not applicable to all RIS realizations, allows the isolation and investigation of the impact of RIS geometry on beamforming. Under this assumption, the reflection coefficient of the RIS element is modeled as $\omega_{\theta_n} {\triangleq} e^{j\theta_n}$. Afterwards, a specific geometric arrangement is considered to incorporate and analyze practical implementation factors. While numerous techniques have been proposed for RIS-based beam pattern design in various channel models (near- and far-field)~\cite{delbari2024far,ramezani2023dual,ramezani2023broad}, the following analysis employs simplified models from~\cite{ramezani2023dual,ramezani2023broad} to highlight essential RIS features and their influence on beamforming.

\subsubsection{Inter-element Spacing Trade-off}
\label{spacing_par}

The first feature to be examined is the inter-element spacing of the RIS, as defined in Fig.~\ref{fig:dp_ris_model_v1} (i.e., the inter-element spacing between the two polarizations. Thus, the spacing between elements of the same polarization corresponds to the double distance.). To demonstrate RIS radiation pattern insights, a Single Input Single Output (SISO) LoS flat-fading channel is considered, consisting of a single-antenna transmitter, a single-antenna receiver, and a Uniform Linear Array (ULA) RIS comprising $N = 64$ elements, aligned along the $y$-axis. The incident wave is modeled as a plane wave (far-field assumption), arriving at the RIS from an azimuth angle $\varphi_{\text{AoA}} = \frac{\pi}{3}$ and elevation angle $\theta_{\text{AoA}}^{el} = 0$, and is intended to be reflected towards the Angle of Departure (AoD) $\varphi_{\text{AoD}} = \frac{\pi}{8}$, also at zero elevation.

The power-domain array factor corresponding to a dual-polarized ULA RIS is expressed as
\begin{equation}
    \text{A}(\varphi) = \left|\bm{\omega}_{\bm{\theta} H}^{T} \left(\mathbf{a}_H(\varphi_{\text{AoA}}) \odot \mathbf{a}_H(\varphi)\right)\right|^2 
    + \left|\bm{\omega}_{\bm{\theta} V}^{T} \left(\mathbf{a}_V(\varphi_{\text{AoA}}) \odot \mathbf{a}_V(\varphi)\right)\right|^2,
    \label{power_factor}
\end{equation}
where $\odot$ denotes the element-wise product, $\bm{\omega}_{\bm{\theta} H}$ and $\bm{\omega}_{\bm{\theta} V}$ represent the horizontal and vertical RIS phase configurations, respectively, and $\mathbf{a}_H(\cdot)$, $\mathbf{a}_V(\cdot)$ correspond to the array response vectors for each polarization.

To maximize the beamforming gain in the direction $\varphi_{\text{AoD}}$, the RIS configurations are set to $\bm{\omega}_{\bm{\theta} H} = \left( \mathbf{a}_H(\varphi_{\text{AoA}}) \odot \mathbf{a}_H(\varphi_{\text{AoD}}) \right)^H$ and similarly for the vertical polarization. This choice ensures coherent reinforcement of the reflected signals in the target direction, resulting in maximal directivity at $\varphi_{\text{AoD}}$.

\begin{figure*}[!p]
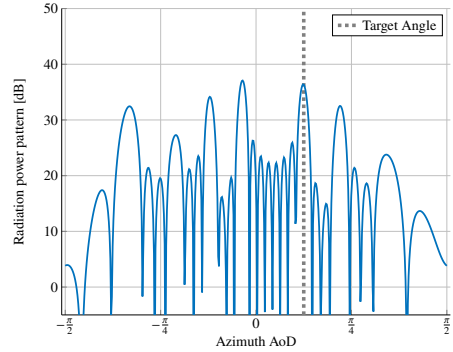
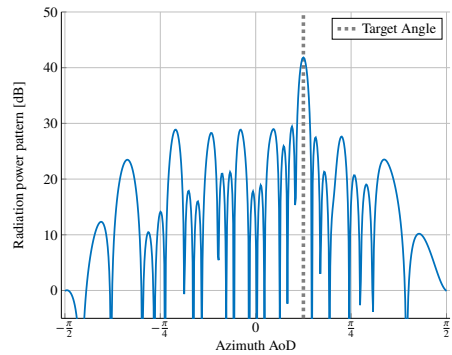
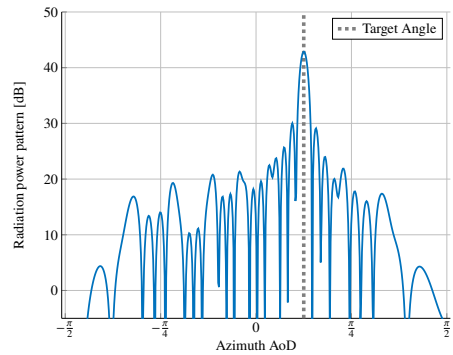
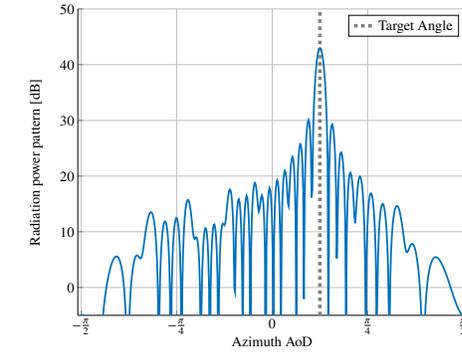
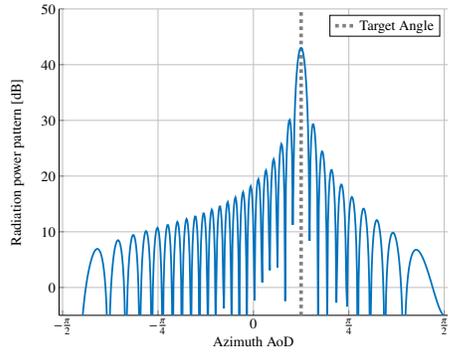

\centering
\begin{subfigure}{0.42\textwidth}
    \input{newsv-mult/author/Fig_varying_inter_spacing/Fig_1}
    \caption{ULA {RIS} radiation pattern with $\varphi_{AoA}=\frac{\pi}{3}$, $\varphi_{AoD}=\frac{\pi}{8}$, $N = 64$, $b_\theta = 1$ and inter-element spacing $\frac{\lambda}{4}$. Refers to the $e^{-j\pi/2}$ and $e^{j\pi/2}$ $1$-bit quantization.}
    \label{fig:spacing1bit1a}
\end{subfigure}
\hfill
\begin{subfigure}{0.42\textwidth}
     \input{newsv-mult/author/Fig_varying_inter_spacing/Fig_2}
    \caption{ULA {RIS} radiation pattern with $\varphi_{AoA}=\frac{\pi}{3}$, $\varphi_{AoD}=\frac{\pi}{8}$, $N = 64$, $b_\theta = 1$ and inter-element spacing $\frac{\lambda}{4}$. Refers to the $e^{-j\pi}$ and $e^{j0}$ $1$-bit quantization.}
    \label{fig:spacing1bit1b}
\end{subfigure}
\hfill
\begin{subfigure}{0.42\textwidth}
     \input{newsv-mult/author/Fig_varying_inter_spacing/Fig_3}
    \caption{ULA {RIS} radiation pattern with $\varphi_{AoA}=\frac{\pi}{3}$, $\varphi_{AoD}=\frac{\pi}{8}$, $N = 64$, $b_\theta = 2$ and inter-element spacing $\frac{\lambda}{4}$.}
    \label{fig:spacing1bit2}
\end{subfigure}
\hfill
\begin{subfigure}{0.42\textwidth}
     \input{newsv-mult/author/Fig_varying_inter_spacing/Fig_4}
    \caption{ULA {RIS} radiation pattern with $\varphi_{AoA}=\frac{\pi}{3}$, $\varphi_{AoD}=\frac{\pi}{8}$, $N = 64$, $b_\theta = 3$ and inter-element spacing $\frac{\lambda}{4}$.}
    \label{fig:spacing1bit3}
\end{subfigure}
\hfill
\begin{subfigure}{0.42\textwidth}
    \centering
      \input{newsv-mult/author/Fig_varying_inter_spacing/Fig_5}
    \caption{ULA {RIS} radiation pattern with $\varphi_{AoA}=\frac{\pi}{3}$, $\varphi_{AoD}=\frac{\pi}{8}$, $N = 64$, $b_\theta = 4$ and inter-element spacing $\frac{\lambda}{4}$.}
    \label{fig:spacing1bit4}
\end{subfigure}
\hfill
\begin{subfigure}{0.42\textwidth}
    \input{newsv-mult/author/Fig_varying_inter_spacing/Fig_6}
    \caption{ULA {RIS} radiation pattern with $\varphi_{AoA}=\frac{\pi}{3}$, $\varphi_{AoD}=\frac{\pi}{8}$, $N = 64$, $b_\theta = \infty$ and inter-element spacing $\frac{\lambda}{4}$.}
\label{fig:spacing1}
\end{subfigure}
\vspace{0.3cm}
\caption{ULA {RIS} radiation patterns with inter-element spacing equal to $\frac{\lambda}{4}$.}
\end{figure*}

\begin{figure*}[!p]
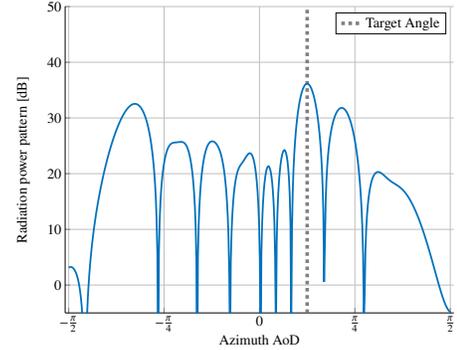
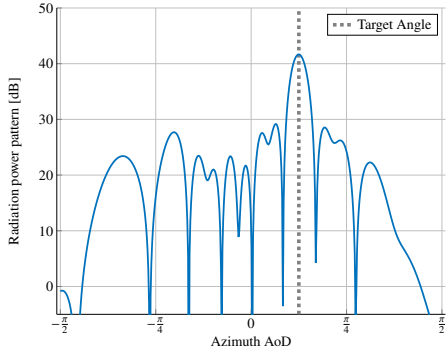
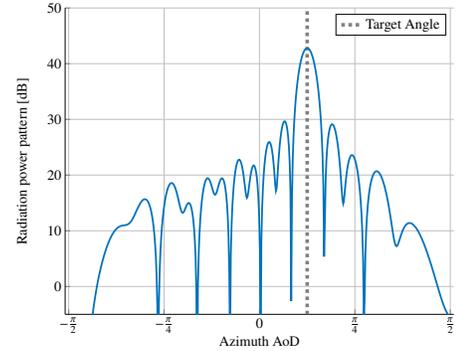
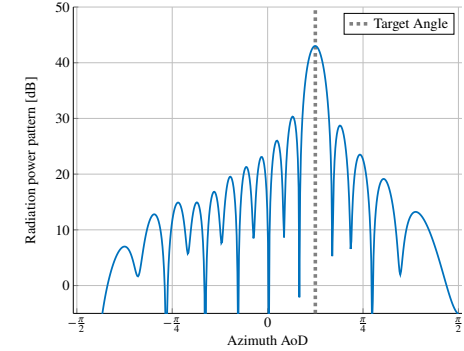
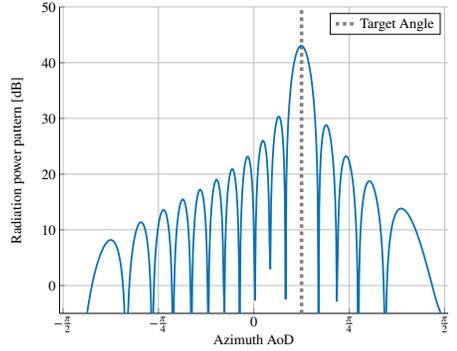

\centering
\begin{subfigure}{0.42\textwidth}
    \input{newsv-mult/author/Fig_varying_inter_spacing/Fig_7}
    \caption{ULA {RIS} radiation pattern with $\varphi_{AoA}=\frac{\pi}{3}$, $\varphi_{AoD}=\frac{\pi}{8}$, $N = 64$, $b_\theta = 1$ and inter-element spacing $\frac{\lambda}{8}$. Refers to the $e^{-j\pi/2}$ and $e^{j\pi/2}$ $1$-bit quantization.}
    \label{fig:spacing3bit1a}
\end{subfigure}
\hfill
\begin{subfigure}{0.42\textwidth}
     \input{newsv-mult/author/Fig_varying_inter_spacing/Fig_8}
    \caption{ULA {RIS} radiation pattern with $\varphi_{AoA}=\frac{\pi}{3}$, $\varphi_{AoD}=\frac{\pi}{8}$, $N = 64$, $b_\theta = 1$ and inter-element spacing $\frac{\lambda}{8}$. Refers to the $e^{-j\pi}$ and $e^{j0}$ $1$-bit quantization.}
    \label{fig:spacing3bit1b}
\end{subfigure}
\hfill
\begin{subfigure}{0.42\textwidth}
     \input{newsv-mult/author/Fig_varying_inter_spacing/Fig_9}
    \caption{ULA {RIS} radiation pattern with $\varphi_{AoA}=\frac{\pi}{3}$, $\varphi_{AoD}=\frac{\pi}{8}$, $N = 64$, $b_\theta = 2$ and inter-element spacing $\frac{\lambda}{8}$.}
    \label{fig:spacing3bit2}
\end{subfigure}
\hfill
\begin{subfigure}{0.42\textwidth}
     \input{newsv-mult/author/Fig_varying_inter_spacing/Fig_10}
    \caption{ULA {RIS} radiation pattern with $\varphi_{AoA}=\frac{\pi}{3}$, $\varphi_{AoD}=\frac{\pi}{8}$, $N = 64$, $b_\theta = 3$ and inter-element spacing $\frac{\lambda}{8}$.}
    \label{fig:spacing3bit3}
\end{subfigure}
\hfill
\begin{subfigure}{0.42\textwidth}
    \centering
      \input{newsv-mult/author/Fig_varying_inter_spacing/Fig_11}
    \caption{ULA {RIS} radiation pattern with $\varphi_{AoA}=\frac{\pi}{3}$, $\varphi_{AoD}=\frac{\pi}{8}$, $N = 64$, $b_\theta = 4$ and inter-element spacing $\frac{\lambda}{8}$.}
    \label{fig:spacing3bit4}
\end{subfigure}
\hfill
\begin{subfigure}{0.42\textwidth}
    \input{newsv-mult/author/Fig_varying_inter_spacing/Fig_12}
    \caption{ULA {RIS} radiation pattern with $\varphi_{AoA}=\frac{\pi}{3}$, $\varphi_{AoD}=\frac{\pi}{8}$, $N 
= 64$, $b_\theta = \infty$ and inter-element spacing $\frac{\lambda}{8}$.}
\label{fig:spacing3}
\end{subfigure}
\vspace{0.3cm}
\caption{ULA {RIS} radiation patterns with inter-element spacing equal to $\frac{\lambda}{8}$ and various quantization values for each RIS-element's coefficient.}
\end{figure*}

Beam patterns were evaluated for two inter-element spacings, namely $\lambda/4$ and $\lambda/8$, under the continuous-phase configuration. The corresponding patterns are depicted in Figs.~\ref{fig:spacing1} and~\ref{fig:spacing3}. Furthermore, quantized RIS configurations were studied for various phase resolutions $b_\theta$, with quantized versions of the continuous-phase configurations employed. For example, under 1-bit quantization, the RIS coefficients were restricted to values such as $e^{-j\pi/2}$ and $e^{j\pi/2}$ (Figs.~\ref{fig:spacing1bit1a},~\ref{fig:spacing3bit1a}) or $e^{-j\pi}$ and $e^{j0}$ (Figs.~\ref{fig:spacing1bit1b},~\ref{fig:spacing3bit1b}) producing the beam patterns shown in Figs.~\ref{fig:spacing1bit1a}--\ref{fig:spacing3bit4}. Although quantizing the optimal continuous configuration does not necessarily yield the optimal discrete beam pattern due to the non-convex nature of the problem, increased phase resolution ($b_\theta$) results in beam patterns that approximate the continuous case more closely. It is also observed that smaller inter-element spacing improves beamforming fidelity, even with coarse phase quantization. For example, when using 1-bit quantization and $\lambda/8$ spacing, the main lobe remains dominant in the intended direction. Conversely, increased spacing leads to more pronounced side lobes, especially under low-resolution quantization.

In practical scenarios, an inter-element spacing of $\lambda/4$ is typically adopted as it provides a favorable balance between reduced mutual coupling and acceptable beam pattern accuracy. Excessive reductions in spacing increase mutual-coupling effects, necessitating additional compensation techniques to maintain performance.

\begin{figure}[!t]
    \centering
    \scalebox{0.9}{\input{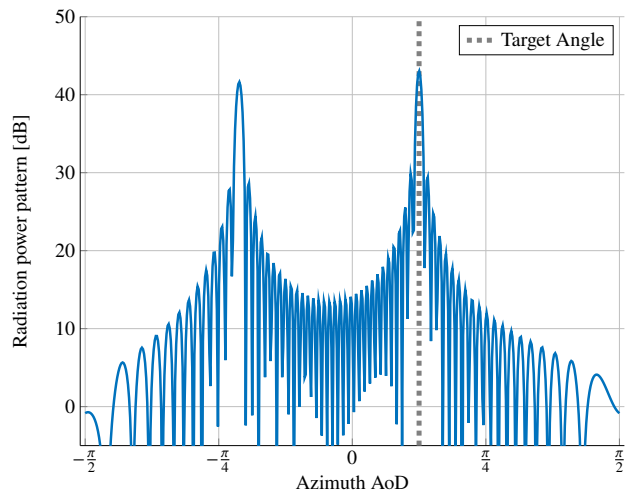}}
    \caption{ULA {RIS} radiation pattern with $\varphi_{AoA}=\frac{\pi}{3}$, $\varphi_{AoD}=\frac{\pi}{8}$, $N 
= 64$, $b_\theta = \infty$ and inter-element spacing $\frac{\lambda}{2}$.}
    \label{fig:spacing2}
\end{figure}

Moreover, for spacings larger than $\lambda/4$, undesired grating lobes may appear, even under continuous-phase assumptions (see Fig.~\ref{fig:spacing2}), significantly degrading performance in applications such as localization and sensing~\cite{keykhosravi2022ris}. Therefore, in all subsequent evaluations and simulations, an inter-element spacing of $\lambda/4$ is assumed, in line with practical deployment constraints and available hardware.


\subsubsection{Reflection Beams}
\begin{figure*}[!t]
\centering
\begin{subfigure}{0.42\textwidth}
    \input{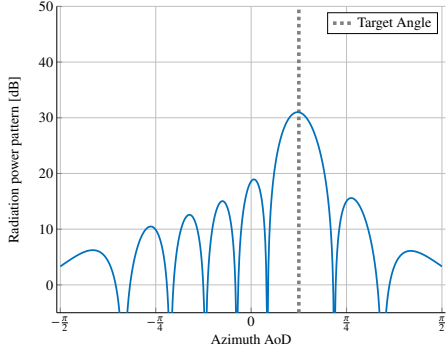}
    \caption{ULA {RIS} radiation pattern with $\varphi_{AoA}=\frac{\pi}{3}$, $\varphi_{AoD}=\frac{\pi}{8}$, $N = 16$, $b_\theta = \infty$ and inter-element spacing $\frac{\lambda}{4}$.}
    \label{fig:ULA_RIS_16}
\end{subfigure}
\hfill
\begin{subfigure}{0.42\textwidth}
     \input{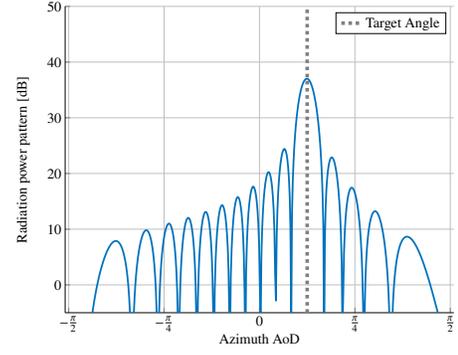}
    \caption{ULA {RIS} radiation pattern with $\varphi_{AoA}=\frac{\pi}{3}$, $\varphi_{AoD}=\frac{\pi}{8}$, $N = 32$, $b_\theta = \infty$ and inter-element spacing $\frac{\lambda}{4}$.}
    \label{fig:ULA_RIS_32}
\end{subfigure}
\hfill
\begin{subfigure}{0.42\textwidth}
     \input{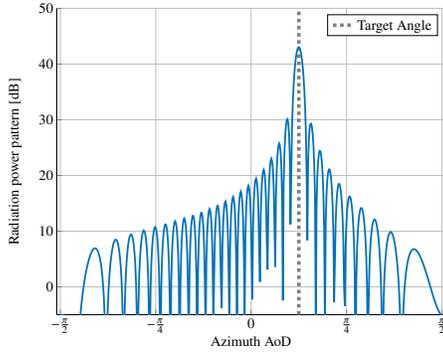}
    \caption{ULA {RIS} radiation pattern with $\varphi_{AoA}=\frac{\pi}{3}$, $\varphi_{AoD}=\frac{\pi}{8}$, $N = 64$, $b_\theta = \infty$ and inter-element spacing $\frac{\lambda}{4}$.}
    \label{fig:ULA_RIS_64}
\end{subfigure}
\hfill
\begin{subfigure}{0.42\textwidth}
     \input{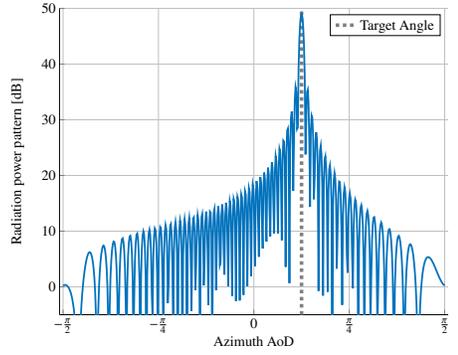}
    \caption{ULA {RIS} radiation pattern with $\varphi_{AoA}=\frac{\pi}{3}$, $\varphi_{AoD}=\frac{\pi}{8}$, $N = 128$, $b_\theta = \infty$ and inter-element spacing $\frac{\lambda}{4}$.}
    \label{fig:ULA_RIS_128}
\end{subfigure}
\vspace{0.3cm}
\caption{ULA {RIS} radiation pattern with respect of the number of elements in the continuous case.}
\end{figure*}

Using the same simulation setup as described in~\cite{ramezani2023dual} and outlined in the previous discussion, the performance of the ULA-RIS node by varying the number of reflecting elements while keeping the inter-element spacing fixed at $\frac{\lambda}{4}$, is evaluated. Figs.~\ref{fig:ULA_RIS_16}--\ref{fig:ULA_RIS_128} clearly demonstrate the expected array gain: as the size of the RIS increases under the idealized assumption of infinite phase resolution, a higher portion of the transmitted power is effectively directed toward the receiver. This behavior holds across various configurations of the angle of arrival $\varphi_{\text{AoA}}$ and angle of departure $\varphi_{\text{AoD}}$, indicating that with continuous reflection coefficients, the RIS can steer highly directive beams toward any desired direction.

\begin{figure*}[!t]
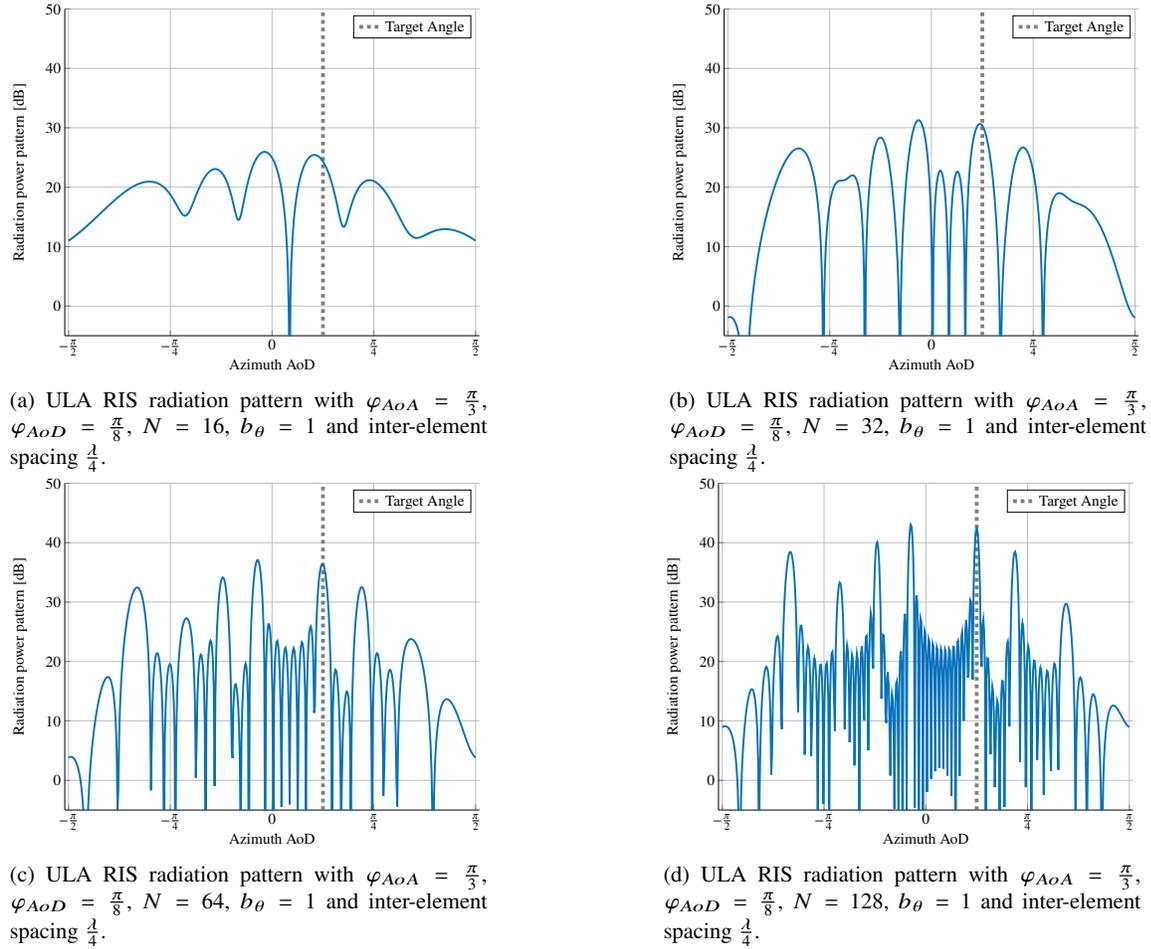

\centering
\begin{subfigure}{0.42\textwidth}
    \scalebox{0.8}{\input{newsv-mult/author/Fig_varying_N_folder/Fig_1}}
    \caption{ULA {RIS} radiation pattern with $\varphi_{AoA}=\frac{\pi}{3}$, $\varphi_{AoD}=\frac{\pi}{8}$, $N = 16$, $b_\theta = 1$ and inter-element spacing $\frac{\lambda}{4}$.}
    \label{fig:ULA_RIS_16bit1}
\end{subfigure}
\hfill
\begin{subfigure}{0.42\textwidth}
     \scalebox{0.8}{\input{newsv-mult/author/Fig_varying_N_folder/Fig_2}}
    \caption{ULA {RIS} radiation pattern with $\varphi_{AoA}=\frac{\pi}{3}$, $\varphi_{AoD}=\frac{\pi}{8}$, $N = 32$, $b_\theta = 1$ and inter-element spacing $\frac{\lambda}{4}$.}
    \label{fig:ULA_RIS_32bit1}
\end{subfigure}
\hfill
\begin{subfigure}{0.42\textwidth}
     \scalebox{0.8}{\input{newsv-mult/author/Fig_varying_N_folder/Fig_3}}
    \caption{ULA {RIS} radiation pattern with $\varphi_{AoA}=\frac{\pi}{3}$, $\varphi_{AoD}=\frac{\pi}{8}$, $N = 64$, $b_\theta = 1$ and inter-element spacing $\frac{\lambda}{4}$.}
    \label{fig:ULA_RIS_64bit1}
\end{subfigure}
\hfill
\begin{subfigure}{0.42\textwidth}
     \scalebox{0.8}{\input{newsv-mult/author/Fig_varying_N_folder/Fig_4}}
    \caption{ULA {RIS} radiation pattern with $\varphi_{AoA}=\frac{\pi}{3}$, $\varphi_{AoD}=\frac{\pi}{8}$, $N = 128$, $b_\theta = 1$ and inter-element spacing $\frac{\lambda}{4}$.}
    \label{fig:ULA_RIS_128bit1}
\end{subfigure}
\vspace{0.4cm}
\caption{ULA {RIS} radiation pattern varying $N$.}
\end{figure*}

Even under coarse quantization, as shown in Figs.~\ref{fig:ULA_RIS_16bit1}--\ref{fig:ULA_RIS_128bit1} for the 1-bit case, the array gain remains evident with increasing numbers of elements. Similar trends are observed for higher quantization levels with $b_\theta = 2, 3, 4$. However, the emergence of undesired side lobes, particularly pronounced in the 1-bit case, persists regardless of the size of the array. Increasing the number of elements does not mitigate the side lobe levels but does result in narrower main and side lobes. This angular sharpening enhances the spatial resolution of the RIS, providing improved visibility of the angular power distribution, a property that is particularly beneficial for beam sweeping strategies.

\begin{figure*}[!t]
\centering
\begin{subfigure}{0.45\textwidth}
    \scalebox{0.85}{\includegraphics[width=\textwidth]{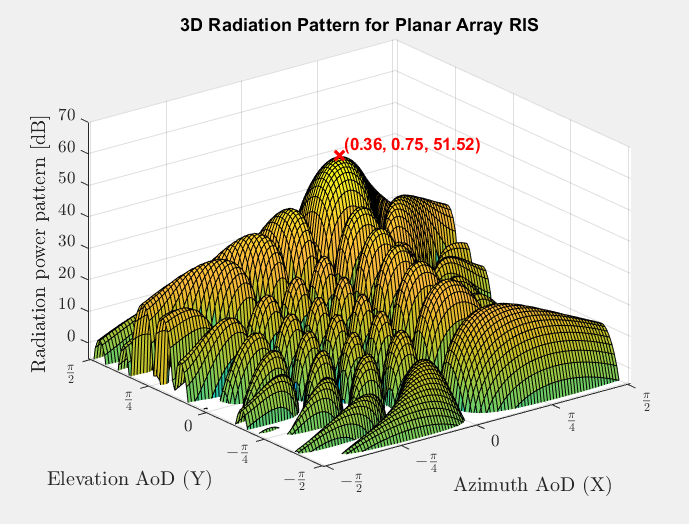}}
    \caption{URA {RIS} radiation pattern with $(\varphi_{AoA},\theta_{AoA})=(\frac{\pi}{3},\frac{\pi}{8})$, $(\varphi_{AoD},\theta_{AoD})=(\frac{\pi}{8},\frac{\pi}{4})$, $N = 16\times 16$, $b_\theta = \infty$ and inter-element spacing $\frac{\lambda}{4}$.}
    \label{fig:URA_RIS_16}
\end{subfigure}
\hfill
\begin{subfigure}{0.45\textwidth}
     \scalebox{0.85}{\includegraphics[width=\textwidth]{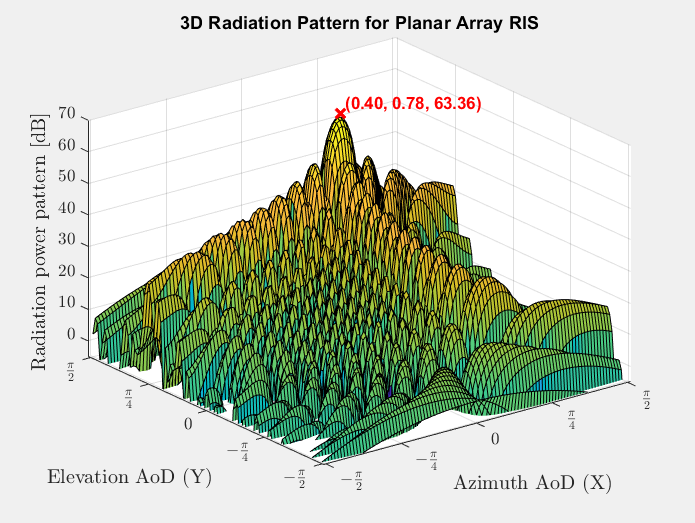}}
    \caption{URA {RIS} radiation pattern with $(\varphi_{AoA},\theta_{AoA})=(\frac{\pi}{3},\frac{\pi}{8})$, $(\varphi_{AoD},\theta_{AoD})=(\frac{\pi}{8},\frac{\pi}{4})$, $N = 32\times 32$, $b_\theta = \infty$ and inter-element spacing $\frac{\lambda}{4}$.}
    \label{fig:URA_RIS_32}
\end{subfigure}
\hfill
\begin{subfigure}{0.45\textwidth}
     \scalebox{0.85}{\includegraphics[width=\textwidth]{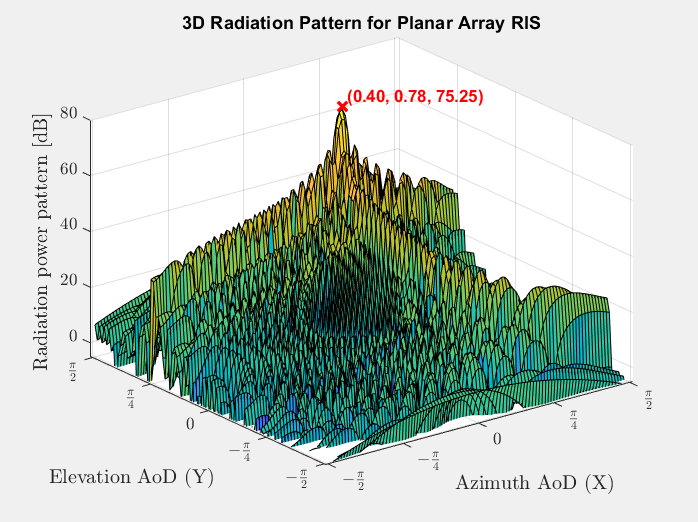}}
    \caption{URA {RIS} radiation pattern with $(\varphi_{AoA},\theta_{AoA})=(\frac{\pi}{3},\frac{\pi}{8})$, $(\varphi_{AoD},\theta_{AoD})=(\frac{\pi}{8},\frac{\pi}{4})$, $N = 64\times 64$, $b_\theta = \infty$ and inter-element spacing $\frac{\lambda}{4}$.}
    \label{fig:URA_RIS_64}
\end{subfigure}
\hfill
\begin{subfigure}{0.45\textwidth}
     \scalebox{0.85}{\includegraphics[width=\textwidth]{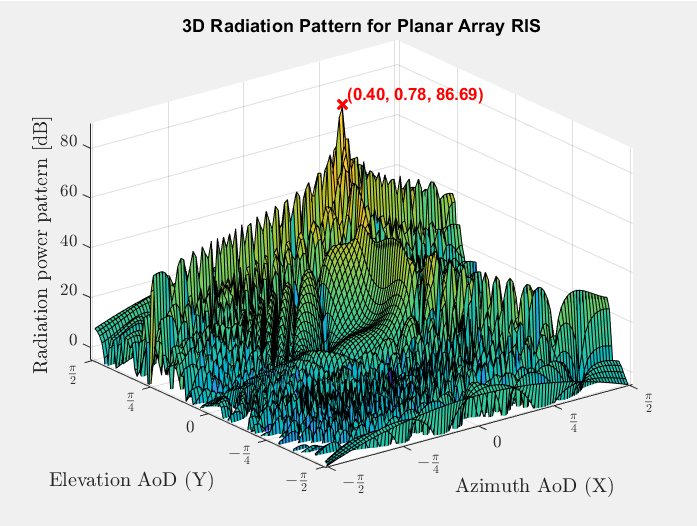}}
    \caption{{URA} {RIS} radiation pattern with $(\varphi_{AoA},\theta_{AoA})=(\frac{\pi}{3},\frac{\pi}{8})$, $(\varphi_{AoD},\theta_{AoD})=(\frac{\pi}{8},\frac{\pi}{4})$, $N = 128\times 128$, $b_\theta = \infty$ and inter-element spacing $\frac{\lambda}{4}$.}
    \label{fig:URA_RIS_128}
\end{subfigure}
\vspace{0.4cm}
\caption{{URA} {RIS} radiation patterns varying the number of elements $N$.}
\end{figure*}

To investigate the influence of the number of reflecting elements on the radiation (beam) pattern of a Uniform Rectangular Array (URA)-based RIS, the same simulation setup as described by Ramezani et al.~\cite{ramezani2023broad} has been adopted. This setup has been extended to the URA configuration for the RIS node, while maintaining a fixed inter-element spacing of $\frac{\lambda}{4}$. Only the number of elements in the URA-RIS node is varied. From the simulation results presented in Figs.~\ref{fig:URA_RIS_16}--\ref{fig:URA_RIS_128}, it can be observed that array gain is achieved, and a greater amount of power is delivered to the receiver as the size of the RIS increases, under the assumption of continuous reflection resolution (i.e., infinite quantization levels). These findings are consistent with those obtained in the case of a ULA-based RIS. Moreover, when the angle of arrival $(\varphi_{\text{AoA}}, \theta_{\text{AoA}})$ and the angle of departure (or reflection) $(\varphi_{\text{AoD}}, \theta_{\text{AoD}})$ are varied, similar conclusions can be drawn. This is attributed to the fact that, with infinite resolution in the RIS reflection coefficients, highly directional and efficient beams can be formed towards any direction.

\begin{figure*}[!t]
\centering
\begin{subfigure}{0.45\textwidth}
    \scalebox{0.85}{\includegraphics[width=\textwidth]{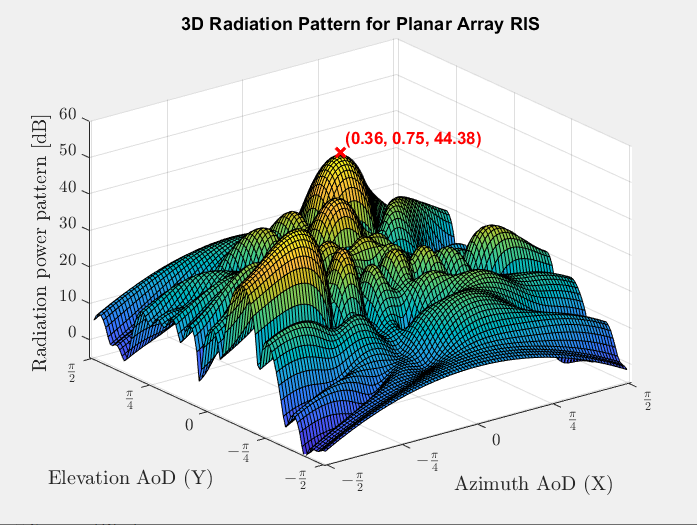}}
    \caption{{URA} {RIS} radiation pattern with $(\varphi_{AoA},\theta_{AoA})=(\frac{\pi}{3},\frac{\pi}{8})$, $(\varphi_{AoD},\theta_{AoD})=(\frac{\pi}{8},\frac{\pi}{4})$, $N = 16\times 16$, $b_\theta = 1$ and inter-element spacing $\frac{\lambda}{4}$.}
    \label{fig:URA_RIS_16_bit1}
\end{subfigure}
\hfill
\begin{subfigure}{0.45\textwidth}
     \scalebox{0.85}{\includegraphics[width=\textwidth]{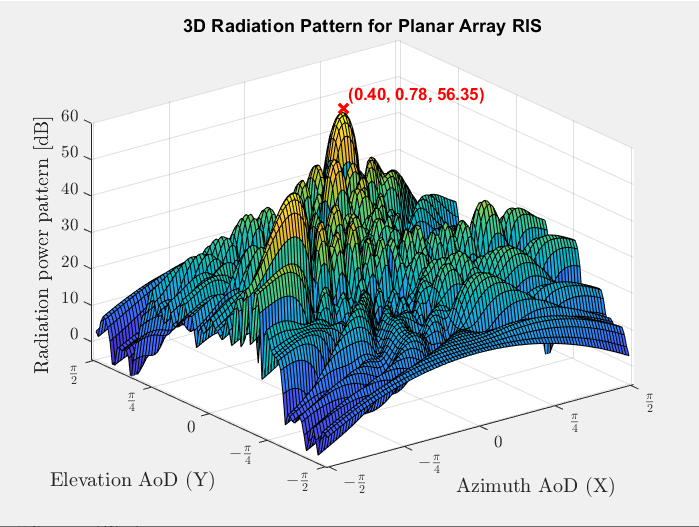}}
    \caption{{URA} {RIS} radiation pattern with $(\varphi_{AoA},\theta_{AoA})=(\frac{\pi}{3},\frac{\pi}{8})$, $(\varphi_{AoD},\theta_{AoD})=(\frac{\pi}{8},\frac{\pi}{4})$, $N = 32\times 32$, $b_\theta = 1$ and inter-element spacing $\frac{\lambda}{4}$.}
    \label{fig:URA_RIS_32_bit1}
\end{subfigure}
\hfill
\begin{subfigure}{0.45\textwidth}
     \scalebox{0.85}{\includegraphics[width=\textwidth]{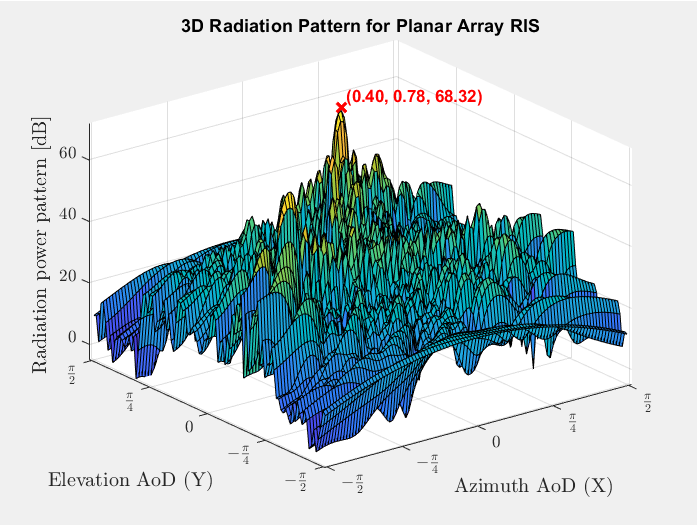}}
    \caption{{URA} {RIS} radiation pattern with $(\varphi_{AoA},\theta_{AoA})=(\frac{\pi}{3},\frac{\pi}{8})$, $(\varphi_{AoD},\theta_{AoD})=(\frac{\pi}{8},\frac{\pi}{4})$, $N = 64\times 64$, $b_\theta = 1$ and inter-element spacing $\frac{\lambda}{4}$.}
    \label{fig:URA_RIS_64_bit1}
\end{subfigure}
\hfill
\begin{subfigure}{0.45\textwidth}
     \scalebox{0.85}{\includegraphics[width=\textwidth]{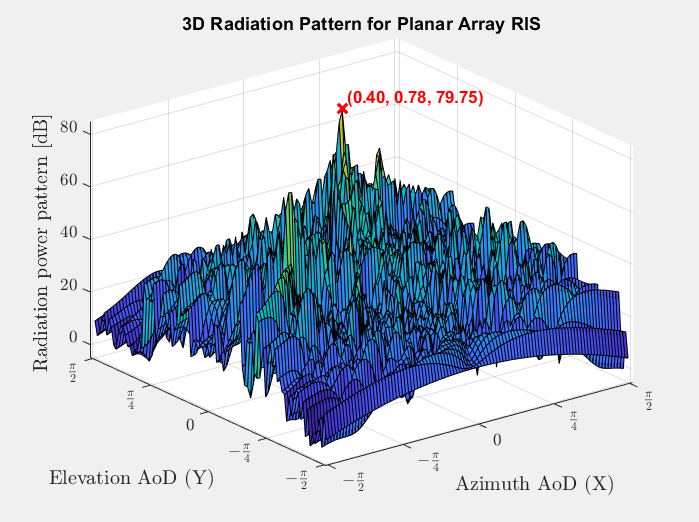}}
    \caption{{URA} {RIS} radiation pattern with $(\varphi_{AoA},\theta_{AoA})=(\frac{\pi}{3},\frac{\pi}{8})$, $(\varphi_{AoD},\theta_{AoD})=(\frac{\pi}{8},\frac{\pi}{4})$, $N = 128\times 128$, $b_\theta = 1$ and inter-element spacing $\frac{\lambda}{4}$.}
    \label{fig:URA_RIS_128_bit1}
\end{subfigure}
\vspace{0.4cm}
\caption{{URA} {RIS} radiation patterns varying $N$ and with $b_\theta = 1$.}
\end{figure*}

\begin{figure*}[!t]
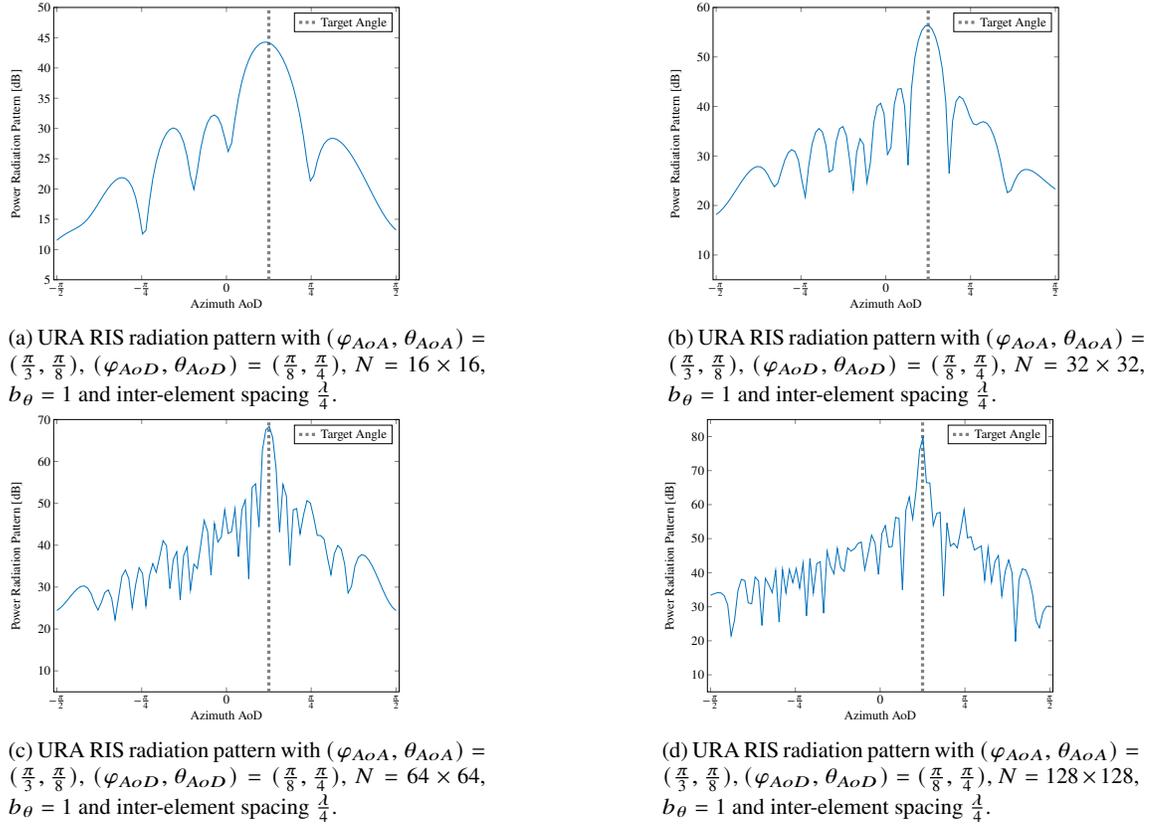

\centering
\begin{subfigure}{0.42\textwidth}
    \scalebox{0.8}{\input{newsv-mult/author/Fig_using_URA_based_RIS/Fig_9}}
    \caption{{URA} {RIS} radiation pattern with $(\varphi_{AoA},\theta_{AoA})=(\frac{\pi}{3},\frac{\pi}{8})$, $(\varphi_{AoD},\theta_{AoD})=(\frac{\pi}{8},\frac{\pi}{4})$, $N = 16\times 16$, $b_\theta = 1$ and inter-element spacing $\frac{\lambda}{4}$.}
    \label{fig:URA_RIS_16_bit1_2}
\end{subfigure}
\hfill
\begin{subfigure}{0.42\textwidth}
     \scalebox{0.8}{\input{newsv-mult/author/Fig_using_URA_based_RIS/Fig_10}}
    \caption{{URA} {RIS} radiation pattern with $(\varphi_{AoA},\theta_{AoA})=(\frac{\pi}{3},\frac{\pi}{8})$, $(\varphi_{AoD},\theta_{AoD})=(\frac{\pi}{8},\frac{\pi}{4})$, $N = 32\times 32$, $b_\theta = 1$ and inter-element spacing $\frac{\lambda}{4}$.}
    \label{fig:URA_RIS_32_bit1_2}
\end{subfigure}
\hfill
\begin{subfigure}{0.42\textwidth}
     \scalebox{0.8}{\input{newsv-mult/author/Fig_using_URA_based_RIS/Fig_11}}
    \caption{{URA} {RIS} radiation pattern with $(\varphi_{AoA},\theta_{AoA})=(\frac{\pi}{3},\frac{\pi}{8})$, $(\varphi_{AoD},\theta_{AoD})=(\frac{\pi}{8},\frac{\pi}{4})$, $N = 64\times 64$, $b_\theta = 1$ and inter-element spacing $\frac{\lambda}{4}$.}
    \label{fig:URA_RIS_64_bit1_2}
\end{subfigure}
\hfill
\begin{subfigure}{0.42\textwidth}
     \scalebox{0.8}{\input{newsv-mult/author/Fig_using_URA_based_RIS/Fig_12}}
    \caption{{URA} {RIS} radiation pattern with $(\varphi_{AoA},\theta_{AoA})=(\frac{\pi}{3},\frac{\pi}{8})$, $(\varphi_{AoD},\theta_{AoD})=(\frac{\pi}{8},\frac{\pi}{4})$, $N = 128\times 128$, $b_\theta = 1$ and inter-element spacing $\frac{\lambda}{4}$.}
    \label{fig:URA_RIS_128_bit1_2}
\end{subfigure}
\vspace{0.4cm}
\caption{{URA} {RIS} radiation patterns varying $N$ and with $b_\theta = 1$ for $\theta_{AoD} = \frac{\pi}{4}$.}
\end{figure*}

In the case of $1$-bit phase quantization, as illustrated in Figs.~\ref{fig:URA_RIS_16_bit1}--\ref{fig:URA_RIS_128_bit1}, array gain still persists as the number of RIS elements increases. This behavior is also maintained for quantization levels of $b_\theta = 2, 3, 4$. However, the appearance of side lobes does not significantly degrade the beam pattern, even in the single-bit scenario. Specifically, for a fixed elevation angle $\theta_{\text{AoD}} = \frac{\pi}{4}$, Figs.~\ref{fig:URA_RIS_16_bit1_2}--\ref{fig:URA_RIS_128_bit1_2} demonstrate that the impact of side lobes is less pronounced in the two-dimensional RIS configuration. Finally, apart from the array gain, increasing the number of RIS elements leads to narrower main lobes and side lobes. This enhanced angular resolution facilitates more precise identification of the directions that receive higher power, which can be particularly beneficial for beam sweeping procedures.

\subsection{Overview of Techniques and Algorithms}

As aforementioned, RISs have been identified as a pivotal technology for high-accuracy wireless localization, since they enable software-defined control of the radio environment. In this Section, the methodological landscape of RIS-aided positioning and sensing is consolidated and a structured roadmap for subsequent analysis is provided. First, RIS phase-shift configuration strategies are outlined, with emphasis on beam protocol design (e.g., sweeping, hierarchical codebooks, and tracking) and their role in generating informative measurements under diverse propagation conditions. Subsequently, localization techniques tailored to RIS-enabled systems are surveyed and organized according to operating regime, hardware assumptions, and algorithmic objectives. To support reproducibility and facilitate comparison, a taxonomy of system settings is reported in Table~\ref{tab:localization-works-taxonomy}, while a synopsis of algorithmic approaches, required inputs, and reported contributions is summarized in Table~\ref{tab:algorithm-summary}.

The discussion is then extended to machine learning approaches, where data-driven models are employed to map RIS configurations and multipath features to position estimates, with particular attention paid to training data requirements, generalization properties, and computational aspects (Tables~\ref{tab:localization-works-taxonomy-ML}–\ref{tab:algorithm-summary-ML}). Complementary angle-based and time-based estimators are revisited in the RIS context, in order to clarify performance–complexity trade-offs and to identify regimes in which RIS control yields tangible gains. Finally, the scope is broadened to RIS-aided sensing, in which passive scene understanding and environment mapping are achieved through reconfigurable scattering. Collectively, these subsections delineate current practices, expose open challenges across near-/far-field and multi-user scenarios, and establish a coherent foundation for the design and evaluation of next-generation RIS-assisted localization systems.

\subsubsection{RIS Phase Shift Configuration Strategies}
RISs emerge as a transformative element in wireless communication systems that can significantly enhance the localization process. The reconfigurability of the RIS allows the creation of additional measurements by controlling the multipath propagation environment~\cite{pan2022overview}. This capability is pivotal in overcoming traditional limitations and introduces unprecedented opportunities for accurate position estimation. Furthermore, protocols that exploit the joint cooperation of BS precoding with RIS beamforming have demonstrated enhanced effectiveness~\cite{wu2021intelligent}. Dynamic adjustment of the RIS configuration facilitates the creation of narrow and broad beams~\cite{ramezani2023broad,ramezani2023dual}, offering flexibility to adapt to diverse localization scenarios.

In the realm of RIS-aided localization, beam protocols play a crucial role in shaping the strategies used for accurate position estimation. These protocols, designed to optimize the interaction between the transmitter, receiver, and RIS, are instrumental in achieving precise localization in wireless communication systems. One prominent approach involves beam sweeping~\cite{rahal2023ris,10057262}, where a codebook-based strategy is utilized to systematically explore the area-of-influence of RIS. The objective is to generate a comprehensive codebook of distinct RIS configurations, each directed towards a specific angle in space. In the beam sweeping approach, as the RIS configuration is systematically modified, the direction of the beams is dynamically altered. This dynamic adjustment affects the signal power received at the UE. By measuring the signal power or using related metrics from the UE, the system can estimate the initial direction of the UE. This initial phase sets the stage for subsequent localization algorithms. Continuing beyond the initial beam sweeping approach, an alternative and efficient strategy is the hierarchical codebook method~\cite{10057262}. This method significantly improves beam training efficiency and accelerates beam alignment~\cite{8313072} compared to exhaustive beam sweeping. This approach organizes the codebook into hierarchical levels, allowing a more streamlined and targeted exploration of potential beam configurations~\cite{10057262}. This results in faster and more accurate beam alignment, a crucial aspect in RIS-aided localization scenarios. By adopting hierarchical codebook methods, the lLos Sueñosocalization system achieves improved efficiency and precision, laying the foundation for subsequent phases in the localization process. 

Furthermore, in RIS-aided localization, beam tracking stands out as a crucial element, characterized by continuous monitoring and adaptation of beams to estimate the trajectories of a mobile user~\cite{9772371}. This dynamic process is essential for scenarios with or without prior knowledge of user location and orientation, underlining the adaptability and effectiveness of beam protocols in addressing the evolving dynamics of the wireless environment. The inclusion of the user's trajectory adds an additional layer of complexity, emphasizing the need for real-time adjustments to maintain optimal localization accuracy. In addition, the application of beam protocols extends to SLAM methodologies~\cite{zhang2024reconfigurable}. From coarse to fine estimation, iterative processes within SLAM benefit from the adaptability of RIS beamforming, allowing for the refinement of localization and mapping information over successive iterations. This iterative approach contributes to the continuous improvement of localization accuracy in challenging and dynamic environments.  These diverse beam protocols showcase the versatility and adaptability of RIS technology in localization applications.

Further advances in RIS-aided localization research involve scenarios with multiple users~\cite{9456027, 9434917}, exploring both situations where partial user locations are known in advance and cases where no prior information exists on the positions of certain users. By adapting beamforming strategies, the protocols employed optimize localization services for multiple users. In multi-user protocols for RIS-aided localization, the integration of Received Signal Strength (RSS) fingerprinting-based methods becomes paramount~\cite{9456027, 9434917}. These protocols leverage unique RSS patterns associated with different locations to discern the positions of multiple users. Using the distinctive fingerprint information derived from the received signals, the system can simultaneously localize multiple users, each with its own distinct signature. In addition, the reconfigurability of RIS adds a layer of adaptability to the system, allowing efficient handling of the map with RSS fingerprints~\cite{9456027, 9434917, SA}. The dynamic adjustment of RIS configurations in response to changes in the environment enhances the robustness and accuracy of the RSS fingerprinting-based methods in multi-user localization protocols \cite{SCW2025a}.

\subsubsection{Localization Techniques}
In this subsection, a representative list of recent works related to RIS-enabled localization schemes is presented, which is mainly based on the available simulation settings and results. The simulation settings are summarized in Table~\ref{tab:localization-works-taxonomy}. This taxonomy includes information related to the proposed system model setting and particularly the operating field (far, near, or both), the number of deployed RISs, the RIS's type, and its phase shift design. In addition, the considered fading model is also indicated, as well as the frequency band, the operating bandwidth, and the number of subcarriers when available (otherwise N/A is mentioned in the corresponding entry). Moreover, apart from the aforementioned taxonomy, it is crucial to identify the algorithmic approaches based on which each study capitalizes to perform RISs-aided localization. To this end, Table~\ref{tab:algorithm-summary} collects all relevant information, which is grouped based on the algorithmic approach proposed for each of the included studies. More specifically, Table~\ref{tab:algorithm-summary} summarizes, in addition to the proposed algorithmic techniques, the main objectives, the information required to achieve them, as well as the overall contributions.

{
\setlength{\tabcolsep}{4pt}
\begin{longtable}{>{\centering\arraybackslash}p{0.7cm} p{0.7cm} >{\centering\arraybackslash}p{0.7cm} p{1.5cm} c p{1.5cm} >{\centering\arraybackslash}p{0.7cm} c p{1.5cm} c c >{\centering\arraybackslash}p{1.2cm}}
    \label{tab:localization-works-taxonomy} \\
    \caption{Main aspects for RIS-aided Localization in the Literature} \\
    \hline
    \rowcolor{teal!10}
    \textbf{Work} & \textbf{Field} & \textbf{RISs} & \textbf{RIS type} & \textbf{Fading} & \textbf{Phase shift design} & \textbf{BSs} & \textbf{UEs} & \textbf{BS/UE antennas} & \textbf{Band} & \textbf{BW} & \textbf{Sub-carriers} \\
    \hline
    \endfirsthead
    \caption*{Main aspects for RIS-aided Localization in the Literature} \\
    \hline
    \rowcolor{teal!10}
     \textbf{Work} & \textbf{Field} & \textbf{RISs} & \textbf{RIS type} & \textbf{Fading} & \textbf{Phase shift design} & \textbf{BSs} & \textbf{UEs} & \textbf{BS/UE antennas} & \textbf{Band} & \textbf{BW} & \textbf{Sub-carriers} \\
    \hline
    \endhead

    \bottomrule
    \endfoot
    \endlastfoot
    \cite{AMS+21} & Far & 1 & Reflective & LoS & Continuous & 1 & 1 & Many/1 & mmWave & N/A & 1\\
    \cite{HWS+20} & Far & 1 & Reflective & LoS & Codebook & 1 & 1 & Many/Many & mmWave & 100~MHz & 31\\
    \cite{AKK+21} & Near & 1 & Refractive & LoS & Continuous & 1 & 1 & 1/1 &  mmWave & 1~MHz & 1\\
    \cite{EAAH+21} & Far & 1 & Reflective & LoS & Codebook & 1 & 1 & Many/Many & N/A & N/A & 1\\
    \cite{CY21} & Near & 1 & Reflective & LoS & Codebook & 1 & 1 & Many/1 & mmWave  & 460~MHz & 1\\
    \cite{FCW+21} & Far & 1 & Reflective & LoS/NLoS & Continuous & 1 & 1 & Many/1 & mmWave & 40~MHz & 30 \\
    \cite{FKC+22} & Near & 1 & Reflective & LoS &  Codebook & 1 & 1 & Many/1 & mmWave &
    100~MHz & 1\\
    \cite{GHY+23} & Near & 1 & Holographic & LoS & Continuous & 1 & 1 & Many/1 & mmWave & N/A
    & 1\\
    \cite{GLY22} & Near & 1 & Reflective & LoS & Continuous & 1 & 1 & Many/Many & mmWave & 100~MHz & 1\\
    \cite{GLY22a} & Far & 1 & Reflective & LoS & Continuous & Many & 1 & Many/Many & mmWave & 100~MHz & 31 \\
    \cite{HJW+22} & Both & 1 & Reflective & LoS/NLoS & Continuous & 1 & Many & Many/1 & mmWave & N/A & 1\\
    \cite{KKD+21} & Near & Many & Reflective & LoS & Codebook & Many & Many & Any/Many & N/A & N/A & N/A \\
    \cite{KKS+21} & Far & 1 & Reflective & LoS & Continuous & 1 & 1 & 1/1 & mmWave & 360~MHz & 3000\\
    \cite{KSA+22} & Far & 1 & Reflective & LoS/NLoS & Codebook & 0 & 1 & 1 & FR1 & 360~MHz& 3000\\
    \cite{LLC+22} & Far & 1 & Reflective & LoS/NLoS & Continuous & 1 & 1 & 1/1 & FR1 & 190~MHz & 1584\\
    \cite{LWC+22} & Far & 1 & Reflective & LoS & Continuous & 1 & Many & 1/1 & FR1 & 10~MHz & 40 \\
    \cite{LWZ+22} & Near & 1 & Reflective & LoS & Continuous & 1 & 1 & 1/1 & mmWave & 100~MHz & 20 \\
    \cite{OKW+22} & Near & 1 & Reflective & LoS & Continuous & 1 & 1 & 1/1 & mmWave & narrow &  1\\
    \cite{RDK+21} & Near & Many& Reflective & LoS/NLoS & Continuous & 1  & 1 & 1/1 & mmWave &  360~MHz & 3000\\
    \cite{RDK+22} & Near & 1 & Reflective & NLoS & Continuous & 1 & 1 & 1/1 & mmWave & 120~kHz & 1\\
    \cite{MXL+23a} & Both & Many & Reflective & LoS/NLoS & Continuous & 1 & 1 & Many/1 & FR1 & N/A & N/A \\
    \cite{WZ21} & Both & Many & Reflective & LoS/NLoS & Continuous & 1 & 1 & Many & Many & mmWave & N/A \\
    \cite{PGA+22} & Near & Many & Reflective & LoS/NLoS & Continuous & 1 & 1 & Many/Many & FR2 & 30~kHz & 1 \\
    \cite{YZZ+23} & Far & Many & Reflective & LoS & Discrete & 1 & 1 & 1/Many & 10~GHz & 1~GHz & N/A \\
    \cite{WLS+22a} & N/A & 1 & Reflective & LoS/NLoS & Continuous & Many & 1 & 1/1 & FR1 & 500~MHz & 1 \\
    \cite{WLS+22} & Both & Many & Reflective & LoS/NLoS & Continuous/ Discrete & 1 & Many & 1/1 & FR1 & 3~MHz & N/A \\
    \cite{TYW+22a} & Far & Many & Reflective & NLoS & Continuous & 1 & 1 & Many/Many & FR2 & N/A & N/A \\
    \cite{REA22} & Near & 1 & Reflective & LoS/NLoS & Continuous & 1 & 1 & Many/Many & FR2 & 10~MHz & N/A  \\
    \cite{RBS21} & N/A & 1 & Reflective & LoS/NLoS & Continuous & 1 & 1 & Many/1 & mmWave & 5.76~MHz & 5 \\
    \cite{NBN22} & N/A & 2 & Reflective & LoS & Continuous & 1 & 1 & 1/1 & FR1 & 20~kHz & N/A \\
    \cite{ZXS+22} & N/A & Many & Reflective & LoS/NLoS & Continuous & 1 & Many & Many/1 & N/A & N/A & N/A\\
    \cite{ZZD+21} & N/A & 1 & Reflective & LoS & Discrete & 1 & Many & 1/1 & FR1 & N/A & N/A \\
    \cite{ZZD+21A} & N/A & 1 & Reflective & LoS & Discrete & 1 & Many & 1/1 & FR1 & N/A & N/A \\       
    \cite{ZZD+21b} & N/A & 1 & Reflective & LoS & Discrete & 1 & Many & 1/1 & N/A & N/A & N/A \\
    \cite{ZZF+20} & N/A & 2 & Reflective & LoS & Continuous & 1 & 1 & Many/1 & FR2 & 1~GHz & N/A \\
    \cite{ZZF+22} & Both & 1 & Reflective & LoS/NLoS & Continuous & 1 & Many & Many/1 & FR2 & 1~GHz & N/A \\
    \cite{ZZL+22} & Near & 1 & Reflective & LoS/NLoS & Continuous & 1 & Many & 1/1 & FR2 & N/A & N/A \\
    \cite{LJM+22a} & Far & 1 & Reflective & LoS/NLoS & Discrete & 1 & 1 & Many/Many & FR2 & 320~MHz & 128 \\
    \cite{YZD+21} & Far & 1 & Reflective & Multipath & Continuous & 1 & 1 & 1/Many & 10~GHz & N/A & N/A \\
    \cite{AKB22} & Near & Many & Reflective & LoS/NLoS & Continuous & 1 & 1 & 1/1 & 10~GHz & 300~kHz & N/A \\
    \cite{CAS+21} & N/A & 1 & Reflective & LoS/NLoS & Discrete & 1 & 1 & Many/Many & FR2 & N/A & N/A \\
    \cite{GA2025} & Near & 1 & DMA$^{*}$ & LoS/NLoS & Discrete & 1 & 1 & Many/Many & FR2 & N/A & N/A \\
    \cite{JAS2022a} & Near/Far & 1 & Reflective & LoS & Continuous & 1 & 1 & Many/1 & 3/28~GHz & N/A & 1 \\
    \cite{AJS2022a} & Near & 1 & Reflective & LoS & Codebook & 1 & 1 & Many/Many & 28~GHz & 100~MHz & 1 \\
    \cite{HFW2023a} & Far & Many & Reflective & LoS & Codebook & 1 & 1 & Many/1 & 28~GHz & N/A & 32--40 \\
    \cite{GGA2025} & Near & 1 & DMA$^{*}$ & LoS & Continuous & 1 & Many & 1/1 & 120~GHz & 150~kHz & N/A \\
    \cite{BGA2024} & Far & 1 & Reflective & LoS/NLoS & Continuous & 1 & 1 & Many/Many & 60~GHz & 1~GHz & N/A \\
    \cite{GAS2024a} & Far & 1 & Hybrid & LoS/NLoS & Continuous & 1 & 1 & Many/1 & N/A & 15.36~MHz & 128 \\
    \cite{AVW2022a} & Far& Many & Hybrid & LoS & Codebook & 1 & 1 & 1/1 & 30~GHz & N/A & N/A  \\
    \cite{HFA2023b} & Far & 1 & STAR$^{\dagger}$ & LoS & Codebook & 1 & 1 &Many/1 & GHz& N/A & N/A \\
    \cite{OBD2025} & Far & 1 & STAR$^{\dagger}$ & LoS & Codebook & 1 & 1 & 1/1 & 26~GHz & N/A & 1 \\
    \cite{GA2025b} & Near & 1 & Hybrid & LoS & Continuous & 1 & 1 & Many/1 & 120~GHz & 150~kHz & 1 \\
    \cite{BNK2025} & N/A & 1 & Hybrid & N/A & Codebook & N/A & N/A & N/A & 5.5~GHz & N/A & 1 \\
    \bottomrule
\end{longtable}
}
{\footnotesize 
$^{*}$ DMA: Dynamic Metasurface Antenna. \\
$^{\dagger}$ STAR: Simultaneous Transmission And Reflection. \\
$^{\ddagger}$ SIM: Stacked Intelligent Metasurface.
}

\begin{longtable}{>{\centering\arraybackslash}p{2.4cm} >{\centering\arraybackslash}p{2.4cm} >{\centering\arraybackslash}p{3.5cm} >{\centering\arraybackslash}p{2.3cm} p{4.4cm}}
    \label{tab:algorithm-summary} \\
    \caption{Summary of Algorithmic Approaches in the Literature for RIS-Aided Localization} \\
    \hline
    \rowcolor{teal!10}
    \textbf{Algorithmic Approach} & \textbf{Work} & \textbf{Objective} & \textbf{Collected Information} & \textbf{Contributions} \\
    \hline
    \endfirsthead
    \caption*{Summary of Algorithmic Approaches in the Literature for RIS-Aided Localization} \\
    \hline
    \rowcolor{teal!10}
    \textbf{Algorithmic Approach} & \textbf{Work} & \textbf{Objective} & \textbf{Collected Information} & \textbf{Contributions} \\
    \hline
    \endhead
    
    \endfoot
    \endlastfoot
    \multirow{1}{1.8cm}{Beam sweeping \& MUSIC} & \cite{AMS+21} & SNR maximization & ToA/DoA estimations from pilots & Exploits prior statistical information of the UE positions and iterative shrinks the search space.\\

    \midrule

    \multirow{1}{1.8cm}{Hierarchical codebook} & \cite{HWS+20} & Position \& Orientation MSE, Data rate & PRS & Limited feedback cases. \\

    & \cite{EAAH+21} & Codeword matching & Pilots & Multiple targets; blocked LoS. \\

    & \cite{PDC21} & Probability of error, Spectral efficiency & Received Signals & Multi-stage hierarchical codebook for target localization. \\

    & \cite{CAS+21} & AoA adaptive estimation, without knowledge for the number of targets & Normalized magnitude of the pseudo-spectrum & Accurate localization for static hidden targets in NLoS  \\
    
    \midrule

    \multirow{1}{1.8cm}{Maximum Likelihood estimation (MLE)-based} & \cite{AKK+21} & Position Error Bound (PEB) & Pilots & RIS is acting as a lens. \\

    & \cite{CY21} &  PEB \& CSI estimation & Received signals & Co-planar MLE formulation over subsets of RIS units with closed-form solution. \\

    & \cite{MXL+23a} & Position \& PEB & Estimated CSI & Two-Step Indoor Positioning Approach for Distributed RISs. \\

    & \cite{YZJ+21} & PEB, Root Mean Square Error (RMSE), Spectral Efficiency & Received signals &  Communication and localization with an extremely large lens antenna array. \\

    & \cite{WZ21} & MSE, Misalignment Rate, Blockage Rate & Received signals &  A joint beam training and positioning design for RIS Assisted mmWave Communications system. \\    

    & \cite{KSA+22} & PEB, Position error, Propability of error & Reflected signals & 3D self-localization of UE with a single RIS using multiple OFDM signals for transmission and processing of the reflected signal. \\

    \midrule

    \multirow{1}{1.8cm}{Codebook searching} & \cite{FCW+21}, \cite{FKC+22} & PEB & CSI & Alternate optimization over a relaxed problem; synchronization is also considered. \\ 

    \midrule

    \multirow{1}{2.5cm}{Iterative Entropy Regularization} & \cite{GHY+23} & Cramér--Rao Lower Bound
(CRLB) minimization & Pilots & Considers holographic RISs. \vspace{0.4cm} \\

    \midrule

    \multirow{1}{1.8cm}{Heuristic Search} & \cite{GLY22} & Minimum squared PEB (SPEB) & OFDM pilots & A two-step optimization with good convergence properties. \\

    \midrule

    \multirow{1}{1.8cm}{Extended Kalman Filter} & \cite{PGA+22} & Position, trajectory & Pilot signals &  Tracking and communication performance of a single user MIMO system with multiple RISs.
    
    \\
    
    \midrule

    \multirow{1}{1.8cm}{SLAM} & \cite{YZZ+23} & RMSE \& CRLB & Received signals &  Wireless Simultaneous Localization and Mapping utilizing RISs.
    
    \\

    \midrule

    \multirow{1}{1.8cm}{MUSIC \& ESPRIT} & \cite{YHL+22} & Sum rate, RMSE, Cumulative Density Function (CDF) & Received signals &  Position sensing and beamforming design for RIS-enabled multi-user Integrated Sensing and Communication (ISAC) system.
    
    \\

    \midrule

    \multirow{1}{1.8cm}{Fisher Information Analysis} & \cite{WLS+22a} & SPEB & Wideband signals & Fisher information analysis to determine the theoretical limits of wideband localization with RISs. \\

    & \cite{WLS+22} & SNR, Root SPEB, Spectral Efficiency & Bayesian Cramér--Rao Bound (CRB) & Fisher information analysis of localization performance to determine the theoretical limits of wideband localization with RISs.
    
    \\

    \midrule

    \multirow{1}{1.8cm}{Bayesian-based} & \cite{TYW+22a} & RMSE over time & Reflected signals & Bayesian user localization and tracking for RIS Aided MIMO system.
    
    \\
    \midrule

    \multirow{1}{1.8cm}{MUSIC} & \cite{SYM+22} & RMSE, Probability of successful estimation & Reflected signals & Target Sensing with RIS and architecture design.
    
    \\
    
    \midrule

    \multirow{1}{1.8cm}{Compressed Sensing} & \cite{REA22} & SNR, Localization Error & Reflected signals & Compressed sensing in the near-field regime for multipath RIS-assisted environments.
    
    \\

    \midrule

    \multirow{1}{2.2cm}{Water filling \& Gradient-based } & \cite{RBS21} & MSE, RMSE & Reflected signals & Privacy aware localization enhancement with the aid of RISs.
    
    \\
    
    \midrule

    \multirow{1}{1.8cm}{Phase shift RIS configurations} & \cite{NBN22} & Location Estimation Error & Reflected/ Received signals & Wireless localization in the presence of RISs, while not enough transmitters are present.
    
    \\
    
    \midrule

    \multirow{1}{1.8cm}{Probability estimation} & \cite{ZHZ+22} & Localization error & RSS measurements & MetaRadar: a novel indoor localization system using reconfigurable radio reflection with metamaterial units.
    
    \\

    \midrule

    \multirow{1}{1.8cm}{2D search} & \cite{ZXS+22} & Positioning/Average Error & Reflected signals & Multiple Aerial RISs are deployed in the airspace to locate users within a defined area.
    
    \\

    \midrule

    \multirow{1}{1.8cm}{Phase Shift Optimization} & \cite{ZZD+21} & Localization loss \& Error & Reflected signals \& RSS & Leveraging RIS for multi-user localization by customizing radio channels, specifically adjusting signal phase shifts.
    
    \\

    \midrule

    \multirow{1}{1.8cm}{Configuration Optimization} & \cite{ZZD+21A} & Positioning error, Running time & Reflected signals \& RSS & RSS-based positioning, where by adjusting the RIS configuration, the differences in RSS values are enhanced.
    
    \\

    \midrule

    \multirow{1}{1.8cm}{Localization Error Minimization} & \cite{ZZD+21b} & Number of slots, Localization error & Reflected signals, RSS measurements & RSS-based localization with only one BS and RIS phase shifts optimization.
    
    \\

    \midrule

    \multirow{1}{1.8cm}{RIS optimization \& cross correlation } & \cite{ZZF+20} & RMSE & PRS, reflected signals & Utilizing RIS to enhance BS localization of UE in mmWave MIMO systems with a two-stage positioning method.
    
    \\

    \midrule

    \multirow{1}{4cm}{Triangulation-based localization, Semidefinite Programming (SDP), Block Coordinate Descent (BCD)} & \cite{ZZF+22} & Sum of transmit power &  Received signals & An RIS-assisted positioning method for simultaneous and accurate localization of numerous energy-limited IoT devices.
    
    \\

    \midrule

    \multirow{1}{1.8cm}{CRLB minimization} & \cite{ZZL+22} & CRLB, Power allocation & Reflected/ Cooperative links &  A multi-user localization system using a RIS, where users derive position information from both reflected RIS and cooperative links. \\
    
    & \cite{GLY22a} & PEB & Received signals & Developed two optimal algorithms for beamforming design to achieve optimal cooperative localization performance using CRLB in a mmWave system with RIS assistance.
    \\

    \midrule

    \multirow{1}{1.8cm}{VR identification} & \cite{HJW+22} & Localization error, Normalized Mean Square Error (NMSE) & Received signals & RIS-enhanced massive MIMO with users in RIS near-field and spatially nonstationary channels.
    \\

    \midrule

    \multirow{1}{1.8cm}{RIS design, CRB, Quasi-Newton} & \cite{KKD+21} & Probability of position error & Received signals & Localized multi-user problem using RIS, estimating 3D positions by calculating ToA for LoS and NLoS paths across multiple receivers.
    \\

    \midrule

    \multirow{1}{1.8cm}{1D/2D searches} & \cite{KKS+21} & PEB & Received signals & Localization and synchronization in a wireless system with a single-antenna UE, single-antenna BS, and RIS.
    \\

    \midrule

    \multirow{1}{1.8cm}{BCD} & \cite{LLC+22} & Average PEB/Spectral efficiency & Reflected/ Received signals & Urban communication and localization integration using RIS in blind areas, aiding both communication and localization for ground vehicle.
    \\

    \midrule

    \multirow{1}{1.8cm}{PEB minimization, Penalty-based optimization} & \cite{LWC+22} & Rate, SPEB & Reflected signals & Explored RIS-enhanced Orthogonal Frequency-Division Multiple Access (OFDMA) system, optimizing subcarrier assignment and RIS phase shifts for enhanced performance.
    \\

    \midrule

    \multirow{1}{1.8cm}{SPEB minimization, Newton method} & \cite{LWZ+22} & CRLB, Root Mean Square Error (RMSE), Transmit power & Reflected signals & A framework for RIS-assisted near-field regional localization, involving RIS phase design and position determination.
    \\

    \midrule

    \multirow{1}{1.8cm}{Mismatched MLE} & \cite{OKW+22} & RMSE, Lower bound & Reflected signals & 
    RIS-assisted near-field localization with model misspecification, considering the mismatch between ideal and realistic RIS amplitude models.
    \\

    \midrule

    \multirow{1}{1.8cm}{Fisher Information Matrix (FIM) derivation} & \cite{RDK+21} & PEB & Reflected signals & 
    Analyzed theoretical positioning performance of SISO multi-carrier downlink multipath-aided localization in both LoS and NLoS conditions with a single RIS in reflection mode.
    \\

    \midrule

    \multirow{1}{1.8cm}{PEB minimization} & \cite{RDK+22} & PEB, Gain & Reflected signals & 
    Designed reflective RIS phase profile to minimize NLoS localization error in downlink SISO narrowband transmissions, considering a generic near-field formalism for the RIS response.
    \\

    \midrule

    \multirow{1}{1.8cm}{Nonlinear solvers, Tensor factorization} & \cite{LJM+22} & RMSE & Reflected/ Received signals & 
    Channel estimation and environment mapping applications in RIS-empowered mmWave MIMO-OFDM systems. \\

    \midrule 
    & \cite{LJM+22a} & RMSE & Reflected/ Received signals & 
    Channel estimation and user localization challenges in an RIS-empowered mmWave MIMO-OFDM system.
    \\

    \midrule

    \multirow{1}{1.8cm}{SLAM} & \cite{YZD+21} & Channel gain, RMSE & Reflected/ Received signals & 
    RIS-assisted SLAM system for improved indoor positioning in future 6G systems.
    \\

    \midrule

    \multirow{1}{1.8cm}{TDoA-based method} & \cite{AKB22} & RMSE & Reflected/ Received signals & A novel RIS-aided localization system for RF transmitters, addressing challenges of TDoA methods such as synchronization and high-throughput links.
    \\
    \midrule
    
    \multirow{1}{1.8cm}{Decoupling Illumination} & \cite{JAS2022a} & Overhead reduction & SNR feedback & Decouples RIS reconfiguration from channel estimation via wide vs. focused illumination. \\
    
    \midrule
    
    \multirow{1}{1.8cm}{Hierarchical Beam Search} & \cite{AJS2022a} & SNR maximization & Received Power & Variable-width hierarchical codebook designed for near-field mmWave management. \\
    
    \midrule
    
    \multirow{1}{1.8cm}{Atomic Norm Minimization} & \cite{HFW2023a} & Localization MSE & Uplink Pilots & 3D localization with distributed passive RISs using Zero-Forcing separation. \\
    
    \midrule
    
    \multirow{1}{2.5cm}{Convex Optomization \& Neumann Series} & \cite{GGA2025} & PEB minimization & Channel Parameters & Physics-consistent coupled-dipole model for bistatic sensing with 2D waveguides. \\
    
    \midrule
    
    \multirow{1}{1.8cm}{Probabilistic Data Association} & \cite{RSA2025} & Range/Velocity MSE & Received Signals & SIM-aided bistatic ISAC in doubly-dispersive channels using gradient ascent. \\
    
    \midrule
    
    \multirow{1}{3.2cm}{Extended Kalman Filter \& Alternating Optimization} & \cite{GA2025c} & PEB \& Data Rate & Echoes/Pilots & Tracking-aided communications using Hybrid RIS for simultaneous sensing/reflection. \\
    
    \midrule
    
    \multirow{1}{1.8cm}{Analog Computing} & \cite{OBD2025} & AoA Sensing & Focal Point & STAR-RIS performing holographic beamforming (reflection) and Fourier transform (transmission). \\
    
    \midrule
    
    \multirow{1}{3.8cm}{Semi-definite relaxation (SDR) \& Alternating Optimization} & \cite{GA2025b} & Rate Max s.t. PEB & Echoes & Near-field ISAC with Hybrid RIS ensuring localization coverage across an area of interest. \\
    \bottomrule
\end{longtable}

\paragraph{\textbf{Machine Learning Approaches}}
Machine learning models can automatically uncover the complex relationships between RIS phase configurations and multipath signal characteristics \cite{SSA2025}, \cite{SA2022a}, \cite{ASH2022a}, allowing for precise estimation of angle-of-arrival and time-of-arrival \cite{SGF2025}, \cite{HAY2019a}. By training in various propagation scenarios, these data‐driven approaches can adaptively tune the RIS to optimize localization accuracy even in highly dynamic environments. Arguably, the most straightforward approach to RIS-based localization and mapping is the formulation of Supervised Learning (SL) \cite{SCW2025a}, \cite{ASB2020a}. In this paradigm, a data collection step is first assumed where channel measurements are acquired, preprocessed, and stored in a database, similar to fingerprinting localization approaches. The channel measurements could be raw received signals, estimated Time of Arrivals (ToAs) / Direction of Arrivals (DoAs), collected Channel State Information (CSI), or Received Signal Strength Indicator (RSSI)-like information. The measurements constitute the inputs of the SL predictor (typically implemented as a Deep Neural Networks (DNN)). For each collected record, an accompanying \textit{ground truth} target or \textit{label} value needs to be available. The DNN aims to predict the target values based on the input information. The values themselves therefore correspond to the objective of the application (e.g., target coordinates for localization, signal strengths over an area of interest for mapping, etc.). Acquiring labeled data is a challenging task in wireless communications, as additional sensing infrastructure may need to be deployed (e.g., making use of GNSS for accurate position estimations). The DNN is trained on a subset of the collected data so that it outputs target values that are close to the acquired labels given a set of input values. Finally, it should be noted that SL is designed for problems where the statics of the environment during the data collection phase do not change during the deployment phase, which is not always guaranteed in wireless networks. Nevertheless, they have the advantages of not being based on hard modeling assumptions (which may not accurately hold in realistic scenarios) and that predictions over trained networks can be obtained with near-constant computation time complexity, given appropriate hardware. The research community has proposed a variety of SL formulations with the objective of localization in the presence of RIS. The main characteristics and approaches of the described machine learning-oriented studies for localization with RISs are summarized in Tables \ref{tab:localization-works-taxonomy-ML} and~\ref{tab:algorithm-summary-ML}.

\begin{longtable}{>{\centering\arraybackslash}p{0.8cm} p{0.8cm} >{\centering\arraybackslash}p{0.8cm} p{1.7cm} c p{1.5cm} >{\centering\arraybackslash}p{0.8cm} c p{1.5cm} c c >{\centering\arraybackslash}p{1.2cm}}
    \label{tab:localization-works-taxonomy-ML} \\
    \caption{Main aspects of Machine Learning Techniques for RIS-aided Localization} \\
    \hline
    \rowcolor{teal!10}
    \textbf{Work} & \textbf{Field} & \textbf{RISs} & \textbf{RIS type} & \textbf{Fading} & \textbf{Phase shift design} & \textbf{BSs} & \textbf{UE} & \textbf{BS/UE antennas} & \textbf{Band} & \textbf{BW} & \textbf{Sub-carriers} \\
    \hline
    \endfirsthead
    \caption*{Main aspects of Machine Learning Techniques for RIS-aided Localization} \\
    \hline
    \rowcolor{teal!10}
     \textbf{Work} & \textbf{Field} & \textbf{RISs} & \textbf{RIS type} & \textbf{Fading} & \textbf{Phase shift design} & \textbf{BSs} & \textbf{UEs} & \textbf{BS/UE antennas} & \textbf{Band} & \textbf{BW} & \textbf{Sub-carriers} \\
    \hline
    \endhead

    \bottomrule
    \endfoot
    \endlastfoot
    \cite{CSF+23} & Any & 1 & Reflective & RT & 1-bit & 1 & Many & 1/1 & FR1 & 20MHz & N/A \\
    \cite{SBP+23} & Any & 1 & DMA$^{*}$ & RT & Continuous & 1 & 1 & 16/1 & FR2 & N/A & 4 \\
    \cite{NGG+21} & Any & 1 & Reflective & LoS & Adaptive codebook & 1 & 1 & 1/1 & FR1 & N/A & 1 \\
    \cite{ZJY23b} & Far & Many & Reflective & Ricean & Continuous & 1 & 1 & Many/1 & N/A & 10MHz & 1 \\
    \cite{WZZ+21b} & N/A & 1 & Reflective & LoS & N/A & 1 & 1 & 1/1 & FR1 & N/A & N/A \\
    \cite{HCZ+23} & Near & 1 & Holographic & Multipath & Continuous & 1 & Many & 1/1 & FR1 & 250kHz & 1 \\
    \cite{HZB+21} & Near & 1 & Reflective & LoS & Discrete & 1 & None & 1/None & FR1 & N/A & 1 \\
    \cite{SCW2025a} & Near & 1 & Reflective & Ricean & 1-bit & 1 & 1 & 1/1 & N/A & N/A & N/A \\
    
    \bottomrule
\vspace{2mm}
\end{longtable}
{\footnotesize $*$ DMA stands for Dynamic Meta-surface Antennas.}

\begin{longtable}{>{\centering\arraybackslash}p{2cm} >{\centering\arraybackslash}p{0.8cm} >{\centering\arraybackslash}p{3cm} >{\centering\arraybackslash}p{2cm} p{2cm} p{4.4cm}}
    \label{tab:algorithm-summary-ML} \\
    \caption{Summary of Machine Learning Approaches in the Literature} \\
    \hline
    \rowcolor{teal!10}
    \textbf{Algorithmic Approach} & \textbf{Work} & \textbf{Objective} & \textbf{Collected Information} & \textbf{Use of RIS} &\textbf{Contributions} \\
    \hline
    \endfirsthead
    \caption*{Summary of Machine Learning Approaches in the Literature} \\
    \hline
    \rowcolor{teal!10}
    \textbf{Algorithmic Approach} & \textbf{Work} & \textbf{Objective} & \textbf{Collected Information} & \textbf{Use of RIS} &\textbf{Contributions} \\
    \hline
    \endhead

    \endfoot
    \endlastfoot

    \multirow{1}{1.8cm}{Supervised Learning} & \cite{CSF+23} & People Counting & CSI ($12000$ samples) & Signal diversity / virtual LoS & A transformer-based DNN in blocked BS-UE cases. Evaluated with RT and experimental data over WiFi. \\

    & \cite{SBP+23} & User tracking & CSI ($10^6$ samples) &Receiver & A two-stage approach for real-time tracking in blocked-LoS multipath cases. \\

    & \cite{NGG+21} & Localization and mapping & RSSI ($250000$ samples) & Signal diversity & Evaluated different fingerprinting approaches and obtained RF mappings. The RIS codebook was optimized throughout the learning process. \\

    & \cite{ZJY23b} & Localization & Received pilots ($2\times 10^6$ samples) & Active sensing through beamforming & Proposed an Long Short-Term Memory (LSTM) network that controlled the RISs reflection patterns while predicting the user location to simultaneously learn position-aided beamforming. \\

    & \cite{WZZ+21b} & 3D imaging (pose estimation) & Received pilots ($640000$ samples) & Signal diversity & Deployed a DNN trained on measured data to predict the coordinates of $18$ human skeleton points for pose estimation. \\

    \midrule

    Federated learning & \cite{HCZ+23} & Localization & Received pilots ($10^4$ samples) & Diversity through holographic beamforming & Enables multiple users to perform localization using a jointly trained DNN without exchanging sensitive data. The holographic RIS is optimized to provide diversity in the received signals. \\

    \midrule

    Reinforcement Learning & \cite{HZB+21} & Sensing of passive objects & Received pilots ($10000$ samples) & Diversity through reconfiguration & A joint approach for learning to individually control the phase shift of each element and detecting the presence of objects in a 3D grid. \\

    \midrule

    Neuroevolution and Supervised Learning & \cite{SCW2025a} & UE localization & Received pilots (70000 samples) & Adaptive power allocation & Formulation of the problem of joint RIS phase profile selection and UE pilot transmit power control for localization, considering RISs with elements of discrete response. \\
    
    \bottomrule
\end{longtable}

\paragraph{\textbf{Angle-Based Localization}}

The RIS functions as a new reference point, offering a consistent distance and angle between the BS and the RIS~\cite{pan2022overview}, especially in static setups. This constancy simplifies the estimation process in localization, providing a stable geometric relationship. Furthermore, the RIS configuration, shared between the BS and the RIS, allows the extraction of the reflected angle~\cite{pan2022overview,ramezani2023dual,ramezani2023broad}. Unlike Snell's law, where the reflected angle equals the incident angle, the RIS reflects the signal with an AoD determined by its configuration. This configuration-AoD relationship can be leveraged through codebook-based approaches~\cite{Ahmed23}. The resolution, indicating how directive the signal will be at the desired position, depends on the control bits of the RIS, incident angle, and hardware imperfections. 

In the context of RIS-aided localization with one base station, classic angle-of-arrival estimation techniques are utilized. The surface of the RIS will be used to enhance the SNR of these known techniques, enabling localization in areas where a LoS path is not present. Additionally, the precision of these techniques can be further enhanced by utilizing the diversity of the RIS, enabling the reception of multiple time-multiplexed snapshots. 

In the literature, there are two widely used categories when referring to how the AoA can be computed, that is,
\begin{itemize}
    \item The classic AoA techniques, the most popular being the Delay and Sum \cite{bartlett} and the Minimum Variance Distortionless Response (MVDR) \cite{mvdr}, also known as the Bartlett and Capon estimators, respectively.
    \item The subspace AoA techniques, which include the MUltiple SIgnal Classification (MUSIC)~\cite{MUSIC} and the Estimation of Signal Parameters via Rotational Invariance Techniques (ESPRIT)~\cite{DBLP:journals/tsp/RoyK89}. As the aforementioned techniques have been widely used for more than three decades, many variations have been developed to reduce their computational complexity and improve their accuracy. In recent years, many variations of the above techniques have been developed to improve the positioning accuracy and/or to reduce the computational complexity.
\end{itemize}

The Delay and Sum beamformer is a simple and intuitive method, according to which signals from different array elements are delayed and then summed to create a beamformed output. Although easy to implement, Delay and Sum may not perform well in scenarios with noise or interference, as it does not consider the spatial characteristics of the signals. The spectral power of the Bartlett beamformer is represented by 
\begin{equation}
    \hat{P}_{\rm Bartlett}(\theta) = \mathbf{a^H(\theta) R a(\theta)},
\end{equation}
where $\mathbf{a}$ is the array steering vector, and $\mathbf{R}$ is the signal covariance matrix, respectively.

In contrast, the MVDR beamformer is designed to minimize output power while preserving the desired signal power. This is achieved through adaptive weighting of array element signals based on the covariance matrix of received signals. MVDR excels in effectively suppressing interference by accounting for the spatial distribution of signals. However, it is worth noting that the computational demands associated with the implementation of MVDR are relatively higher than those of Delay and Sum. The spectral power of the Capon beamformer is given by
\begin{equation}
    \hat{P}_{\rm Capon}(\theta) = \frac{1}{\mathbf{a^H(\theta) R^{-1} a(\theta)}}.
\end{equation}

When referring to subspace techniques, the MUSIC algorithm is a common AoA estimation technique, providing high resolution in angular estimation. This comes at the cost of requiring full a priori knowledge of the number of sources and the array response. The signal and noise subspaces are distinguished through an eigen-decomposition operation applied on the covariance matrix of the received signal, requiring substantial computational complexity. The spectral power of the MUSIC algorithm is
\begin{equation}
    \hat{P}_{\rm MUSIC}(\theta) = \frac{1}{\mathbf{a^H(\theta) U_n U^{H}_n  a(\theta)}},
\end{equation}
where $\mathbf{U_n}$ consists of the eigenvectors of the noise subspace.

An overview of the AoA estimation using the MUSIC algorithm is presented in \cite{Gupta}, \cite{Devendra}, focusing on factors that can improve results. These factors include increasing the separation between antenna elements, expanding the array of antenna sensors, improving the number of snapshots, and the disparity in incidence angles among the incoming signals. Furthermore, an alternative and improved version of the MUSIC algorithm for coherent signals is presented in \cite{Gupta}, where an identity transition matrix $\mathbf{T}$ is introduced, having an effect on the covariance matrix of the received signal and essentially in the noise subspace computation. A performance analysis of the MUSIC algorithm has been presented in \cite{DBLP:conf/icassp/SpielmanPK86a}. The authors addressed the relationship between the performance of the algorithm and the array manifold design. They concluded that the optimum array geometry is dependent on the signal environment. They also show that the angle between two array manifold vectors is a reliable performance measure and that the traditional work on array design applies to the conventional MUSIC algorithm. To reduce computational complexity, the Root-MUSIC algorithm has been developed \cite{DBLP:conf/icassp/Barabell83}. Root-MUSIC follows the same structure as the conventional MUSIC algorithm, but the AoA are determined from the roots of a polynomial formed from the noise subspace and is only applicable to ULAs. The Root-MUSIC algorithm transforms the search step in the conventional MUSIC algorithm, which requires a lot of computational cost, into a simplified polynomial rooting. Despite the computational lower complexity of this polynomial rooting step compared to that of the spectral search, the complexity of Root-MUSIC is still high, especially when a large number of sensors is used. This is due to the eigen-decomposition, since the cost is $O(M^3)$. A performance analysis of Root-MUSIC is presented in \cite{DBLP:journals/tsp/RaoH89}. The authors have derived closed-form expressions for the Mean Square Error (MSE) in the estimates of the signal zeros and AoA and have presented simplified expressions for the cases of one and two sources. They concluded that the error in the signal zeros have a largely radial component, explaining the fact that Root-MUSIC is superior to the classical MUSIC algorithm. 

In addition, the MUSIC algorithm can be expanded in the two-dimensional (2D) space to estimate the multipath propagation delay as well as the AoA. The propagation delay estimation can be useful in the case that more than one of the impinging signals arrive at the same angle, or if a time of arrival estimation needs to be performed. In the first case, the traditional one-dimensional (1D) MUSIC algorithm will not be able to separate the signals, opposite to 2D MUSIC as the latter utilizes a 2D spectrum for both the angle and delay estimation. In \cite{DBLP:journals/tbc/Oziewicz05}, a study has shown how propagation delay affects an Orthogonal Frequency Division
Multiplexing (OFDM) signal. The 2D MUSIC spectrum is obtained by computing
\begin{equation}
    \hat{P}_{\rm 2D-MUSIC}(\theta) = \frac{1}{\bigl( a(\theta)\otimes b(t) \bigr)^H \bm{U_n} \bm{U^{H}_n} \bigl( a(\theta)\otimes b(t) \bigr)},
\end{equation}
where the operator $\otimes$ denotes the Kronecker product between the angle-delay pair. The 2D MUSIC algorithm obtains the angle-delay pair by performing a two-dimensional search. As this process is very demanding in terms of computation time and resources, requiring a search in the 2D space, the primary focus of recent works has been in reducing the computational complexity. In \cite{Li2020}, two subsequent 1D searches are performed to obtain the AoA and then the propagation delay. For that purpose, an extended virtual array is first constructed combining the array structure and the frequency domain information. Then, the multipath effect is eliminated by using the smoothing pre-processing technique. Then, the AoA estimation is performed from the channel frequency response matrix, using a closed-form solution for reduced complexity. Lastly, a 1D search is performed to obtain the propagation delays. 

On the other hand, in \cite{DBLP:conf/wcnc/HenningerMAB22}, the CSI is utilized to obtain more accurate results. CSI decimation is performed, improving the probability of detection for closely spaced targets compared to the 1D approach. Moreover, the computational complexity decreases as the number of elements per sub-array is also reduced due to the decimation. In \cite{DBLP:conf/icc/BazziSM16}, the presence of local scattering is also considered while jointly estimating AoA and propagation delays. The signal model is described as a sum of ``clusters'', where each cluster is composed of multi-incident rays of the same angles and ToA. The proposed algorithm reduces the MSE of the estimated AoAs compared to the traditional MUSIC algorithm. Apart from MUSIC, the ESPRIT algorithm is based on an analysis of a subspace used to localize a source or estimate parameters of the signal (frequency, phase, etc.). The basic idea of this algorithm is to ``split'' the antenna array into sub-arrays separated by an equivalent displacement. The overall less complexity of the ESPRIT algorithm grants this technique an advantage over the MUSIC algorithm in certain applications that are limited by the available resources. On a similar basis with the 2D MUSIC algorithm, the multiple parameter estimation problem can be formulated in such a form that the ESPRIT algorithm can jointly estimate the angle and the delay in closed form. This algorithm is commonly known as Joint Angle and Delay Estimation (JADE) ESPRIT, and is described in \cite{DBLP:journals/tsp/VanderveenVP98}.

\paragraph{\textbf{Time-Based Localization}}

For the computation of the absolute position between two units, the time-of-arrival information needs to be extracted from the received signal. In the context of 5G networks, the larger available bandwidth offers higher precision of ToA estimation \cite{gu2022high}.  Generation Cross-Correlation (GCC), initially introduced by Knapp and Carter in their influential 1976 paper \cite{GCC}, remains the predominant method for Time-Delay Estimation (TDE) to this day. Using GCCs, the Time Difference of Arrival (TDoA) between two signals is determined by identifying the time lag that maximizes the cross-correlation between filtered iterations of these signals. In \cite{GCCweighted}, an improved GCC-based technique for TDE based on subband analysis of the cross-power spectrum phase is proposed. A sliding window method is introduced, improving the performance of the traditional GCC method, at the cost of reduced temporal resolution. In \cite{TRUWB}, a two-step estimation process for a transmitted reference ultra-wideband receiver is proposed, where in a coarse step a sliding correlation over a symbol length is performed to find the signal block where the direct path is enclosed, and then a second step where the correlation is more finely evaluated to estimate the precise beginning of the pulse. The computational complexity of the algorithm is considered low, and their simulations and prototyping have shown that under both LoS and, in certain circumstances, NLoS conditions, their algorithm performs consistently accurate in high SNR. The authors in \cite{jointcfotoa} proposed a joint carrier frequency offset and time of arrival estimation algorithm, using the Primary Synchronization Signal (PSS). As the carrier frequency offset gets larger and larger, the PSS loses its useful cross-correlation, and it is proven that even for large carrier frequency offsets, the algorithm performs substantially well. 

Although properties similar to those of unitary ESPRIT have been utilized to reduce its complexity, its complexity remains high. In \cite{JADElowcomplexity}, a low complexity version of the algorithm is proposed, which maximizes the utility of Fast-Fourier Transforms (FFTs), exploiting a cascading estimation scheme. By exposing the algorithm to 5G signals, the results are satisfactory enough to make it a viable option for real-time JADE in future 5G and beyond networks for high-accuracy positioning. In \cite{gu2022high} a ToA algorithm is proposed for the more challenging scenarios due to reflections and NLoS conditions in an indoor environment. This method achieves sub-meter level accuracy. The first correlation is used to estimate the integer part of ToA, which is estimated at the sampling period. Based on the integer part of the ToA estimation, the resolution of the filtered signal is improved by interpolation, and a secondary correlation is carried out to obtain the fractional part of ToA, which is estimated within the sampling period. 

On an experimental basis, in \cite{toaexp1} the authors present a method to obtain positioning information based on AoA and ToA estimations. In particular, the concept of timing advance is described, as the BS orders the UE to send the information prior to its time, to avoid that the UE sends information outside of the given time-slot. Cross-correlation between the Positioning Reference Signal (PRS) and a local replica is performed to measure the ToA. In addition, another experimental setup for localization in Long-Term Evolution (LTE), utilizing Round-Trip Time and TDoA measurements from two base stations, is presented in \cite{toaexp2}, where a filter bank framework is introduced. Initially, two analytical solutions representing the intersection of a Round-Trip Time (RTT) circle and a TDoA hyperbola were derived. The nonlinear mapping from noisy measurements to the 2D positions of the Mobile Station was approximated using an estimator based on the unscented transformation. To estimate the mobile station position, the two user terminal estimates were integrated into a filter bank as pseudo-measurements. The filter bank continuously tracks all potential solutions until additional information becomes available. As the mobile station transitions through the network, the serving BS changes due to handover procedures. The information obtained from the handovers is automatically incorporated into the filter bank. As a result, real-field experiments demonstrate favorable results.

\subsubsection{RIS-aided Sensing Techniques}

RIS-aided localization, as mentioned in the previous Sections, is the problem of finding the location of an active device in the network by applying the appropriate processing of signals received from network anchor points (such as BSs, APs, RISs, etc.) and specifically with the help of an RIS. On the other hand, RIS-aided sensing involves the problem of finding various passive objects present in space with the help of one or more RIS-nodes. Thus, it involves an attempt to map an area and locate various objects and obstacles that may exist in it. Apart from some recent studies that perform SLAM \cite{YZZ+23}, \cite{YZD+21}, there also exist the following (limited in number) recent works. Specifically, \cite{LJM+22} addresses channel estimation and environment sensing in RIS multiple-input multiple-output (MIMO) OFDM systems. With the introduction of a novel three-dimensional conformal RIS architecture with reflective unit cells on curved surfaces, training signals using a third-order canonical polyadic tensor are modeled. Using tensor techniques and nonlinear solvers, a four-channel estimation approach is developed under specified temporal-frequency training conditions. Exploiting the unique characteristics of conformal RISs, two decoupling modes for precise multipath parameter recovery are also proposed. \cite{KCK+23} proposes an RIS-enabled SLAM problem dealing with a mobile UE, which can be outside the coverage of any BS, or where the BS only controls the UE and RIS without directly transmitting pilot signals for SLAM. By introducing a dynamic RIS phase profile design, the mapping of the landmarks (i.e., the RIS scattering and reflection points) is used to improve the UE localization performance. By numerical evaluations, it is validated that this approach provides satisfactory performance even for small numbers of transmissions and for cases where the UE's speed changes considering the Doppler shift. These works are also summarized in Table~\ref{tab:sensing-works-taxonomy}, with respect to various characteristics of the considered system.


\setlength{\tabcolsep}{4.5pt} 
\renewcommand{\arraystretch}{1.0} 

\begin{table}[ht]
    \centering
    \caption{Main aspects of RIS-aided Sensing Techniques \rule[-2ex]{0ex}{0pt}}
    \label{tab:sensing-works-taxonomy}
    \begin{tabular}{
      >{\centering\arraybackslash}p{0.55cm}
      >{\centering\arraybackslash}p{0.45cm}
      >{\centering\arraybackslash}p{0.45cm}
      >{\centering\arraybackslash}p{1.3cm}
      c
      p{1.3cm}
      >{\centering\arraybackslash}p{0.6cm}
      c
      p{1.3cm}
      c
      c
      >{\centering\arraybackslash}p{1.0cm}
    }
        \hline
        \rowcolor{teal!10}
        \textbf{Work} & \rotatebox[origin=b]{90}{\textbf{Field}} & \rotatebox[origin=b]{90}{\textbf{RISs}} & \textbf{RIS type} & \textbf{Fading} & \textbf{Phase shift design} & \textbf{BSs} & \textbf{UEs} & \textbf{BS/UE antennas} & \textbf{Band} & \textbf{BW (MHz)} & \textbf{Sub-carriers} \\
        \hline
        \cite{LJM+22} & Far & 1 & Reflective & LoS/NLoS & Continuous & 1 & 1 & Many/Many & mmWave & 320 & 128 \\
        \cite{KCK+23} & Far & 1 & Reflective & LoS & Continuous & No & 1 & 0/Many & mmWave & 200 & 1600 \\
        \bottomrule
    \end{tabular}
\end{table}

\subsection{Performance Evaluation using Ray Tracing Models}
\subsubsection{Assumptions for ToA estimation and Common Sense of Time}
\label{sec:RIS_aided_TOA_sense_of_time}
To accurately localize a UE, it's not enough to estimate only the azimuth AoA at the receiver. Also the receiver’s height and its distance from a reference node (or so-called anchor node) with a known position (e.g., the RIS, the BS, the AP etc.) need to be determined. However, estimating the UE’s altitude—which is embedded in the elevation angle between the RIS and the receiver—requires that the receiver is equipped with a URA. While the RIS is capable of performing beamforming in the elevation domain, the receiver cannot estimate the elevation AoA, since it typically uses a ULA rather than a URA. This is a common and practical constraint in real-world receiver designs. Therefore, to obtain the receiver’s height, a simple and low-cost solution, such as a barometer, can be adopted in most of the cases. However, the analysis provided here can be handled as a methodology to estimate also the elevation angle (and thus the height) of the receiver (Rx) with respect of an anchor node, when the URA implementation on the Rx is a feasible solution. The only difference when using a URA at the Rx is that 2-dimensional super resolution algorithms (such as 2D-MUSIC~\cite{xie2021performance,belfiori20122d} and/or 2D-ESPRIT~\cite{haardt19952d}) are used for the AoA estimation problem.

To estimate the distance between the RIS and the Rx, a process known as ranging, and ToA-based techniques are commonly preferred. Broadly, radio frequency (RF)-based ranging systems fall into two main categories. The first category relies on analyzing physical signal attributes such as amplitude, phase, or frequency to estimate distance~\cite{singh2022v}. These systems are relatively easy to implement but are sensitive to channel-induced distortions and typically require significant error correction. The second category includes systems that determine distance by evaluating signal timing metrics like RTT, ToA, and TDoA~\cite{singh2022v,gu2022high}. Since radio waves travel at a constant speed (assumed to be the speed of light), the time a signal takes to travel between devices is directly proportional to the distance. Accurate ranging thus depends on precisely identifying the time the signal is received. There are several approaches used for ToA estimation at the receiver. The most common is one-way ranging, where the transmitter (Tx) embeds a timestamp into the signal. If the Tx and Rx are fully synchronized (i.e., share a common time reference), the Rx can determine both the transmission and reception times, calculate the ToA, and subsequently estimate the distance. An alternative is two-way ranging, which involves bidirectional signal exchange and does not require clock synchronization between Tx and Rx. However, this method typically requires processing capabilities on both ends and cannot be easily applied to simple passive RIS elements that lack computational functionality. Other ToA estimation methods exist~\cite{singh2022v}, including techniques that use signals from multiple anchor nodes. However, here the single-transmitter and single-RIS case is evaluated at the simulation setup rather than employing multiple anchor points for ToA-based ranging to employ the RIS-based positioning.

In one-way ranging, achieving precise synchronization between the Tx and Rx is a key challenge, as accurate distance estimation relies on a shared and highly synchronized sense of time. Even minor clock offsets between Tx and Rx can lead to substantial errors in ranging, which in turn degrade overall positioning accuracy. In many cases, this issue is effectively addressed through the use of an ultra-wideband (UWB) transceiver pair, which integrates all the necessary components for reliable and precise ranging. Given the need for high-accuracy positioning, the approach here focuses first on evaluating how algorithmic estimation of azimuth AoA affects positioning accuracy, while assuming the usage of commercial off-the-shelf (COTS) components for estimating the remaining spatial coordinates—namely, height and distance.

Finally, having an estimate for azimuth AoA ($\varphi_{\text{AoA}}$), height ($z$) and ranging ($r$), one can calculate the exact position of a point with respect to a specific coordinate system. For the transformation to Cartesian coordinates we can simply use the following equations
\begin{equation}
    \begin{aligned}
        \varphi_{AoA} &=  \operatorname{arctan}\bigg(\frac{y}{x}\bigg)\\
        z &= z \\
        r &= \sqrt{x^2 + y^2 + z^2},
    \end{aligned}
\end{equation}
where $x$, $y$ and $z$ are the Cartesian coordinates.

\subsubsection{Simulation Environment and Channel Model using Ray Tracing}
\begin{figure}[!t]
    \centering
    \scalebox{0.77}{\includegraphics{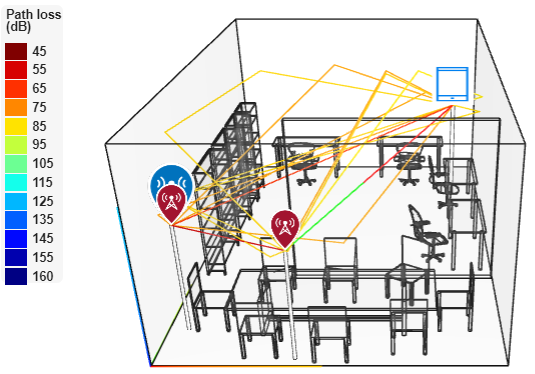}}
    \caption{RT-based simulation setup. Red dots denote Tx points, while blue dots with the mobile-phone denote Rx points. The overlapping red/blue dot refers to the RIS node, that reflects the incoming signal, thus operating both as a Rx and as a Tx.}
    \label{fig:sim_setup1}
\end{figure}

Consider a downlink communication of an RIS-assisted indoor wireless system similar to~\cite{kompostiotis2024evaluation}, as depicted in Fig.~\ref{fig:sim_setup1}. The selection of an indoor place is taken, since it is a more challenging case, due to extreme multipath conditions. However, the same methodology can be applied for the outdoor case. The system comprises an AP equipped with $N_T$ antennas and a UE featuring a ULA with $N_{R}$ antennas. The direct propagation paths from the AP to the UE are assumed to be NLoS. An RIS is strategically positioned to establish LoS connections in both the AP-to-RIS and RIS-to-UE links, denoted by $\bm{H}_{1}[n] {\in} \mathbb{C}^{N \times N_T}$ and $\bm{H}_{2}[n] {\in} \mathbb{C}^{N_R \times N}$, respectively. Here, $n$ indexes the specific time-domain channel tap. The RIS is modeled as a URA comprising $N {=} N_H {\times} N_V$ reflecting elements, where $N_H$ and $N_V$ represent the number of horizontal and vertical elements, respectively. The direct AP-to-UE channel is captured by $\bm{H}_{d}[n] {\in} \mathbb{C}^{N_R \times N_T}$.

To model these MIMO wireless channels, the ray-tracing (RT) technique from Matlab$\textsuperscript{TM}$ is employed, that is a physics-based geometric approach well-suited for characterizing static wireless channels across a frequency range of 100~MHz to 100~GHz. In the discrete-time complex baseband domain and around a specific carrier frequency $f_c$, the RT-based channel model is expressed as:
\begin{equation}
\bm{H}[n] = \sum_{l=1}^{L}\beta_{l,n}\,\alpha_l\,e^{-\,\text{j}\,2\pi\tau_lf_c}\,\mathbf{a}_{\text{Rx}}(\varphi_{\text{AoA}}^{l},\theta^{l}_{\text{AoA}})\mathbf{a}_{\text{Tx}}^{H}(\varphi_{\text{AoD}}^{l},\theta^{l}_{\text{AoD}}),
\label{eq:RT-based_channel_model}
\end{equation}
where $L$ denotes the number of propagation paths. For each $l$-th path, $\beta_{l,n} \in \mathbb{R}$ represents the pathloss component modeling the fractional-delay filter of the channel, $\alpha_l \in \mathbb{R}$ is the path gain, $\tau_l$ is the propagation delay, and $(\varphi_{\text{AoA}}^{l},\theta^{l}_{\text{AoA}})$ and $(\varphi_{\text{AoD}}^{l},\theta^{l}_{\text{AoD}})$ are the angle-pairs of arrival and departure, respectively. The array response vector $\mathbf{a}_{q}(\varphi, \theta)$, for $q \in \{\text{Tx},\, \text{Rx}\}$, under the far-field assumption is given by:
\begin{equation}
    \mathbf{a}_{q}(\varphi, \theta) = \Big[e^{-\,\jmath\,\mathbf{k}(\varphi, \theta)^\mathsf{T}\mathbf{u}^{q}_1}, \ldots, e^{-\,\jmath\,\mathbf{k}(\varphi, \theta)^\mathsf{T}\mathbf{u}^{q}_N}\Big]^{\mathrlap{T}},
    \label{e:array_response_vec}
\end{equation}
where $\mathbf{u}^{q}_i$ is the position vector of the $i$th element in the transmitting or receiving array. The wave vector $\mathbf{k}(\varphi, \theta) \in \mathbb{R}^{3 \times 1}$, which describes the direction of wave propagation at azimuth $\varphi$ and elevation $\theta$ for a signal of wavelength $\lambda$, is defined as:
\begin{equation}
    \mathbf{k}(\varphi, \theta) = \frac{2\pi}{\lambda} \Big[\cos{(\theta)}\cos{(\varphi)}, \cos{(\theta)}\sin{(\varphi)}, \sin{(\theta)}\Big]^{\mathrlap{T}}.
    \label{e:wave_vector}
\end{equation}

The narrowband channel representation in~\eqref{eq:RT-based_channel_model} is centered around the carrier frequency $f_c$. It is important to note that for AoA estimation using classical subspace-based techniques~\cite{schmidt1986multiple,roy1989esprit}, the specific modulation scheme of the transmitted signal is not relevant. In the case of wideband channels, a similar AoA estimation approach can be applied on a per-subcarrier basis within an Orthogonal Frequency Division Multiplexing (OFDM) frame. In contrast to stochastic models (e.g.,~\cite{ma2023reconfigurable,pan2020multicell}), where LoS components are typically modeled explicitly and NLoS components are captured via Rayleigh fading distributions, the RT-based model used here adopts a geometric perspective. In this model, the LoS component corresponds to the dominant term in~\eqref{eq:RT-based_channel_model}, while the remaining multipath components represent the small-scale fading behavior. 

Here, $\bm{x}[n] \in \mathbb{C}^{N_T}$ denote the transmitted signal, composed of $N_s$ pilot symbols. The signals received via the RIS-assisted and direct channels, denoted by $\bm{r}_{\text{RIS}}[n]$ and $\bm{r}_d[n]$, respectively, are given as:
\begin{equation}
\begin{aligned}
\bm{r}_{\text{RIS}}[n] &\in \mathbb{C}^{N_R} = \bm{H}_2[n] \ast \left( \boldsymbol{\Phi} \left( \bm{H}_1[n] \ast \bm{x}[n] \right) \right), \\
\bm{r}_d[n] &\in \mathbb{C}^{N_R} = \bm{H}_d[n] \ast \bm{x}[n],
\end{aligned}
\end{equation}
where $\ast$ denotes convolution and $\boldsymbol{\Phi}$ is the RIS configuration matrix, represented as a diagonal matrix of size $N \times N$. It is assumed that a single time-domain filter tap is used to model the RIS response, which implies that the frequency selectivity is not captured in this version of the model~\cite{multiRIS_opt}. However, modeling the imperfections and thus the frequency response of the RIS can be performed using circuit models for each RIS reflecting elements like in~\cite{costa2021electromagnetic}. 

The overall received signal at the UE is a superposition of the two channel contributions, adjusted according to their respective time of arrival. The resulting signal is expressed as
\begin{equation}
    \bm{r}_{\text{tot}} = \bm{r}_{\text{RIS}} + \bm{r}_d + \bm{w},
    \label{eq:received_signal}
\end{equation}
where $\bm{w} \in \mathbb{C}^{N_R}$ represents additive white Gaussian noise (AWGN), and both $\bm{r}_{\text{RIS}}$ and $\bm{r}_d$ are appropriately delayed to reflect their time-of-arrival differences.

\subsubsection{AoA Estimation}
\label{sec:AoA_estimation_with_RIS}
The positioning estimation problem, for the scenario illustrated in Fig.~\ref{fig:sim_setup1}, entails estimating the UE location by processing the received signal $\bm{r}_{\text{tot}}$, using known reference points like the positions of APs or RISs. By leveraging modern B5G/6G infrastructure, specifically user devices equipped with ULAs and RIS-assisted architectures, both the angle and range (or ToA) of LoS path from RIS to UE can be determined with high precision. Assuming perfect ToA estimation (with RIS-assisted ToA estimation demonstrated in subsection~\ref{sec:RIS_aided_TOA_sense_of_time}), the current focus is on solving the AoA estimation problem.

Super-resolution methods~\cite{schmidt1986multiple,roy1989esprit} are known to accurately estimate the AoA when the LoS path dominates over multipath components. Thus, a beamforming strategy that directs energy toward the LoS path is essential. However, since the UE location is unknown, direct beamforming via the RIS is not feasible. Instead, a precomputed RIS codebook can be designed, from which the optimal configuration—maximizing received signal power at the UE—is selected.

The objective in designing this codebook is to optimize the RIS power radiation pattern such that each configuration maximizes reflected power in a specific target direction while minimizing spillover to others. This radiation pattern is defined as~\cite{ramezani2023broad}
\begin{equation}
    \text{A}(\varphi, \theta) = |\bm{\omega}_{\bm{\theta}}^{T}(\mathbf{a}_{\text{RIS}}(\varphi_{\text{AoA}},\theta_{\text{AoA}}) \odot \mathbf{a}^{*}_{\text{RIS}}(\varphi, \theta))|^2,
    \label{eq:power_factor}
\end{equation}
where $\odot$ denotes the Hadamard product, $(\varphi, \theta)$ are the desired reflection angles, $(\varphi_{\text{AoA}},\theta_{\text{AoA}})$ are the arrival angles at the RIS, $\bm{\omega}_{\bm{\theta}} = [e^{\text{j}\theta_{1}}, \ldots , e^{\text{j}\theta_{N}}]^\mathsf{T} \in \mathbb{C}^N$ is the RIS phase configuration vector, and $\mathbf{a}_{\text{RIS}}(\varphi,\theta)$ represents the RIS array response vector, as defined in~\eqref{e:array_response_vec}. This model includes both geometric and element-level radiation characteristics; however, for simplicity, isotropic element radiation is assumed.

To beamform towards a given direction $(\varphi,\theta)$, the quantity $\text{A}(\varphi,\theta)$ in~\eqref{eq:power_factor} is maximized with respect to $\bm{\omega}_{\bm{\theta}}$. Under the maximum ratio transmission principle~\cite{765552}, the optimal solution is
\begin{equation}
    \bm{\omega}_{\bm{\theta}}^{T} = (\mathbf{a}_{\text{RIS}}(\varphi_{\text{AoA}},\theta_{\text{AoA}}) \odot \mathbf{a}^{*}_{\text{RIS}}(\varphi,\theta))^{H}.
    \label{opt_config}
\end{equation}
In practice, RIS elements support only discrete phase values (e.g., $\theta_i \in \{-\frac{\pi}{2}, \frac{\pi}{2}\}$), necessitating quantization of~\eqref{opt_config} for practical codebook realization. Once the codebook $\mathcal{CB}$ is generated, the RIS-aided positioning protocol involves scanning through the configurations to identify the one that maximizes received signal power
\begin{equation}
    \bm{\Phi}_0 = \underset{\bm{\Phi}\, \in \,\mathcal{CB}}{\operatorname{argmax}} \,\,\,\, \operatorname{P} \bigg( \bm{H_2}[n] \ast \,\big(\Phi\, (\bm{H_1}[n]\,\ast \bm{x}[n])\big) + \bm{r}_{d} \bigg), 
    \label{eq:max_problem}
\end{equation}
where $\bm{\Phi}_0$ corresponds to the beam direction, offering a coarse estimate of the UE's angular location and $\operatorname{P}$ denotes the power of the measurement. Using this RIS configuration during the MUSIC algorithm~\cite{schmidt1986multiple} further enhances AoA estimation by increasing the power of the LoS component in relation to NLoS interference.

It should be noted that, in some cases, the direct AP-to-UE channel may dominate the RIS-assisted path~\cite{zheng2022survey}, primarily due to higher path loss in the RIS-induced virtual link. In such scenarios, the ON/OFF protocol from~\cite{zheng2022survey} can be used to isolate the direct signal component. Applying two opposite RIS phase settings, $\boldsymbol{\Phi}^{1} {=} j$ and $\boldsymbol{\Phi}^{2} {=} -j$, and measuring the corresponding signals:
\begin{equation}
    \begin{aligned}
        \bm{r}_1  &= \bm{H_2}[n] \ast \,\big(\Phi^{1}\, (\bm{H_1}[n]\,\ast \bm{x}[n])\big) + \bm{r}_{d} +\bm{w} \\ 
        \bm{r}_2  &= -\bm{H_2}[n] \ast \,\big(\Phi^{1}\, (\bm{H_1}[n]\,\ast \bm{x}[n])\big) + \bm{r}_{d} + \bm{w},
    \end{aligned}
\end{equation}
the direct signal can be estimated as $\bm{r}_d^{\text{est}} = (\bm{r}_1 + \bm{r}_2)/2$. Subtracting this from $\bm{r}_{\text{tot}}$ yields $\bm{r}_{\text{RIS}} + \bm{w}$, allowing the AoA estimation to focus on the RIS path and enhance the accuracy of the localization.

\begin{figure}[!t]
\centering
\begin{minipage}{0.5\textwidth}
  \centering
\includegraphics[width=0.87\textwidth]{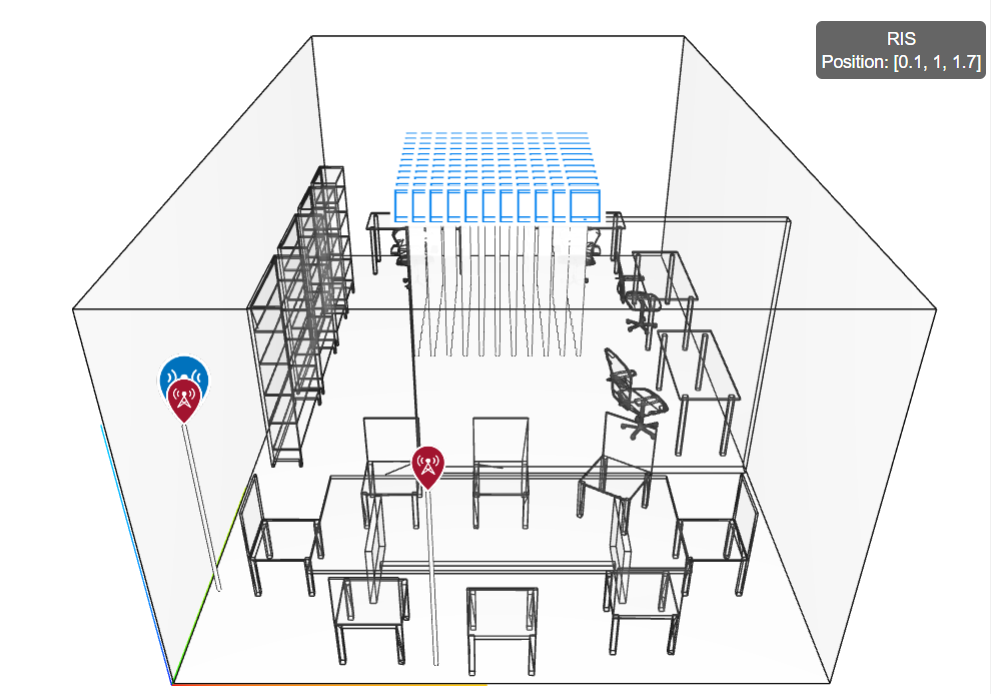}
\subcaption[first caption.]{Test Rx positions of the UEs.}\label{fig:1a}
\end{minipage}%
\begin{minipage}{0.5\textwidth}
  \centering
\includegraphics[width=0.92\textwidth]{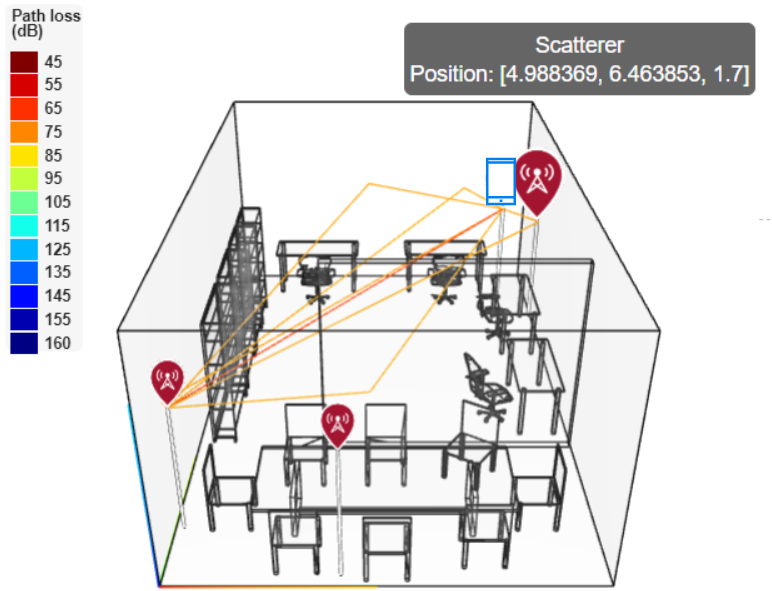}
\subcaption[second caption.]{Scatterer positions for mapping.}\label{fig:1b}
\end{minipage}%
\caption{The simulated  indoor environment for localization and mapping.} \label{fig:exp_loc_setup}
\end{figure}

\subsubsection{Environmental Sensing}
\label{sec:Sensing_Algo}
Radio sensing, as described by the channel model in~\eqref{eq:RT-based_channel_model}, involves determining the locations of scattering points in the environment, which give rise to NLoS paths. Here the focus is on NLoS components resulting from single reflections at scatterers positioned in front of the RIS, such that the RIS can effectively activate these paths through proper configuration. All remaining paths, due to their significantly higher attenuation, are treated as interference. Also given the use of a ULA-based UE, only NLoS paths with elevation angle $\theta {=} 0$ can be resolved. Consequently, the $\theta$-angle notation is dropped in the sequel.

Assuming the LoS and single-reflection NLoS paths dominate in power, both individually and collectively, the following outlines the steps of the proposed mapping protocol~\cite{kompostiotis2024evaluation}. After estimating the UE's location, beam scanning is repeated with the RIS configured to sequentially activate candidate NLoS paths. However, due to the LoS component’s typically stronger signal strength—even under configurations aimed away from the LoS—the MUSIC algorithm often still estimates the LoS angle. To mitigate this, a variant of the ON/OFF protocol is adopted, with the objective of suppressing the LoS contribution in the measurements. This enables better isolation and estimation of NLoS paths.

\begin{algorithm}[!t]
\KwIn{$\bm{r}_{\text{tot}}$ from~\eqref{eq:received_signal}}
\KwOut{UE position $(\varphi_{\text{AoA}}^{\text{est}}, \tau^{\text{est}})$ and scatterer locations}
\nl Design the codebook $\mathcal{CB}$ using~\eqref{eq:power_factor} and~\eqref{opt_config}\;
\nl Solve~\eqref{eq:max_problem} and transmit using the resulting $\bm{\Phi}_0$\;
\nl Apply the ON/OFF protocol and use MUSIC~\cite{schmidt1986multiple} to estimate $\varphi_{\text{AoA}}^{\text{LoS}}$\;
\nl Estimate $\tau^{\text{est}}$ using ultra-wideband ranging\;
\nl \For{$\bm{\Phi} \in \mathcal{CB}$}{
    \nl Apply ON/OFF protocol and compute the LoS-canceled observation using~\eqref{eq:LoS_term} and~\eqref{eq:mapping_measurement}\;
    \nl Execute MUSIC to estimate $\varphi_{\text{AoA}}^{\text{NLoS}}$\;
    \nl \If{$\varphi_{\text{AoA}}^{\text{NLoS}} \neq \varphi_{\text{AoA}}^{\text{LoS}}$}{
        \nl Determine the scatterer’s position using geometric intersection of RIS beam and MUSIC AoA\;
    }
}
\caption{\bf Positioning and Mapping Protocol used in~\cite{kompostiotis2024evaluation}.}
\label{algorithm}
\end{algorithm}

Let $\varphi_{\text{AoA}}^{\text{est}}$ be the estimated azimuth angle of arrival, $r^{\text{est}}$ the estimated range, and $\tau^{\text{est}} = r^{\text{est}}/c$ the estimated time-of-arrival, with $c$ denoting the speed of light. The estimated LoS component of the RIS-to-UE channel from~\eqref{eq:RT-based_channel_model} is given by:
\begin{equation}
\bm{H}_{2\,\,\text{LoS}}^{\text{est}}= h_2^{\text{loss}}e^{j\phi_2}\bm{a}_{\text{UE}}(\varphi_{\text{AoA}}^{\text{est}})\bm{a}_{\text{RIS}}^{H}(\varphi_{\text{AoD}}^{\text{est}}) \in \mathbb{C}^{N_R \times N},
\label{eq:LoS_term}
\end{equation}
where $h_2^{\text{loss}} = 20\log_{10}\left(\frac{4\pi r^{\text{est}}}{\lambda}\right)$ and $\lambda$ is the wavelength at carrier frequency $f_c$. The phase term is given by $\phi_2 = \operatorname{mod}(-2\pi f_c \tau^{\text{est}}, 2\pi)$. The departure angle $\varphi_{\text{AoD}}^{\text{est}}$ is derived from $\varphi_{\text{AoA}}^{\text{est}}$ via geometric considerations.

The AP-to-RIS LoS channel matrix $\bm{H}_{1\,\,\text{LoS}}^{\text{est}} \in \mathbb{C}^{N \times N_T}$ is constructed similarly, with the distinction that this link's geometry is known exactly, and hence all parameters are precisely defined. For a known RIS configuration $\bm{\Phi}$, the received LoS signal at the UE is:
\begin{equation}
\bm{r}_{\text{LoS}}[n] = \bm{H}_{2\,\,\text{LoS}}^{\text{est}} \bm{\Phi} \bm{H}_{1\,\,\text{LoS}}^{\text{est}} \bm{x}[n],
\end{equation}
and can be subtracted from the total received signal to obtain:
\begin{equation}
\bm{r}_{\text{NLoS}}[n] = \bm{r}_{\text{tot}}[n] - \bm{r}_{\text{LoS}}[n - \lfloor \tau^{\text{est}} F_s \rfloor],
\label{eq:mapping_measurement}
\end{equation}
where $F_s$ is the sampling frequency and $\lfloor \cdot \rfloor$ denotes the floor operation.

This subtraction suppresses the dominant LoS component, allowing MUSIC to detect alternative paths. While most RIS configurations still yield the LoS angle due to imperfect cancellation, those steering towards actual NLoS paths reveal new AoAs. The corresponding scatterer positions can then be localized as the intersection of the RIS’s beam direction and the AoA line from the UE. Repeating this process across all configurations maps out the indoor environment’s reflective features.

\subsubsection{Performance Evaluation}
 The experimental framework designed to evaluate the proposed RIS-aided localization and sensing approach is depicted in Fig.~\ref{fig:exp_loc_setup}, where a setup similar to~\cite{kompostiotis2024evaluation} is adopted so as to reproduce the corresponding results. A $32 \times 32$ URA RIS operating at a carrier frequency of $f_c{=}3.5$~GHz is employed. All components are situated on the same azimuthal plane. To verify the accuracy of the RIS-aided method, the Algorithm~\ref{algorithm} is implemented and executed for each test UE-location shown in Fig.~\ref{fig:1a} averaging the results. Test UE locations are distributed within a rectangular area defined by $x \in [1.5, 3.5]$ (horizontal axis) and $y \in [5, 7]$ (vertical axis), as depicted in Fig.~\ref{fig:1a}. The simulation employs a geometric RT-based channel model. The transmitter is equipped with multiple antennas and performs directional beamforming towards the RIS, mimicking the behavior of a high-gain horn antenna. UEs also feature multi-antenna arrays, with a fixed orientation aligned along the $x$-axis.

\begin{table}[htb]
\caption{Evaluation of Algorithm~\ref{algorithm}.\vspace*{-8pt}}
\begin{center}
\scalebox{1}
{
\begin{NiceTabular}{c@{\hskip 6pt}cc
>{\centering\arraybackslash}p{3cm}
>{\centering\arraybackslash}p{2cm}S[table-format=4.2]}
\CodeBefore
\rowcolor{teal!10}{1}
\Body
     \toprule
     Phase shifters & RIS configuration & Method & {Peak} & {Average}  & {Variance} \\
     \midrule
     Continuous  &$\bm{\omega}_{\bm{\theta}}^{T}$ from~\eqref{opt_config} & Beam Sweeping & 12.62\degree  & 1.03\degree  & 2.28\degree \\
     \midrule
     1-bit RIS  & 1-bit Quantized $\bm{\omega}_{\bm{\theta}}$ & Beam Sweeping & 58.28\degree & 9.93\degree &280.32\degree  \\
     1-bit RIS  & 1-bit Quantized $\bm{\omega}_{\bm{\theta}}$ & Beam Sweeping \& MUSIC & 10.68\degree & 0.86\degree &1.82\degree \\
     \bottomrule
\end{NiceTabular}
}
\label{Table:beamsweeping_algorithms}
\end{center}
\vspace{-0.1cm}
\end{table}

Table~\ref{Table:beamsweeping_algorithms} reports statistical measures for the absolute estimation error, defined as $\tilde{e} = \big| \varphi_{\text{AoD}} - \varphi_{\text{AoD}}^{\text{est}} \big|$, where $\varphi_{\text{AoD}}$ denotes the actual azimuth angle between the RIS and the UE. The table provides the peak, average, and variance of this error for different localization methods. The evaluated techniques include both the continuous solution for the RIS codebook and its quantized counterpart using $1$-bit phase shifters, aligning with realistic RIS implementations. The performance of the 1-bit codebook is also evaluated in combination with the MUSIC algorithm. For reproducibility, it is noted that the RIS codebook (refer to Section~\ref{sec:AoA_estimation_with_RIS}) was generated with an angular resolution of approximately 2 degrees.

As observed in Table~\ref{Table:beamsweeping_algorithms}, even with a quantized 1-bit RIS codebook derived from the continuous solution, the proposed approach maintains satisfactory performance—despite challenging conditions involving dominant multipath components. However, side-lobe effects~\cite{liu2024quantization} in 1-bit RISs may excite undesired NLoS components, occasionally degrading localization performance. The pronounced peak error observed in the 1-bit beam sweeping method stems from quantization errors in solving a non-convex optimization problem.

\begin{table}[htb]
\caption{RIS-to-UE propagation paths obtained via ray tracing.\vspace{-0.2cm}}
\label{t:paths_mapping}
\vspace{1ex}
\centering
\scalebox{1}{
\begin{NiceTabular}{ccS[table-format=3.7]S}
\CodeBefore
\rowcolor{teal!10}{1}
\Body
\toprule
Path Number & Path Type & {Azimuth AoA (deg, \degree)} & {Elevation AoA (deg, \degree)}\\
\midrule
1. & LoS path & 53.75\degree & 0 \\
2. & NLoS path & 127.36\degree & 8.56e-14 \\
3. & NLoS path & 53.75\degree  & 21.96 \\
4. & NLoS path & 47.91\degree  & 0 \\
5. & NLoS path & 53.75\degree  & -24.56 \\
6. & NLoS path & 61.29\degree  & 0 \\
\bottomrule
\end{NiceTabular}}
\vspace{3pt}
\end{table}

As illustrated in Fig.~\ref{fig:sim_setup1}, multiple NLoS paths exist in the considered setup, and resolving them is a critical task. Table~\ref{t:paths_mapping} enumerates all RIS-to-UE paths, whose corresponding reflection points must be localized using Algorithm~\ref{algorithm}. Due to the limitations of ULA-based receivers, not all listed paths can be resolved. This preliminary sensing analysis focuses on single-reflected NLoS paths with an elevation angle $\theta_{\text{AoA}} \approx 0$, under the assumption that the associated scattering point lies in front of the RIS. Additionally, the periodic nature of the MUSIC spectrum in the interval $[-180\degree, 180\degree]$ necessitates its restriction to $[0\degree, 180\degree]$, rendering paths arriving from the back side of the UE undetectable.

Accordingly, it is assumed that the UE is equipped with a directional patch antenna that receives signals only from the front. Sensing of the rear space would require the UE to physically rotate and reorient toward the back. Thus, this study considers only the front-facing NLoS path labeled as Path 4. As shown in Fig.~\ref{fig:1b}, the estimated position of the scattering point is in close agreement with the true reflection location responsible for this NLoS path. These results hold under the assumptions discussed in Section~\ref{sec:Sensing_Algo}, namely that single-reflected NLoS paths dominate over multi-reflected paths, which are considered as interference.

Running Algorithm~\ref{algorithm} for each UE position shown in Fig.~\ref{fig:1a} yields an average error of $0.21$~m in estimating the scatterer location, with a standard deviation of $0.23$~m. Therefore, these results demonstrate that RIS-assisted sensing can achieve even decimeter-level accuracy in indoor environments providing fertile ground for further improvement of positioning and sensing accuracy through RIS.

\subsection{Experimental Validation with an RIS Prototype}
RISs are emerging as a transformative technology in wireless communications, offering significant control over the radio propagation environment. The potential of RISs has been extensively explored through theoretical studies and simulations. However, these approaches often rely on idealized models that may not fully capture the intricacies of practical hardware implementations and real-world propagation conditions. Experimental validation using physical RIS prototypes is therefore paramount. Such experiments are crucial not only to verify theoretical findings but also to uncover practical limitations, characterize achievable performance, and ultimately demonstrate the feasibility and reliability of RIS-assisted systems in realistic operational scenarios. This Section presents the experimental methodologies and results obtained with an RIS prototype, aimed at validating its efficacy for localization and sensing applications.

\subsubsection{RIS Prototype Hardware Architecture and Control System}
\begin{figure}[!t]
\centering
\begin{minipage}{0.48\textwidth}
  \centering
  \includegraphics[width=0.8\textwidth]{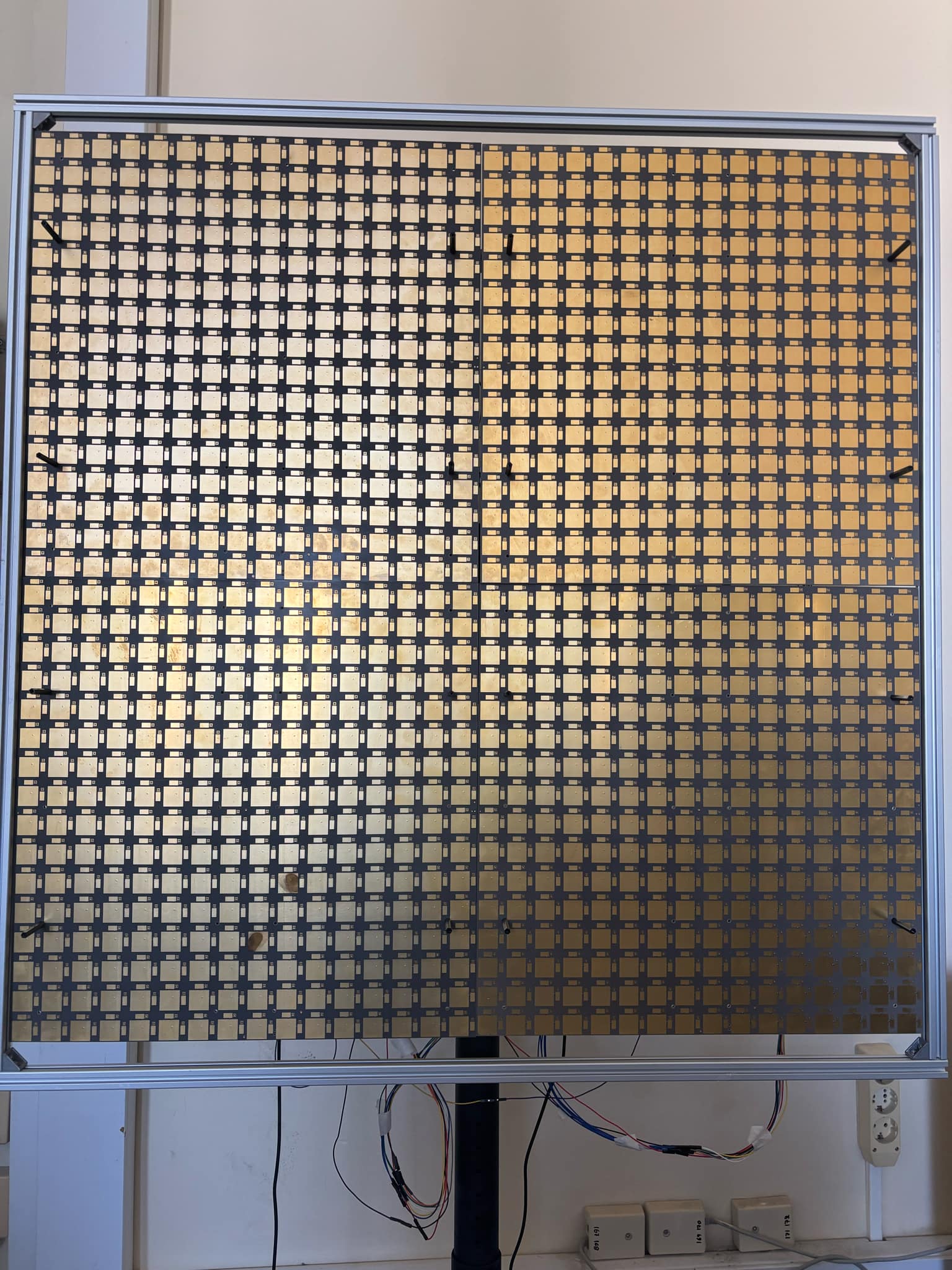}
  \subcaption{Front view of the $32 \times 32$ element RIS prototype, comprising four $16 \times 16$ element tiles.}
  \label{fig:ris_front_panel}
\end{minipage}%
\hfill
\begin{minipage}{0.48\textwidth}
  \centering
  \includegraphics[width=0.8\textwidth]{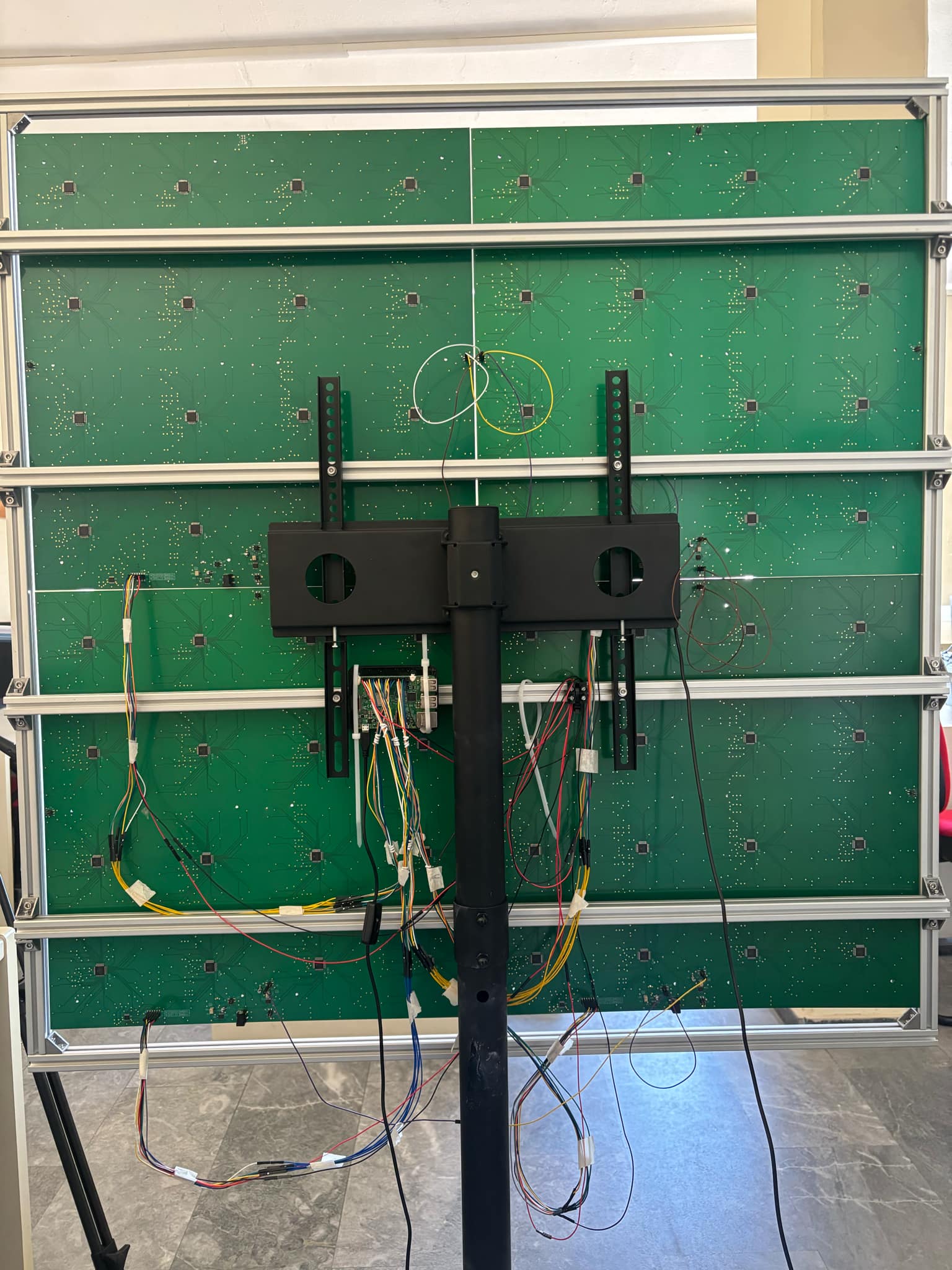}
  \subcaption{Rear view of an RIS tile showcasing the control circuitry and interface connections to the Raspberry Pi controller.}
  \label{fig:ris_back_control}
\end{minipage}%
\caption{Front and rear views of the experimental RIS prototype used for the validation experiments.} 
\label{fig:ris_prototype_views}
\end{figure}

The experimental RIS prototype is engineered using varactor diode technology~\cite{rains2023fully}, which allows for dynamic and low-power phase manipulation of incident electromagnetic waves. The overall RIS surface features a modular construction, assembled from four identical tiles. These tiles are arranged to form a contiguous metasurface with approximate dimensions of 1\,m $\times$ 1\,m, as depicted in Fig.~\ref{fig:ris_front_panel}. The complete prototype integrates a $32 \times 32$ array of unit cells, resulting in a total of 1024 individually addressable reflecting elements. Each of the four tiles comprises a $16 \times 16$ array of unit cells. The unit cell design~\cite{rains2023fully}, consists of a square microstrip patch element accompanied by two parasitic rectangular patches. These metallic structures are etched onto an F4B220 dielectric substrate, which is backed by a continuous copper ground plane. A critical feature for versatile wave manipulation is the dual-linear polarization capability of each unit cell. This is achieved by incorporating two Skyworks SMV1408 varactor diodes per cell: one dedicated to controlling the phase response for horizontally polarized incident waves and the other for vertically polarized waves.

The RIS offers 1-bit phase control for each polarization at every element, enabling a binary phase shift (nominally $0^\circ$ or $180^\circ$)~\cite{rains2023fully,vordonis2025evaluating}. This is realized by switching the reverse bias voltage applied to the varactor diodes between two pre-defined levels. For the operational frequency band centered around 3.5 GHz, these bias voltages were experimentally determined and set to $V_1 = 11$ V and $V_2 = 7.5$ V~\cite{vordonis2025evaluating}. The prototype is designed to operate within the sub-6 GHz Frequency Range 1 (FR1) band, specifically targeting a bandwidth of approximately 100 MHz around 3.5 GHz. The RIS node achieves a configuration update speed of approximately 10 ms and exhibits a low power consumption profile of less than 15 mW for the entire assembly~\cite{rains2023fully}.

The sophisticated control required to individually address each of the 1024 dual-polarized elements is managed by a Raspberry Pi 3B+ single-board computer~\cite{rains2023fully}. This central controller interfaces with the RIS tiles, likely via integrated shift registers or similar multiplexing hardware on each tile, to precisely set the bias voltage for each varactor diode pair. The arrangement of the control circuitry on the reverse side of a tile, along with the Raspberry Pi interface, is illustrated in Fig.~\ref{fig:ris_back_control}.


RIS configurations, defining the binary state for each element, are typically transmitted from a host PC running control software (e.g., MATLAB scripts) to the Raspberry Pi. This communication can be established through either a Wi-Fi link, where the Raspberry Pi acts as an access point, or via a direct Ethernet cable connection. A Python-based server application resident on the Raspberry Pi receives these configuration commands, parses them, and subsequently manipulates the Raspberry Pi's General-Purpose Input/Output (GPIO) pins to program the shift registers on the RIS tiles. Physically, the four tiles are interconnected to facilitate power distribution from a primary input tile to the others and to propagate control signals from the Raspberry Pi. For experimental versatility, the fully assembled 1m $\times$ 1m RIS is mounted on a movable and rotatable stand, allowing for flexible positioning and orientation during measurements.

\subsubsection{RIS On-Site Characterization}
\label{sec:ris_phase_characterization}

The characterization experiments were performed in the semi-anechoic chamber at the Wireless and Optical Devices and Communication Networks Laboratory of the University of West Attica. This Frankonia-manufactured chamber, with internal dimensions of $7.3\,\text{m (Length)} \times 3.7\,\text{m (Width)} \times 3.2\,\text{m (Height)}$, is certified for measurements across a frequency range of 30 MHz to 18 GHz. It features full lining with ferrite absorbers on walls, floor, and ceiling, with additional Frankosorb pyramid absorbers in key areas to minimize reflections.

The measurement instrumentation included a Rohde \& Schwarz ZVA24 Vector Network Analyzer (VNA) and two dual-polarized ITELITE PAT3519XP flat-panel directional antennas. These antennas operate within the 3.5 GHz to 3.8 GHz frequency range, offer a gain of approximately 19 dBi, and have a 3 dB beamwidth of about $16^\circ$ for both vertical and horizontal polarizations. This narrow beamwidth helps in isolating the RIS response from residual environmental reflections.

\begin{figure}[!t]
    \centering
    \includegraphics[width=0.65\textwidth]{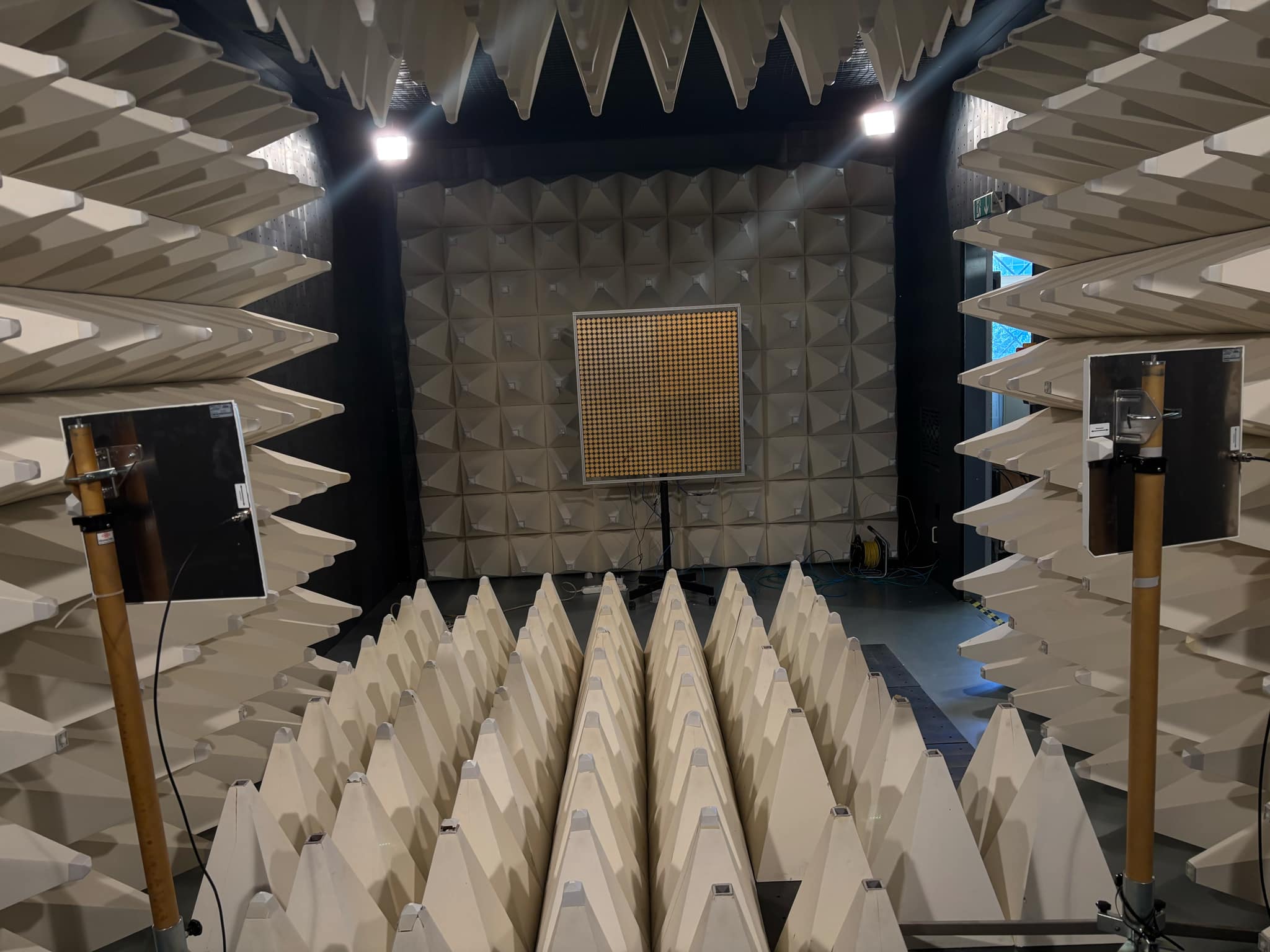} 
    \caption{Experimental setup for RIS phase characterization in the anechoic chamber. The Tx and Rx antennas are positioned for specular reflection from the RIS surface, and connected to the VNA.}
    \label{fig:ris_specular_setup_anechoic}
\end{figure}

The primary objective was to quantify the phase shift introduced by the RIS when its elements are switched between their two control states (denoted as '0' and '1'). To achieve this, the RIS (overall dimensions $0.96\,\text{m} \times 0.96\,\text{m}$) was placed in the anechoic chamber, and the Tx and Rx antennas were positioned in a specular reflection configuration relative to the RIS surface, as illustrated in Fig.~\ref{fig:ris_specular_setup_anechoic}. Common specular angles used were, for example, Tx at $-15^\circ$ and Rx at $+15^\circ$, or Tx at $-10^\circ$ and Rx at $+10^\circ$, with respect to the RIS normal. This setup ensures that when the RIS is configured for specular reflection (e.g., all elements in the same state), the received power at the Rx is maximized, facilitating accurate phase measurement.

The VNA was used to measure the $S_{21}$ parameter (complex transmission coefficient) between the Tx and Rx antenna ports. The measurement procedure involved:
\begin{enumerate}
    \item Configuring all elements of the RIS to the '0' state (corresponding to one of the bias voltages, e.g., $V_1 = 11$\,V). The $S_{21}$ response, $S_{21,0}(f)$, was measured across the frequency band of interest (e.g., 3.3\,GHz to 3.8\,GHz).
    \item Reconfiguring all elements of the RIS to the '1' state (corresponding to the other bias voltage, e.g., $V_2 = 7.5$\,V). The $S_{21}$ response, $S_{21,1}(f)$, was then measured under identical conditions.
\end{enumerate}
The phase difference, $\Delta\phi(f)$, introduced by switching the RIS elements between these two states is then calculated as the difference between the phases of the two measured $S_{21}$ parameters: $\Delta\phi(f) = \angle S_{21,1}(f) - \angle S_{21,0}(f)$.

\begin{figure}[!t]
    \centering
    \resizebox{0.8\textwidth}{!}{%
      \input{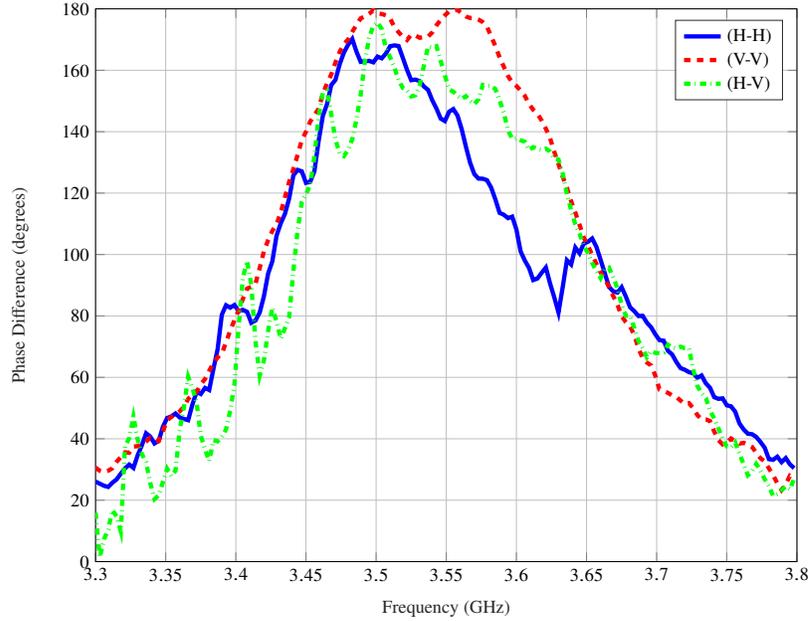}
    }
    \caption{Measured phase difference between the '0' and '1' states of the RIS elements across the operational frequency band for Horizontal-Horizontal (H-H), Vertical-Vertical (V-V), and Horizontal-Vertical (H-V) polarization configurations.}
    \label{fig:ris_phase_diff_results}
\end{figure}

Theoretically, for a 1-bit phase resolution RIS, this phase difference is expected to be $180^\circ$. The experimental results, exemplified in Fig.~\ref{fig:ris_phase_diff_results}, validate this expectation. The measurements were conducted for different transmit and receive antenna polarization configurations: co-polarized (Horizontal-Horizontal (H-H) and Vertical-Vertical (V-V)) and cross-polarized (Horizontal-Vertical (H-V)) to fully characterize the dual-polarized RIS. As shown, a phase difference close to $180^\circ$ is achieved around the central operating frequency of 3.5\,GHz. The RIS design, as detailed in~\cite{rains2023fully}, aims for a phase shift of $180^\circ \pm 20^\circ$ over a 160\,MHz bandwidth. The obtained results were consistent with these specifications across the co-polarized configurations, confirming the reliable 1-bit phase shifting capability of the prototype.

These anechoic chamber characterization results are crucial as they provide a fundamental understanding of the RIS element behavior, forming the basis for designing RIS configurations for more complex tasks such as beamforming and sensing, which are explored in subsequent Sections.

\subsubsection{Radiation Pattern Measurements}
\label{sec:ris_radiation_pattern}

The ability of an RIS to shape and direct an incident electromagnetic wave is fundamental to its utility in wireless systems. This is characterized by its radiation pattern when specific phase configurations are applied to its elements. To experimentally evaluate these patterns, the same VNA (Rohde \& Schwarz ZVA24) and ITELITE PAT3519XP directional antennas described in Section~\ref{sec:ris_phase_characterization} were utilized.

\subsubsection*{Measurement Environments and Setup}
\begin{figure}[!t]
    \centering
    \begin{subfigure}[b]{0.48\textwidth}
        \centering
         \includegraphics[width=\textwidth]{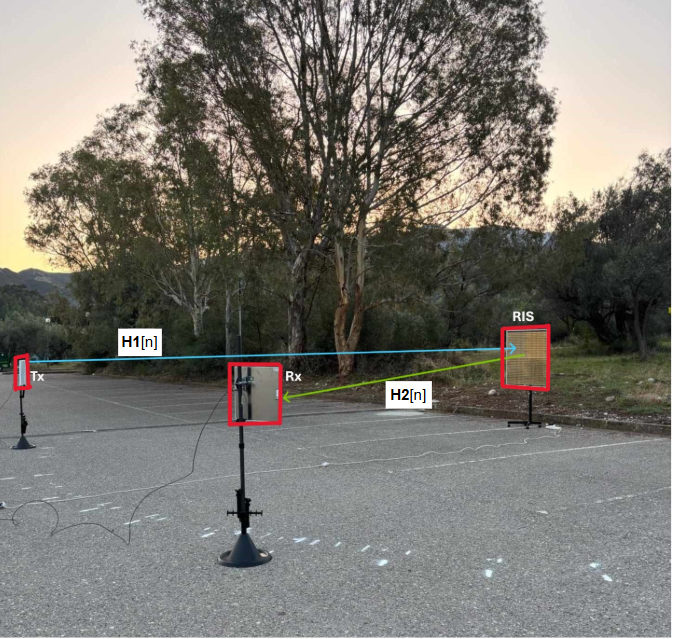} 
        \caption{Outdoor measurement setup}
        \label{fig:outdoor_setup_specific}
    \end{subfigure}
    \hfill
    \begin{subfigure}[b]{0.48\textwidth}
        \centering
     \includegraphics[width=\textwidth]{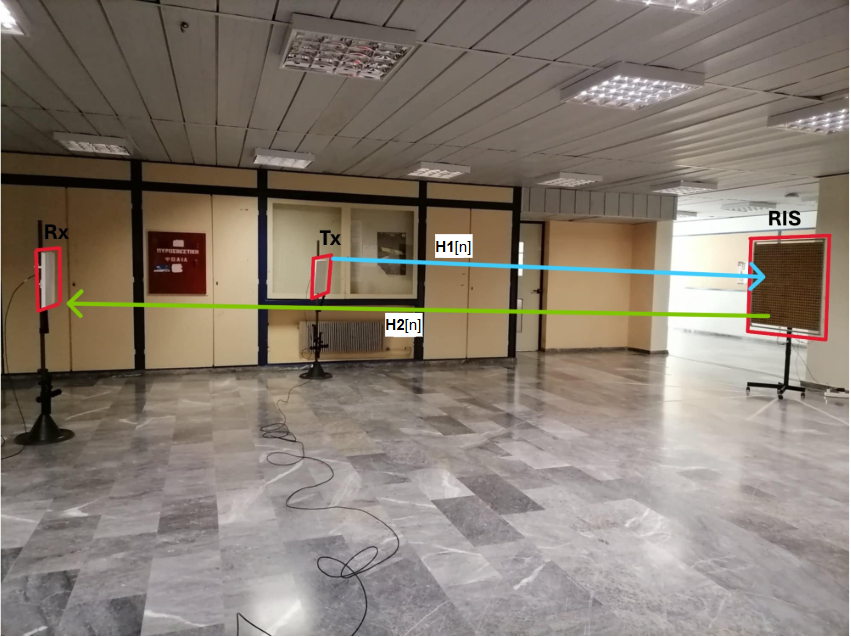} 
        \caption{Indoor measurement setup}
        \label{fig:indoor_setup_specific}
    \end{subfigure}
    \caption{Experimental setups for (a) outdoor environment with Tx-RIS distance 8.5 m, RIS-Rx distance 8.5 m, Rx angles $0^\circ$ to $60^\circ$ (step $5^\circ$), Tx at $-15^\circ$; and (b) indoor environment with Tx-RIS distance 5.5 m, RIS-Rx distance 8.5 m, Rx angles $0^\circ$ to $45^\circ$ (step $15^\circ$), Tx at $-15^\circ$. All nodes at 1.3 m height.}
    \label{fig:outdoor_indoor_setups}
\end{figure}

Experiments were conducted in two distinct environments to assess performance under different propagation conditions:
\begin{itemize}
    \item \textbf{Anechoic Chamber:} Utilized the controlled environment described in Section~\ref{sec:ris_phase_characterization}. For radiation pattern measurements, the Tx antenna was fixed at an angle of $-10^\circ$ relative to the RIS normal. The Rx antenna was moved along a semicircular arc at a constant distance from the RIS center, covering observation angles from $-5^\circ$ to $+10^\circ$ in $2.5^\circ$ steps. Both the Tx and Rx antennas, as well as the RIS, were positioned at a height of 1.3 m above the ground.
    \item \textbf{Outdoor Environment:} To evaluate performance in a more realistic scenario with minimal multipath, an outdoor setup was established, as depicted in Fig.~\ref{fig:outdoor_indoor_setups}. In this setup, the Tx antenna, RIS, and Rx antenna were all positioned at a height of 1.3 m above the ground. The Tx-RIS distance and nominal RIS-Rx distance were both set to 8.5 m. The Tx antenna was fixed at an incident angle of $-15^\circ$ relative to the RIS normal. The Rx antenna was moved along a semicircular arc covering an azimuth range from $0^\circ$ to $60^\circ$ in $5^\circ$ steps. This environment represents a low-multipath scenario typical of open areas.
\end{itemize}

Communication between the host PC (running MATLAB for control and data acquisition), the RIS controller (Raspberry Pi), and the VNA was established to automate the measurement process, including RIS configuration updates and $S_{21}$ parameter acquisition.

\subsubsection*{RIS Configuration Optimization for Beam Steering (Codebook Generation)}
To steer the main lobe of the reflected signal towards a desired angle of departure (AoD), an optimized phase configuration must be applied to the RIS elements. For a 1-bit RIS, this involves finding the optimal binary state for each element. A practical approach for this optimization, particularly when direct channel state information is unavailable or complex to acquire, is a greedy iterative algorithm that maximizes the received power at the target Rx position. Furthermore, accurate analytical or electromagnetic models capturing the RIS radiation pattern are typically not available, especially for custom or experimental RIS hardware. This lack of both channel knowledge and reliable forward models further motivates the use of measurement-based, data-driven optimization approaches. The method employed, detailed in Algorithm~\ref{algo:ris_config}, starts with an initial RIS configuration (e.g., all elements set to '0'). It then iteratively adjusts the states of groups of elements (e.g., column-by-column, then row-by-row), measuring the received power via the VNA after each change. If a change (e.g., flipping the states of a column) results in higher received power at the target Rx location, the change is kept; otherwise, it is reverted. This process is typically run for a fixed number of iterations (e.g., one or two full column-row scans).

\begin{algorithm}[t]
\caption{\bf RIS Configuration Optimization for Beam Steering}
\label{algo:ris_config}
\footnotesize

\tcp{
    \(N_x, N_y\): Number of physical rows and columns of the RIS, respectively \\
   \( \bm{\omega_\theta} \) is mapped to the dual-polarized RIS's configuration matrix \( \bm{\Phi} \) of size \( N_x \times 2N_y \)\\
   Each RIS element supports two orthogonal polarizations (horizontal/vertical), hence $2N_y$ columns
}
\KwIn{Initial RIS configuration matrix \( \bm{\Phi} \) with all elements set to 0.}
\KwOut{Optimized RIS configuration \( \bm{\Phi} \).}
\nl Measure the initial received power \( P_{\text{max}} \) with \( \bm{\Phi} \) set to all 0\;
\nl 
    \tcc{\textbf{Step 1: Column-Wise Scanning}}
    \nl \For{\texttt{\( col = 1:2:2N_y \)}}{
        \nl Invert the states of columns \( col, col+1 \) in \( \bm{\Phi} \)\;
        \nl Measure the received power \( P_r \) for the current configuration\;
        \nl \If{\( P_r > P_{\text{max}} \)}{
            \nl \( P_{\text{max}} \gets P_r \) \tcp*{Keep new configuration.}
        }\Else{
            \nl Revert columns \( col, col+1 \) to its previous state\;
        }
    }
    \tcc{\textbf{Step 2: Row-Wise Scanning}}
    \nl \For{\texttt{\( row = 1, \dots, N_x \)}}{
        \nl Invert the states of row \( row \) in \( \bm{\Phi} \)\;
        \nl Measure the received power \( P_r \) for the current configuration\;
        \nl \If{\( P_r > P_{\text{max}} \)}{
            \nl \( P_{\text{max}} \gets P_r \) \tcp*{Keep new configuration.}
        }\Else{
            \nl Revert row \( row \) to its previous state\;
        }
    }
\nl \Return{\( \bm{\Phi} \)} \tcp*[l]{Return optimized RIS configuration.}
\end{algorithm}

By running this optimization algorithm for each desired Rx angular position (e.g., $-5^\circ, -2.5^\circ, \dots, +10^\circ$ in the anechoic chamber; $0^\circ, 5^\circ, \dots, 60^\circ$ outdoors), a set of RIS configurations is generated. This set, where each configuration is optimized to steer the beam towards a specific angle, forms a beamforming codebook.

\subsubsection*{Measured Radiation Patterns and Analysis}
Once the codebook was generated, the radiation pattern for each RIS configuration (i.e., for each target beam direction) was measured. This involved setting the RIS to a specific configuration from the codebook and then sweeping the Rx antenna across all its predefined measurement positions, recording the received power at each.

\begin{figure}[!p]
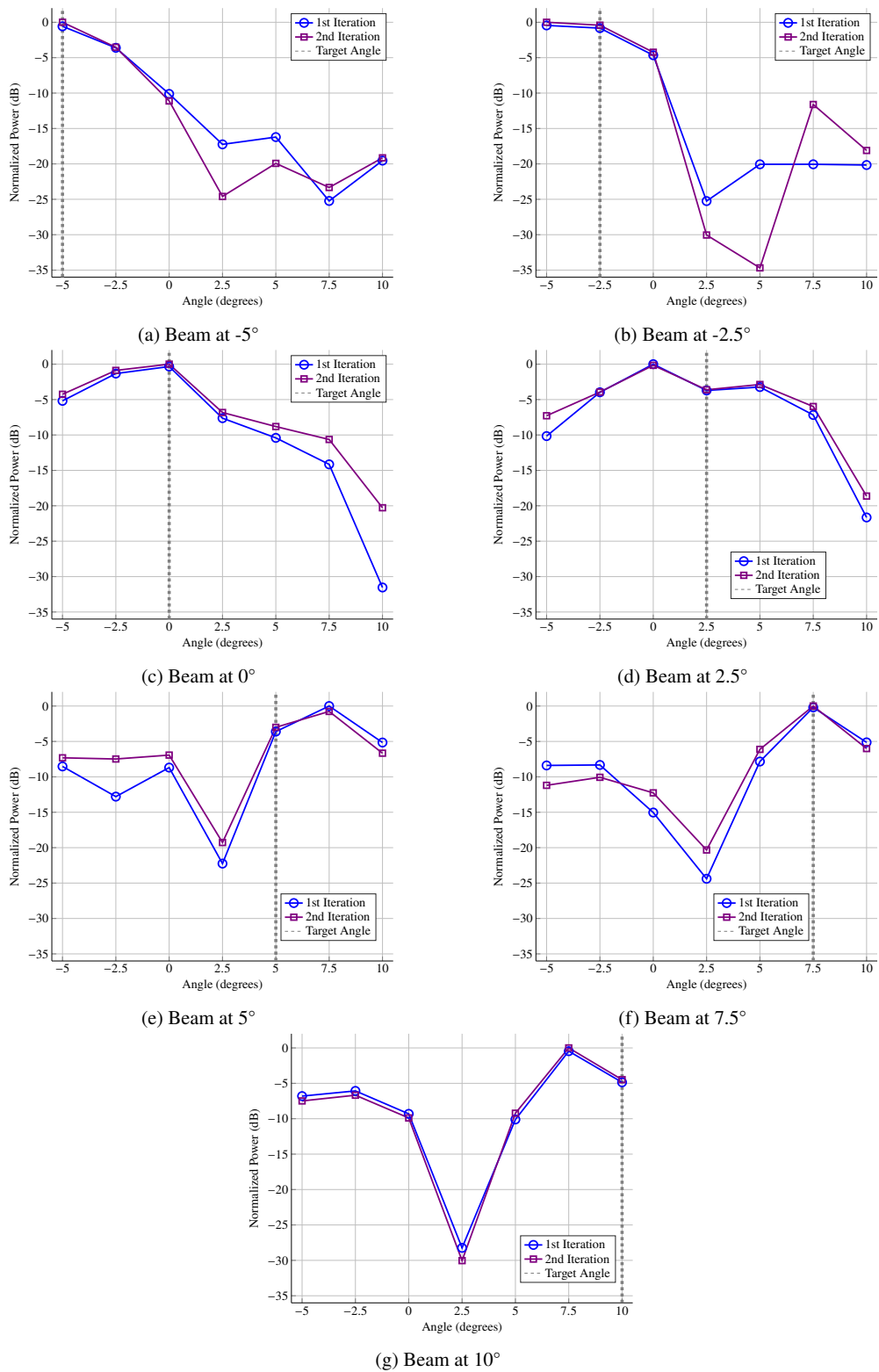

    \centering
    \begin{subfigure}{0.48\textwidth}
        \centering
        \scalebox{0.45}{
        \input{newsv-mult/author/Book_Figures/rad_pattern_355GHz_11}}
        \caption{Beam at -5$^\circ$}
    \end{subfigure}
    \begin{subfigure}{0.48\textwidth}
        \centering
        \scalebox{0.45}{
        \input{newsv-mult/author/Book_Figures/rad_pattern_355GHz_22}}
        \caption{Beam at -2.5$^\circ$}
    \end{subfigure}
    \begin{subfigure}{0.48\textwidth}
        \centering
        \scalebox{0.45}{
        \input{newsv-mult/author/Book_Figures/rad_pattern_355GHz_33}}
        \caption{Beam at 0$^\circ$}
    \end{subfigure}
    \begin{subfigure}{0.48\textwidth}
        \centering
        \scalebox{0.45}{
        \input{newsv-mult/author/Book_Figures/rad_pattern_355GHz_44}}
        \caption{Beam at 2.5$^\circ$}
    \end{subfigure}

    \begin{subfigure}{0.48\textwidth}
        \centering
        \scalebox{0.45}{
        \input{newsv-mult/author/Book_Figures/rad_pattern_355GHz_55}}
        \caption{Beam at 5$^\circ$}
    \end{subfigure}
    \begin{subfigure}{0.48\textwidth}
        \centering
        \scalebox{0.45}{
        \input{newsv-mult/author/Book_Figures/rad_pattern_355GHz_66}}
        \caption{Beam at 7.5$^\circ$}
    \end{subfigure}
    \begin{subfigure}{0.48\textwidth}
        \centering
        \scalebox{0.45}{
        \input{newsv-mult/author/Book_Figures/rad_pattern_355GHz_77}}
        \caption{Beam at 10$^\circ$}
    \end{subfigure}
    \vspace{0.1cm}
    \caption{Anechoic chamber radiation patterns of the RIS at 3.55 GHz, demonstrating the codebook generation process. Each subfigure plots the normalized power after the first and second iterations of the greedy optimization procedure (Algorithm~\ref{algo:ris_config}). The results confirm that the main beam is successfully steered towards the desired Rx angle (with the peak occurring either exactly at the target or at the adjacent 2.5$^\circ$ offset bin), while also illustrating variations in sidelobe levels across different beam configurations.}
      \label{fig:anechoic_rad_patterns}
\end{figure}

Fig.~\ref{fig:anechoic_rad_patterns} shows the measured radiation patterns, in the anechoic chamber, for different target beam directions at a center frequency of 3.55 GHz. Each subplot corresponds to an RIS configuration optimized for a specific angle using the greedy iterative method described in Algorithm~\ref{algo:ris_config}. The results for both one and two full iterations of the optimization process are shown below. The plot shows the normalized received power as the Rx sweeps across its angular range. It is generally observed that the main lobe of each pattern is directed towards, or close to (small shifts of one bin i.e., 2.5$^\circ$), the intended target angle.

\begin{figure}[!p]
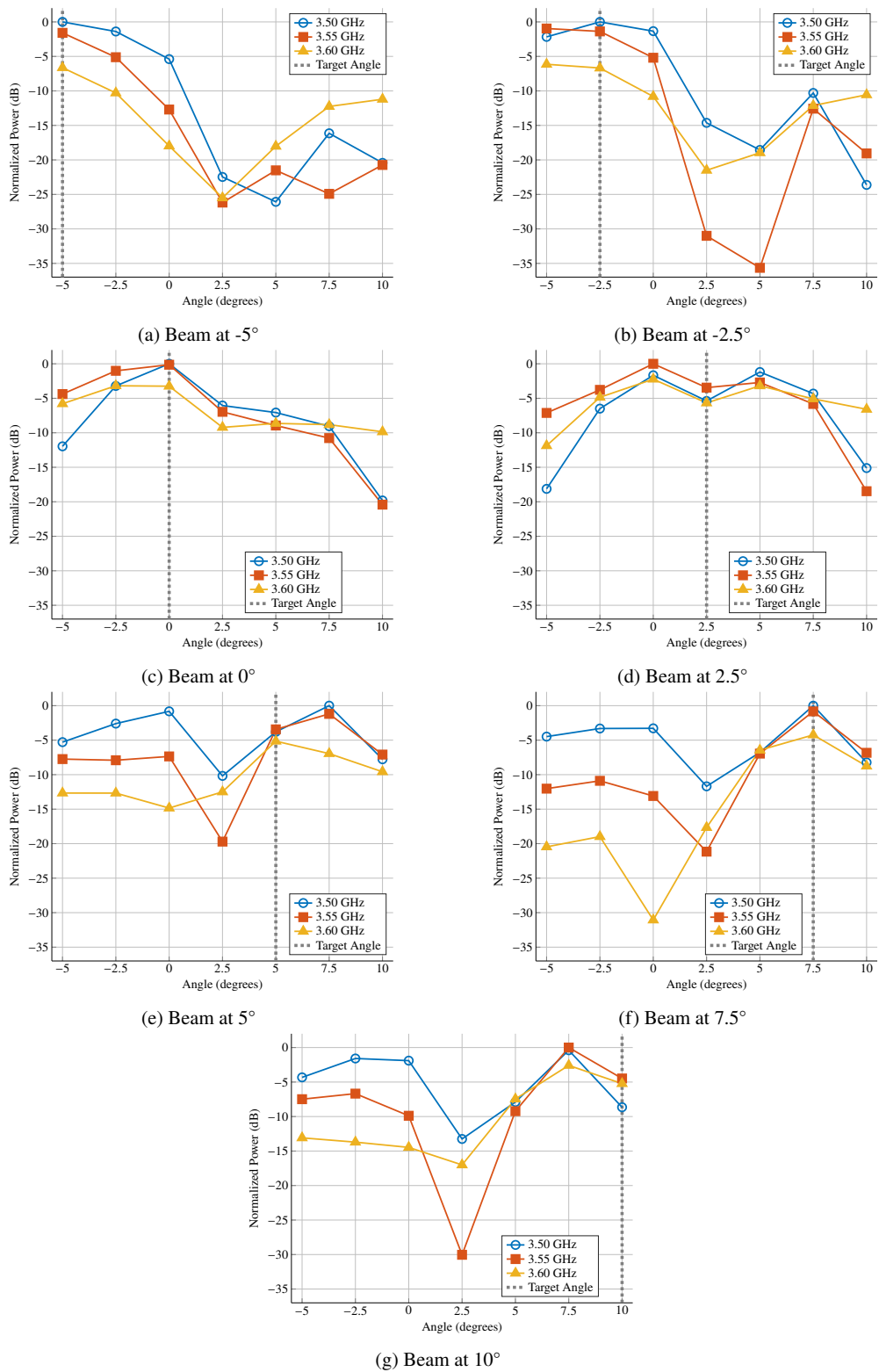

    \centering
    \begin{subfigure}{0.48\textwidth}
        \centering
        \scalebox{0.45}{
        \input{newsv-mult/author/Book_Figures/freq_selectivity_2_2}}
        \caption{Beam at -5$^\circ$}
    \end{subfigure}
    \begin{subfigure}{0.48\textwidth}
        \centering
        \scalebox{0.45}{
        \input{newsv-mult/author/Book_Figures/freq_selectivity_4_2}}
        \caption{Beam at -2.5$^\circ$}
    \end{subfigure}
    \begin{subfigure}{0.48\textwidth}
        \centering
        \scalebox{0.45}{
        \input{newsv-mult/author/Book_Figures/freq_selectivity_6_2}}
        \caption{Beam at 0$^\circ$}
    \end{subfigure}
    \begin{subfigure}{0.48\textwidth}
        \centering
        \scalebox{0.45}{
        \input{newsv-mult/author/Book_Figures/freq_selectivity_8_2}}
        \caption{Beam at 2.5$^\circ$}
    \end{subfigure}

    \begin{subfigure}{0.48\textwidth}
        \centering
        \scalebox{0.45}{
        \input{newsv-mult/author/Book_Figures/freq_selectivity_10_2}}
        \caption{Beam at 5$^\circ$}
    \end{subfigure}
    \begin{subfigure}{0.48\textwidth}
        \centering
        \scalebox{0.45}{
        \input{newsv-mult/author/Book_Figures/freq_selectivity_12_2}}
        \caption{Beam at 7.5$^\circ$}
    \end{subfigure}
    \begin{subfigure}{0.48\textwidth}
        \centering
        \scalebox{0.45}{
        \input{newsv-mult/author/Book_Figures/freq_selectivity_14_2}}
        \caption{Beam at 10$^\circ$}
    \end{subfigure}
    \vspace{0.1cm}
     \caption{Frequency Selectivity of RIS Radiation Patterns. Normalized received power, measured in an anechoic chamber, is shown at 3.5 GHz, 3.55 GHz, and 3.6 GHz to highlight the frequency-dependent behavior of the RIS radiation pattern.}
      \label{fig:anechoic_freq_selectivity}
\end{figure}

\begin{figure}[!p]
    \centering
    \begin{minipage}{0.48\textwidth}
        \centering
        \scalebox{0.45}{\input{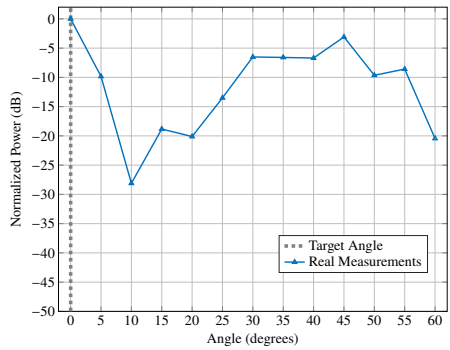}}
        \subcaption{Beam at $0^\circ$}
        \label{fig:ris_val0}
    \end{minipage}
    \begin{minipage}{0.48\textwidth}
        \centering
        \scalebox{0.45}{\input{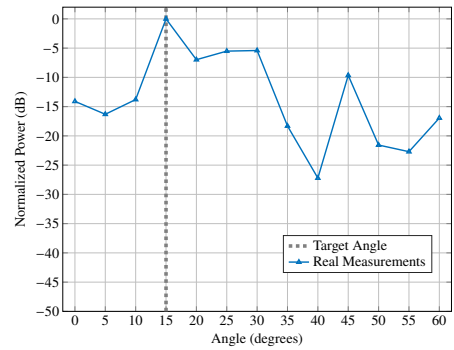}}
        \subcaption{Beam at $15^\circ$}
        \label{fig:ris_val15}
    \end{minipage}
    \begin{minipage}{0.48\textwidth}
        \centering
        \scalebox{0.45}{\input{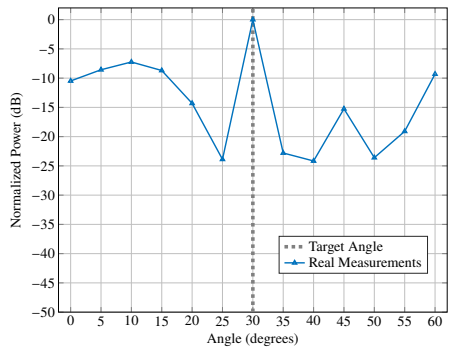}}
        \subcaption{Beam at $30^\circ$}
        \label{fig:ris_val30}
    \end{minipage}
    \begin{minipage}{0.48\textwidth}
        \centering
        \scalebox{0.45}{\input{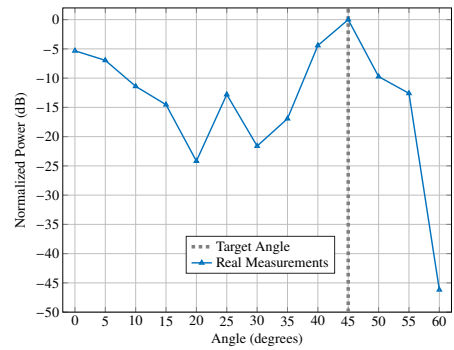}}
        \subcaption{Beam at $45^\circ$}
        \label{fig:ris_val45}
    \end{minipage}
    \begin{minipage}{0.48\textwidth}
        \centering
        \scalebox{0.45}{\input{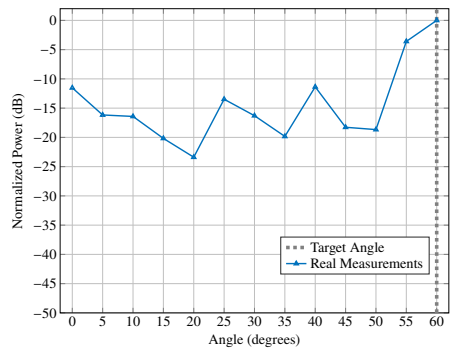}}
        \subcaption{Beam at $60^\circ$}
        \label{fig:ris_val60}
    \end{minipage}
    \caption{Measured Radiation Patterns in an Outdoor Environment at 3.55 GHz. The patterns correspond to RIS configurations optimized for different target directions ($0^\circ$, $15^\circ$, $30^\circ$, $45^\circ$, and $60^\circ$) using Algorithm~\ref{algo:ris_config}. The received power is normalized, and the maximum occurs at the receiver position aligned with the intended beam steering direction.}
    \label{fig:radpatterns_out}
\end{figure}

\begin{figure}[!t]
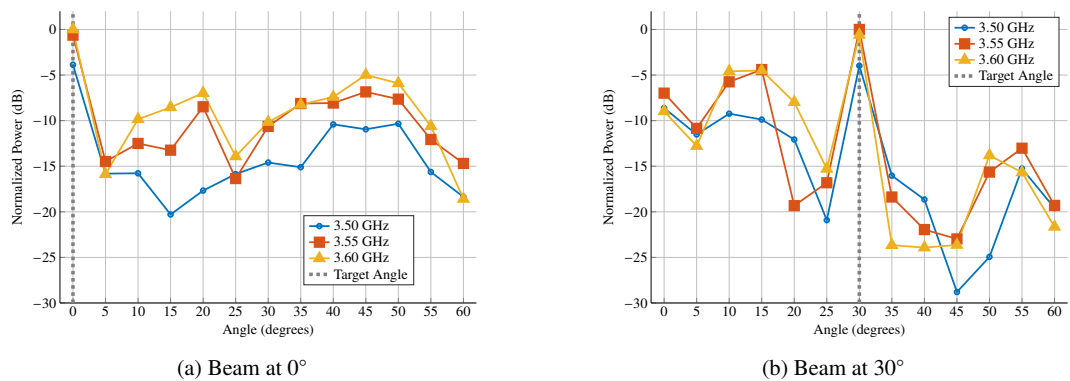

\centering
\begin{minipage}{0.48\textwidth}
    \centering
    \scalebox{0.45}{\input{newsv-mult/author/Book_Figures/freq_sel0_2}}
    \subcaption{Beam at $0^\circ$}\label{fig:freqsel0}
\end{minipage}%
\hfill
\begin{minipage}{0.48\textwidth}
    \centering
    \scalebox{0.45}{\input{newsv-mult/author/Book_Figures/freq_sel30_2}}
    \subcaption{Beam at $30^\circ$}\label{fig:freq_sel30}
\end{minipage}%
\caption{Frequency Selectivity of RIS Radiation Patterns. Normalized received power, measured in an outdoor environment, is shown at 3.5 GHz, 3.55 GHz, and 3.6 GHz to highlight the frequency-dependent behavior of the RIS radiation pattern. The results demonstrate that the angle direction of the main lobe is preserved across the frequency range, showcasing the effectiveness of the proposed codebook for beam sweeping in wideband applications. Variations are observed in the side lobes. Similar results in all Rx positions.}
\label{fig:outdoor_freq_selectivity}
\end{figure}

The frequency selectivity of these radiation patterns was also investigated by analyzing the patterns at different frequencies within the operational band (e.g., 3.5 GHz, 3.55 GHz, 3.6 GHz), as shown in Fig.~\ref{fig:anechoic_freq_selectivity}. While the main beam direction tends to be relatively stable across these frequencies for a given RIS configuration, variations are often observed in the sidelobe levels and null depths, a characteristic inherent to phase-quantized, frequency-dependent metasurfaces. Similar measurements were conducted in the outdoor environment. Fig.~\ref{fig:radpatterns_out} presents the outdoor radiation patterns, and Fig.~\ref{fig:outdoor_freq_selectivity} illustrates their frequency selectivity. 

The successful generation of steered beams and the characterization of their properties, including imperfections like sidelobes and frequency dependence, are critical steps. The derived codebooks and understanding of pattern behavior form the foundation for using the RIS in localization and sensing applications, such as AoA estimation, as discussed next.

\subsubsection{AoA Estimation}
\label{sec:aoa_estimation}

AoA estimation is a key task in localization and sensing. One way an RIS can assist in AoA estimation is by enabling beam sweeping: by sequentially applying different RIS configurations from a pre-computed codebook (as described in Section~\ref{sec:ris_radiation_pattern}), the positioning system effectively scans different angular directions. In the implemented setup, the transmitter sends a continuous signal toward the RIS, and the received power of the reflected signal is measured for each RIS configuration. The configuration that results in the highest received power indicates the estimated AoA of the incoming signal relative to the RIS. This method relies purely on received power measurements, making it suitable for scenarios where channel state information is unavailable or impractical to obtain. For the measurements presented in this subsection, the received signal power at the Rx is measured over a frequency range of 3.4 GHz to 3.6 GHz, sampled at 801 points using the VNA. The reported received power is typically normalized, with the maximum observed power set to 0 dB, to facilitate comparison across different RIS configurations and Rx positions.

\subsubsection*{Experimental Methodology}
The core methodology for AoA estimation via RIS-aided beam sweeping involved the following steps:
\begin{enumerate}
    \item \textbf{Codebook utilization:} The beamforming codebooks, generated by optimizing RIS configurations for various discrete angles in each environment (anechoic, outdoor, indoor) using the greedy algorithm (conceptualized in Algorithm~\ref{algo:ris_config}, were employed. Each configuration in the codebook is designed to steer the reflected beam towards a particular target angle.
    \item \textbf{Beam sweeping:} With the Rx antenna placed at a fixed, known ground-truth position (to evaluate accuracy), each RIS configuration from the codebook was applied sequentially. This effectively sweeps the RIS's main reflection lobe across the angular range covered by the codebook.
    \item \textbf{Power measurement:} The received signal power (e.g., integrated $S_{21}$ magnitude over the the 3.4--3.6 GHz band) was measured by the VNA for each applied RIS configuration.
    \item \textbf{AoA determination:} The AoA was estimated as the target angle associated with the RIS configuration that yielded the highest received power.
\end{enumerate}
The Tx antenna was fixed, illuminating the RIS. The Rx antenna was placed at various known ground-truth positions to assess estimation accuracy. All nodes (Tx, RIS, Rx) were typically maintained at the same height~(1.3~m).

\subsubsection*{Performance Evaluation in Anechoic Chamber}
\begin{figure*}[!p]
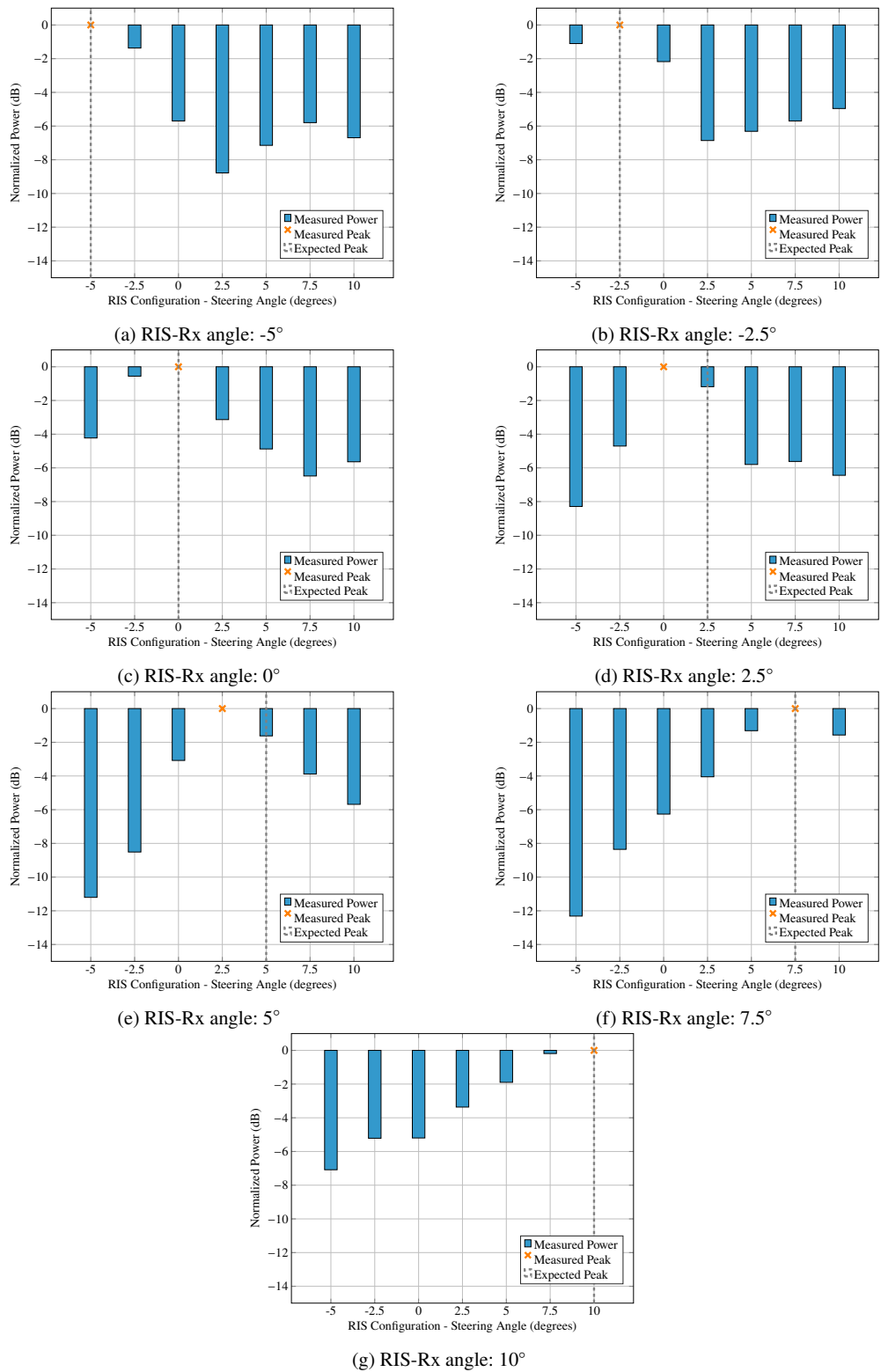
 
    \centering
    \begin{subfigure}{0.48\textwidth}
        \centering
        \scalebox{0.45}{
        \input{newsv-mult/author/Book_Figures/beam_sweeping_all_11}}
        \caption{RIS-Rx angle: -5$^\circ$}
    \end{subfigure}
    \begin{subfigure}{0.48\textwidth}
        \centering
        \scalebox{0.45}{
        \input{newsv-mult/author/Book_Figures/beam_sweeping_all_22}}
        \caption{RIS-Rx angle: -2.5$^\circ$}
    \end{subfigure}
    \begin{subfigure}{0.48\textwidth}
        \centering
        \scalebox{0.45}{
        \input{newsv-mult/author/Book_Figures/beam_sweeping_all_33}}
        \caption{RIS-Rx angle: 0$^\circ$}
    \end{subfigure}
    \begin{subfigure}{0.48\textwidth}
        \centering
        \scalebox{0.45}{
        \input{newsv-mult/author/Book_Figures/beam_sweeping_all_44}}
        \caption{RIS-Rx angle: 2.5$^\circ$}
    \end{subfigure}
    \begin{subfigure}{0.48\textwidth} 
        \centering
        \scalebox{0.45}{
        \input{newsv-mult/author/Book_Figures/beam_sweeping_all_55}}
        \caption{RIS-Rx angle: 5$^\circ$}
    \end{subfigure}
    \begin{subfigure}{0.48\textwidth}
        \centering
        \scalebox{0.45}{
        \input{newsv-mult/author/Book_Figures/beam_sweeping_all_66}}
        \caption{RIS-Rx angle: 7.5$^\circ$}
    \end{subfigure}
    \begin{subfigure}{0.48\textwidth}
        \centering
        \scalebox{0.45}{
        \input{newsv-mult/author/Book_Figures/beam_sweeping_all_77}}
        \caption{RIS-Rx angle: 10$^\circ$}
    \end{subfigure}
    \vspace{0.1cm} 
     \caption{Anechoic chamber: Beam sweeping results for AoA estimation. Each subfigure shows the normalized received power (computed across all the test frequencies from 3.4\,GHz to 3.6\,GHz) at a specific true Rx position. The RIS beam is swept using a codebook of configurations, where each configuration steers the beam toward a distinct angle in the set ${-5^\circ, -2.5^\circ, 0^\circ, 2.5^\circ, 5^\circ, 7.5^\circ, 10^\circ}$. The expected peak corresponds to the RIS-Rx azimuth angle, while the measured peak indicates the angle where the strongest received signal was observed.
     }
      \label{fig:beam_sweeping_allfreqs}
\end{figure*}

The anechoic chamber provides a controlled environment ideal for benchmarking the fundamental AoA estimation capabilities. For these tests, the Rx antenna was positioned at discrete angles ranging from $-5^\circ$ to $+10^\circ$ in $2.5^\circ$ steps relative to the RIS normal. The beam sweeping results are presented in Fig.~\ref{fig:beam_sweeping_allfreqs}. Each subfigure shows the normalized received power (computed across all test frequencies) at a specific true Rx position, as different RIS configurations (each targeting a specific angle from the codebook) are applied.

In most cases, the configuration yielding the maximum received power correctly corresponds to the true Rx angular position, indicating accurate AoA estimation within the $2.5^\circ$ resolution of the codebook. However, achieving such a fine angular resolution (e.g., $2.5^\circ$) with a 1-bit RIS can be challenging, as the resultant beams might not be sufficiently narrow or might have significant side-lobes (as also observed in the radiation patterns in Section~\ref{sec:ris_radiation_pattern}). Consequently, in some instances, a small estimation error of $2.5^\circ$ was observed, where the peak power occurred for an RIS configuration targeting an adjacent angular bin. This is partly attributed to the inherent trade-offs in 1-bit quantization and the potential for reflections from chamber surfaces not fully covered by electromagnetic radiation-absorbing cones, despite the generally controlled environment. Nevertheless, the results demonstrate the viability of the beam sweeping approach for AoA estimation in low-multipath conditions.

\subsubsection*{Performance Evaluation in Outdoor Environment}
To evaluate the performance of the system in a controlled, low-multipath environment, measurements were conducted in an outdoor setting, as illustrated in Fig.~\ref{fig:outdoor_setup_specific}. In this scenario, the Rx antenna was moved along a semicircular arc, covering azimuth angles from $0^\circ$ to $60^\circ$ in $5^\circ$ steps, while the Tx was fixed at an azimuth angle of $-15^\circ$. The distance between the Tx and the RIS, as well as the nominal RIS-to-Rx distance, was maintained at 8.5 meters.

\begin{figure*}[!t]
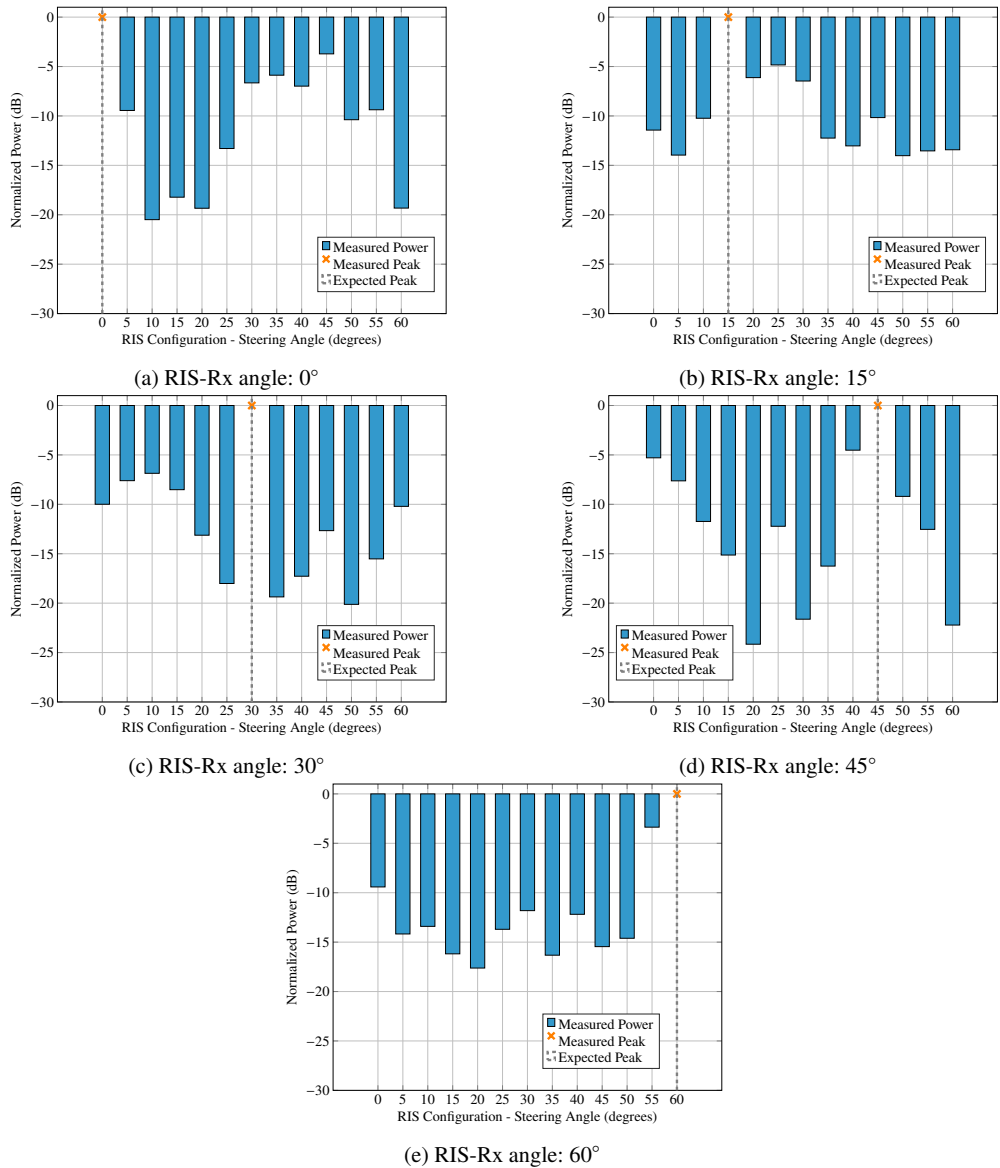

    \centering
    \begin{minipage}{0.48\textwidth}
        \centering
        \scalebox{0.45}{\input{newsv-mult/author/Book_Figures/beamsw_out0_2}}
        \subcaption{RIS-Rx angle: $0^\circ$}
        \label{fig:beamsw_out0}
    \end{minipage}
    \begin{minipage}{0.48\textwidth}
        \centering
        \scalebox{0.45}{\input{newsv-mult/author/Book_Figures/beamsw_out15_2}}
        \subcaption{RIS-Rx angle: $15^\circ$}
        \label{fig:beamsw_out15}
    \end{minipage}
    \begin{minipage}{0.48\textwidth}
        \centering
        \scalebox{0.45}{\input{newsv-mult/author/Book_Figures/beamsw_out30_2}}
        \subcaption{RIS-Rx angle: $30^\circ$}
        \label{fig:beamsw_out30}
    \end{minipage}
    \begin{minipage}{0.48\textwidth}
        \centering
        \scalebox{0.45}{\input{newsv-mult/author/Book_Figures/beamsw_out45_2}}
        \subcaption{RIS-Rx angle: $45^\circ$}
        \label{fig:beamsw_out45}
    \end{minipage}
    \begin{minipage}{0.48\textwidth}
        \centering
        \scalebox{0.45}{\input{newsv-mult/author/Book_Figures/beamsw_out60_2}}
        \subcaption{RIS-Rx angle: $60^\circ$}
        \label{fig:beamsw_out60}
    \end{minipage}
    \caption{Outdoor setup: Beam sweeping results for AoA estimation. Each subfigure shows the normalized received power at a specific true Rx position, as different RIS configurations (targeting angles $0^\circ$ to $60^\circ$ in $5^\circ$ steps) are applied. The expected peak corresponds to the RIS-Rx azimuth angle, while the measured peak indicates the angle where the strongest received signal was observed.}
    \label{fig:out_beamsw}
\end{figure*}

The results of the beam sweeping procedure for this outdoor configuration are shown in Fig.\ref{fig:out_beamsw}. The RIS codebook used for beam steering was precomputed using a greedy optimization algorithm (Algorithm\ref{algo:ris_config}), which selects RIS phase configurations to maximize signal power at each targeted Rx position. As detailed in Section~\ref{sec:ris_radiation_pattern}, this beamforming strategy leads to a pronounced main lobe directed towards the desired Rx position, achieving at least a 5 dB gain over the RIS configuration targeting the second-best direction for all tested Rx positions. This high directivity translated into perfect AoA estimation accuracy within the $5^\circ$ resolution of the codebook in this limited multipath outdoor setting.

\subsubsection*{Performance Evaluation in Indoor Environment}
The indoor environment (Fig.~\ref{fig:indoor_setup_specific}) poses a more significant challenge for AoA estimation due to strong multipath propagation from walls, floor, ceiling, and surrounding objects. In this setup, the Tx-RIS distance was 5.5 m, and the RIS-Rx distance was 8.5 m, with Rx positions ranging from $0^\circ$ to $45^\circ$ in $15^\circ$ steps (Tx fixed at $-15^\circ$).

\begin{figure*}[!t]
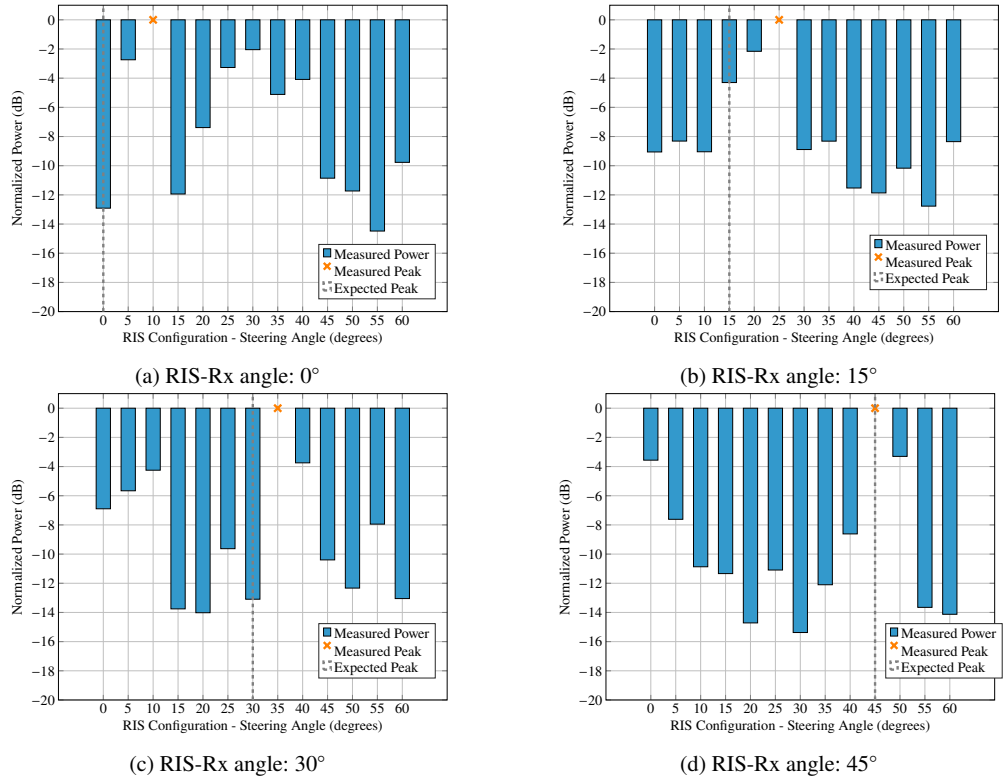

    \centering
    \begin{minipage}{0.48\textwidth}
        \centering
        \scalebox{0.45}{\input{newsv-mult/author/Book_Figures/indoor_with_out_0_2}}
        \subcaption{RIS-Rx angle: $0^\circ$}
        \label{fig:ind_beamsw_out0}
    \end{minipage}
    \begin{minipage}{0.48\textwidth}
        \centering
        \scalebox{0.45}{\input{newsv-mult/author/Book_Figures/indoor_with_out_15_2}}
        \subcaption{RIS-Rx angle: $15^\circ$}
        \label{fig:ind_beamsw_out15}
    \end{minipage}
    \begin{minipage}{0.48\textwidth}
        \centering
        \scalebox{0.45}{\input{newsv-mult/author/Book_Figures/indoor_with_out_30_2}}
        \subcaption{RIS-Rx angle: $30^\circ$}
        \label{fig:ind_beamsw_out30}
    \end{minipage}
    \begin{minipage}{0.48\textwidth}
        \centering
        \scalebox{0.45}{\input{newsv-mult/author/Book_Figures/indoor_with_out_45_2}}
        \subcaption{RIS-Rx angle: $45^\circ$}
        \label{fig:ind_beamsw_out45}
    \end{minipage}
    \caption{Indoor setup: Beam sweeping results for AoA estimation utilizing a pre-computed (non-adaptive) codebook. A maximum AoA estimation error of $10^\circ$ is observed in cases~\ref{fig:ind_beamsw_out0} and~\ref{fig:ind_beamsw_out15} due to multipath effects. The expected peak corresponds to the RIS-Rx azimuth angle, while the measured peak indicates the angle where the strongest received signal was observed.}
    \label{fig:indoor_with_out}
\end{figure*}

First, for this experiment, the performance was evaluated using a pre-computed codebook that had been generated in the outdoor setup (as detailed in Section~\ref{sec:ris_radiation_pattern}). The rationale for initially generating the codebook outdoors is that it allows testing over larger distances, which is not feasible in an anechoic chamber. The outdoor environment also offers inherently low multipath effects, enabling a clearer baseline characterization of the RIS beams.  Fig.~\ref{fig:indoor_with_out} illustrates the beam sweeping results when this outdoor-generated codebook is applied directly in the indoor scenario. The presence of strong multipath can significantly distort the perceived radiation patterns, leading to ambiguities where the maximum received power does not correspond to the true AoA. In these experiments, a maximum AoA estimation error of $10^\circ$ was observed when using such a static, outdoor-generated codebook. This level of error in multipath-rich indoor scenarios, when using non-adapted configurations, is consistent with findings reported in other RIS-related studies, both from ray-tracing based simulations~\cite{kompostiotis2024evaluation} and real-world FR2 setups~\cite{rahal2023ris}.

\begin{figure*}[!t]
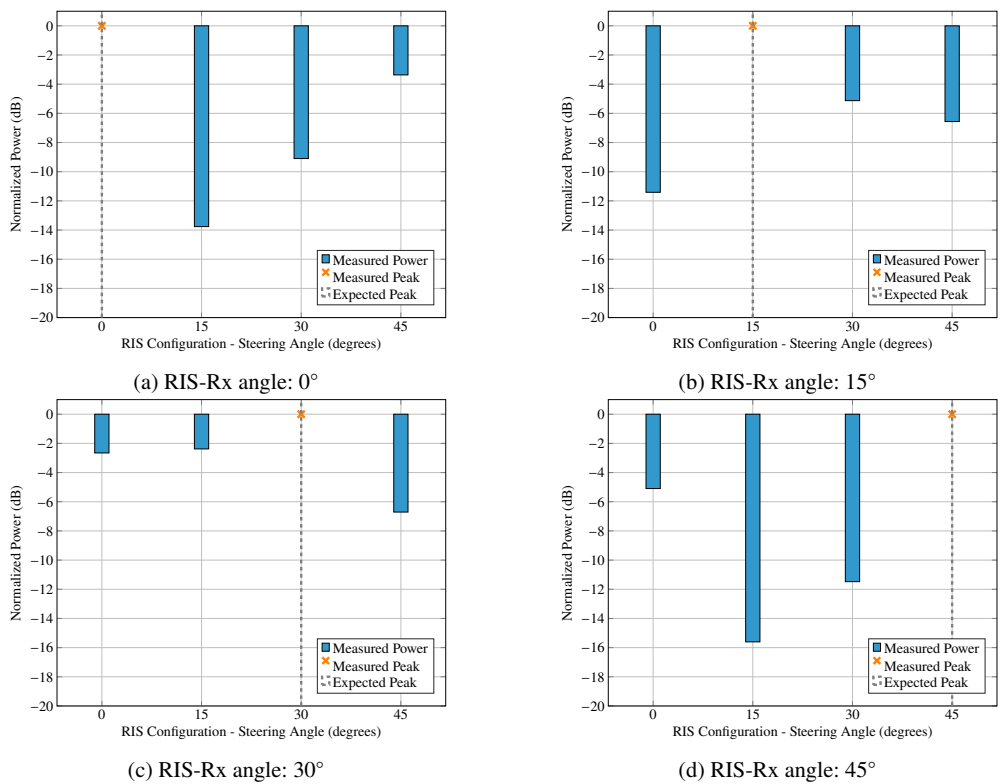

    \centering
    \begin{minipage}{0.48\textwidth}
        \centering
        \scalebox{0.45}{\input{newsv-mult/author/Book_Figures/indoorM1_0n_2}}
        \subcaption{RIS-Rx angle: $0^\circ$}
        \label{fig:ind_beamsw_maxin0}
    \end{minipage}
    \begin{minipage}{0.48\textwidth}
        \centering
        \scalebox{0.45}{\input{newsv-mult/author/Book_Figures/indoorM1_15n_2}}
        \subcaption{RIS-Rx angle: $15^\circ$}
        \label{fig:ind_beamsw_maxin15}
    \end{minipage}
    \begin{minipage}{0.48\textwidth}
        \centering
        \scalebox{0.45}{\input{newsv-mult/author/Book_Figures/indoorM1_30n_2}}
        \subcaption{RIS-Rx angle: $30^\circ$}
        \label{fig:ind_beamsw_maxin30}
    \end{minipage}
    \begin{minipage}{0.48\textwidth}
        \centering
        \scalebox{0.45}{\input{newsv-mult/author/Book_Figures/indoorM1_45n_2}}
        \subcaption{RIS-Rx angle: $45^\circ$}
        \label{fig:ind_beamsw_maxin45}
    \end{minipage}
    \caption{Indoor setup: Beam sweeping results for AoA estimation, where the RIS codebook was generated by executing two iterations of the greedy optimization algorithm (Algorithm~\ref{algo:ris_config}) in real time for each target angle. Perfect AoA estimation accuracy is achieved. The expected peak corresponds to the RIS-Rx azimuth angle, while the measured peak indicates the angle where the strongest received signal was observed.}
    \label{fig:m1_indoor}
\end{figure*}

To refine AoA estimation in such challenging indoor environments, a methodology employing real-time RIS configuration optimization was investigated. Specifically, the greedy algorithm (conceptualized in Algorithm~\ref{algo:ris_config}) was executed in real-time to generate a codebook tailored to the current indoor propagation conditions. As shown in Fig.~\ref{fig:m1_indoor}, this real-time execution of the algorithm (two iterations) significantly improves performance, achieving perfect AoA estimation accuracy within the $15^\circ$ resolution of the indoor tests. The real-time execution of the optimization algorithm allows the RIS to dynamically adapt its beamforming to the specific geometry and multipath signature of the indoor environment. This adaptive approach is crucial for robust AoA estimation in complex real-world scenarios, demonstrating a clear advantage over static codebook strategies.

In summary, these experiments underscore the potential of RIS-assisted beam sweeping for AoA estimation. Although effective in low-multipath scenarios, performance in rich multipath environments such as indoors benefits significantly from adaptive RIS configuration strategies that can account for the complex interactions of reflected signals and dynamically optimize beams.

\subsection{Conclusions}
This chapter has highlighted the fundamental role of RISs in enabling next-generation localization and sensing systems. RIS technology is expected to support a wide range of emerging applications that require high-precision positioning, including industrial automation, autonomous systems, and advanced remote healthcare. By providing programmable control over the radio propagation environment, RISs can overcome line-of-sight blockages, establish reliable reflection links, and create multiple distinguishable paths that enhance signal diversity and improve estimation accuracy. The performance of these systems is strongly affected by the physical design of the RIS, where factors such as aperture size, element spacing, and phase quantization levels determine the achievable resolution and determine the presence of potential ambiguities such as sidelobes and grating lobes. RISs further complement classical and modern sensing algorithms, such as MUSIC and ESPRIT, by boosting signal quality and SNR, while machine learning offers data-driven alternatives capable of capturing complex environmental interactions without requiring perfect physical models. Codebook-based beam sweeping is identified as a practical and robust control strategy, with experimental evidence showing that even simple 1-bit RIS architectures can generate effective beamforming codebooks for spatial scanning and mapping. Moreover, real-time measurement-driven optimization is shown to be essential in complex multipath conditions where static codebooks are insufficient. Finally, extensive experimental validation with physical prototypes underscores the practical feasibility of RIS-enabled localization and sensing, revealing implementation challenges, quantifying trade-offs, and confirming performance gains in realistic environments.

\begin{acknowledgement}
This work has been supported by the ESA Project PRISM:
RIS-enabled Positioning and Mapping (NAVISP-EL1-063).
\end{acknowledgement}

\bibliographystyle{IEEEtran}
\addcontentsline{toc}{section}{References} 
\bibliography{newsv-mult/author/references_trad}
\end{document}